\documentclass[review]{elsarticle}

\usepackage{amssymb}                       
\usepackage{amsmath}                       
\usepackage{amsthm}                        
\usepackage{listings}                      
\usepackage{lineno,hyperref}
\usepackage{enumitem}
\usepackage{caption}
\usepackage{textcase}
\modulolinenumbers[5]

\usepackage{float} 
\usepackage[T1]{fontenc} 

\journal{Computer Physics Communications}

\newcommand{\be}{\begin{equation}}
\newcommand{\ee}{\end{equation}}
\newcommand{\bi}{\begin{itemize}}
\newcommand{\ei}{\end{itemize}}

\newcommand{\beq}{\begin{equation}}
\newcommand{\eeq}{\end{equation}}

\begin{document}

\begin{frontmatter}

\title{Physics-based r-adaptive algorithms for high-speed flows and plasma simulations}

\author[vkiaddress]{Firas Ben Ameur\corref{mycorrespondingauthor}}
\cortext[mycorrespondingauthor]{Corresponding author}
\ead{firas.benameur@vki.ac.be}

\author[kuladdress]{Andrea Lani}
\ead{andrea.lani@kuleuven.be}

\address[vkiaddress]{Von Karman Institute for Fluid Dynamics, Waterloosesteenweg 72, 1640, Sint Genesius Rode, Belgium}
\address[kuladdress]{KU Leuven/Centrum voor mathematische Plasma-Astrofysica, Celestijnenlaan 200B, B-3001 Leuven, Belgium} 

\begin{abstract}
The computational modeling of high-speed flows (e.g. hypersonic) and space plasmas is characterized by a plethora of complex physical phenomena, in particular involving strong oblique shocks, bow shocks and/or shock waves boundary layer interactions. The characterization of those flows requires accurate, robust and advanced numerical techniques. To this end, adaptive mesh algorithms provide an automatic way to improve the quality of the numerical results, by increasing the mesh density where required in order to resolve the most critical physical features. In this work, we propose a r-adaptive algorithm that consists in repositioning mesh nodes as resulting from the solution of a physics-driven pseudo-elastic system of equations.
The developed mesh refinement techniques are based upon spring networks deriving from linear, semi-torsional and ortho-semi-torsional analogies, but driven by a combination of local physical and geometrical properties depending on a user-defined monitoring flow variable. Furthermore, a mesh quality indicator is developed within this work in order to grade and investigate the quality of an adapted mesh. Finally, a refinement stop indicator is proposed and demonstrated in order to further automatize the resulting adaptive simulation. 
All new physics-based mesh motion algorithms are illustrated through multiple examples that emphasize the applicability to different physical models and problems together with the improved quality of the results.

\end{abstract}


\end{frontmatter}


\section{Introduction}
\begin{figure}[H]
\centering
\includegraphics[width=.5\textwidth]{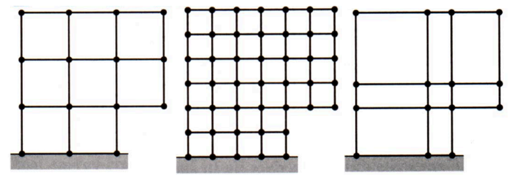}
\caption{Initial mesh (left), after h-refinement (middle), after r-refinement (right).}
\label{fig:AMR}
\end{figure}

High-speed flows (e.g. hypersonic flows \cite{Anderson}) and space plasmas \cite{Poedts} are typically characterized by strong shocks, shock/shock and/or shock/diffusion layers interactions. The numerical simulation of such flow problems may require extremely fine meshes over narrow regions of the physical domain in order to resolve the steep gradients occurring in the flow field. The high-gradient regions are not known to the analyst a priori. Thus, a-posteriori Adaptive Mesh Refinement (AMR) techniques represent a quite effective and established procedure to better capture the relevant flow features and to improve the overall quality of the numerical results. In particular, AMR allows for aligning grid cells with flow discontinuities (e.g. shocks, contact surfaces) in hypersonic flows \cite{Kleb2007} and for tackling the large disparity of scales (ranging from mega-meters to the ion and electron scales) within the same computational domain for space weather simulations \cite{Muller2011} , respectively, at the price of an increased algorithmic complexity. AMR is driven by physics-based sensors and can involve, h-refinement and/or r-refinement.
\begin{itemize}
    \item \textbf{h-refinement}\\
    The method consists of locally increasing the mesh resolution by adding or removing points, for instance via recursive cell subdivision or local re-meshing \cite{r-h-refinement}. This technique is relatively complex to implement, especially on unstructured grids and deeply affects the parallelization, requiring load balancing methods, for e.g the Dynamic Domain Decomposition \cite{Masaharu2013}, to keep a good performance and equidistribute the workload among the involved processors.
    \item \textbf{r-refinement}\\
    The r-refinement consists of repositioning the mesh points while keeping their number and connectivity frozen. This method is much more easily parallelizable than h-refinement and therefore it is highly desirable in large-scale simulations, since it naturally preserves the load balancing among processes \cite{whyr,mario}. While h-refinement is often used in hypersonic flow and astrophysical plasma applications, r-refinement is much less consolidated. This can be due to two main reasons: 
\begin{enumerate}
    \item Most hypersonic flow codes use cartesian meshes with high aspect ratio to improve the heat flux prediction and to reduce spurious entropy \cite{Kleb2007}, while r-refinement performs best on unstructured meshes (with triangles in 2D and tetrahedral in 3D).
    \item State-of-the-art r-refinement typically relies upon the solution of pseudo-elastic systems (associated to the given mesh) \cite{L}, requiring the use of efficient Linear System Solvers (LSS) and increasing the overall complexity of the method.
\end{enumerate}
\end{itemize}
Fig.\ref{fig:AMR} shows a comparison between the two approaches applied on a simple Cartesian grid.

In this work, we developed a novel, robust and efficient r-refinement algorithm in which the local physical characteristics are the main driver of the adaptation method. The resulting algorithm has been implemented into the COOLFluiD platform \cite{Kimpe,COOLFluiDAiaa}, a world-class open source framework for multi-physics modeling and simulations, particularly of hypersonic flows \cite{GaricanoHF,PanesiTCNEQ}, radiation \cite{DuarteMC}, laboratory \cite{ZhangLabo} and space plasmas \cite{lagunatwofluid,laguna2017effect,maneva2017multi,ALVAREZLAGUNA,lani2014gpu}. The selection of different monitor variables can help resolving different features in the final solution, according to the needs of the modeler (e.g. density or pressure). The developed AMR algorithm works on triangles, quadrilateral and tetrahedral cells, is fully parallel, implemented as a standalone module and totally physics-independent, letting the user decide which monitor physical quantity to use for driving the adaptation according to the application.
After giving an overview about the state-of-the-art r-refinement techniques in Sec.\ref{sec::stateoftheart}, a high-level description of the mesh adaptation algorithm is developed in Sec.\ref{sec:prob statement}. 
Details about the definition of the network of fictitious springs upon which the algorithm relies and the corresponding stiffness computations are given in Sec.\ref{sec:math}. Numerical results are presented in Sec.\ref{sec:results}, showing the good performance of the developed method on a variety of application scenarios. Finally, Sec.\ref{sec:MQI} and Sec.\ref{sec:RSI} propose and demonstrate novel mesh quality indicator and refinement stop indicator concepts respectively.

\section{State-of-the-art r-refinement}
\label{sec::stateoftheart}
R-refinement (a.k.a. mesh fitting) techniques are usually developed as error- or geometry-based. Blom \cite{L} investigates the linear spring analogy, first introduced by Batina \cite{Batina} by adding fictitious springs to the grid with stiffness chosen to be inversely proportional to the length of the supporting edge. Yet, he showed that the linear spring analogy frequently produces negative cell volumes and becomes unreliable when the mesh points undergo large displacements. Farhat \cite{T,farhat3D} proposes the torsional spring analogy to upgrade the linear spring analogy concept and to mitigate the appearance of invalid triangulation by adding torsional stiffness attached to each mesh vertex, in order to counterbalance the change of the angle at the vertex. This approach appears to be robust but complex especially in 3D AMR simulations. A simpler model is proposed by Zeng and Ethier \cite{ST}, i.e. the semi-torsional spring analogy for triangular and tetrahedral meshes, where the simplicity of the linear spring implementation is preserved and corrected by a factor reflecting the local geometrical properties of the triangular element. Finally, for 3D test cases, Markou \cite{OST} proposed the ortho-semi-torsional spring analogy forcing the validity of the tetrahedral element by preventing the corner vertex to cross the opposite face. Detailed reviews of multiple mesh deformation methods, advantages, disadvantages, and computational complexity can be found in \cite{joliT}.
\section{Problem statement}
\label{sec:prob statement}
Let $n$ $\in$ $\mathbb{N}$ be the number of the nodes in a mesh $\mathcal{M}$ and let $\textbf{P}=\{\textbf{P}_\textbf{1}, \textbf{P}_\textbf{2}...\textbf{P}_\textbf{n}\}$ be the set of the nodes positions inside $\mathcal{M}$ \footnote{Depending on the dimensions of the problem $\textbf{P}_\textbf{i}$=\{$x_i$; $y_i$\} or $\textbf{P}_\textbf{i}$=\{$x_i$; $y_i$; $z_i$\}}.\\
Let $\textbf{L}$ be the incidence matrix defined as in \cite{firasMS}:
\label{eq:Lij}
$$
    L_{ij}= \left\{
    \begin{array}{ll}
         1,  \mbox {   ~~~~   if nodes \textit{i} and \textit{j} are edge-connected}\\
         0,  \mbox {  ~~~~    otherwise.}
    \end{array}
\right.
$$\\
We want to equidistribute the mesh nodes according to a positive scalar function $W = W(x)$ to achieve an optimal mesh \cite{mario}. For the 1D case \cite{EulerLagrange}, between node positions $x_i$ and $x_{i+1}$ we have:
\begin{equation}
      \int_{x_{i}}^{x_{i+1}} W(x) dx = \text{constant}.
\end{equation}
For the multidimensional case, let \{$\textbf{P}_\textbf{i}$,$\textbf{P}_\textbf{j}$\} be a set of two nodes positions such that $L_{ij}=1$, and let $\textbf{r}(s)$ be the edge parametrization obeying to the Eq.\ref{eq:r}:
\begin{equation}
    \label{eq:r}
    \textbf{r}(s)=\textbf{P}_\textbf{i}+s(\textbf{P}_\textbf{j}-\textbf{P}_\textbf{i}),
\end{equation}
where $s \in [0,1]$.\\
Then, in order to equidistribute the mesh nodes, the line integral $I$, expressed in Eq.\ref{eq:LineIntegral}, must be constant:
\begin{equation}
\label{eq:LineIntegral}
  I=\int_0^1 W(\textbf{r}(s))\cdot r'(s) ds = \text{constant}
\end{equation}
Eq.\ref{eq:LineIntegral} is the solution of the Euler-Lagrange equation to the minimization of the energy which reads:
\begin{equation}
\label{eq:energy}
E_{ij}=L_{ij} \int_0^1 W(\textbf{r}(s)) (\textbf{P}_\textbf{j}-\textbf{P}_\textbf{i})^2 ds,
\end{equation}
where, the incidence matrix $\textbf{L}$ is artificially added ensuring the physical meaning of the energy function $E$.

\begin{proof}
The Euler-Lagrange equation \cite{EulerLagrange} may be written as:
\begin{equation}
   \left( \frac{\partial}{\partial \textbf{r}} - \frac{d}{ds} \left( \frac{\partial}{\partial \textbf{r}'} \right) \right)E =0.
\end{equation}
Using Eq.\ref{eq:r}, we obtain $\textbf{r}' = (\textbf{P}_\textbf{j}-\textbf{P}_\textbf{i})$. Hence, the energy equation may be re-written as: 
\begin{equation}
\label{eq:energy1}
E_{ij}=L_{ij} \int_0^1 W(\textbf{r}(s)) (\textbf{r}')^2 ds,
\end{equation}
and applying the chain rule:
\begin{align}
\begin{split}
 \frac{\partial E}{\partial \textbf{r}} &= \frac{\partial E}{\partial s} \frac{\partial s}{\partial \textbf{r} } \\
 &= \frac{\partial E}{\partial s} \frac{1}{\textbf{r}'} .
\end{split}
\end{align}

Therefore, after dropping the incidence matrix, the Euler-Lagrange equation can be expressed as:
\begin{align}
\begin{split}
 \frac{1}{\textbf{r}'}  \frac{\partial E}{\partial s} - \frac{d}{ds} \left( \frac{\partial E}{\partial \textbf{r}'} \right)  & =     \frac{1}{\textbf{r}'}  \frac{\partial }{\partial s} \left(\int_0^1 W(\textbf{r}(s)) (\textbf{r}')^2 ds \right)  - \frac{d}{ds} \left( \frac{\partial}{\partial \textbf{r}'}\left(\int_0^1 W(\textbf{r}(s)) (\textbf{r}')^2 ds \right)  \right) \\
  &= \frac{\partial}{\partial s} \left(\int_0^1 W(\textbf{r}(s)) (\textbf{r}') ds \right)  - \frac{d}{ds} \left(\int_0^1 2W(\textbf{r}(s)) (\textbf{r}') ds \right)\\
  & = \frac{d}{d s} \left(\int_0^1 W(\textbf{r}(s)) (\textbf{r}') ds \right)  - 2\frac{d}{ds} \left(\int_0^1 W(\textbf{r}(s)) (\textbf{r}') ds \right)\\
    & = - \frac{d}{d s} \left(\int_0^1 W(\textbf{r}(s)) (\textbf{r}') ds \right) = 0,
\end{split}
\end{align}
hence, $ \int_0^1 W(\textbf{r}(s)) (\textbf{r}') ds$  is a constant.
\end{proof}

Since we are considering a cell-centered Finite Volume method, the weight function $W$ can be considered constant between two edge-connected nodes, such that $W = W_{ij}$. Hence, the energy equation can be simplified into:
\begin{equation}
    \label{eq:itttt}
    E_{ij}=L_{ij} W_{ij} (\textbf{P}_\textbf{j}-\textbf{P}_\textbf{i})^2,
\end{equation}
which is analogous to the spring potential energy equation:
\begin{equation}
    \label{potentialEnergy}
    V =c^{t} k |\Delta \textbf{x}|^2,
\end{equation}
where $V$ is the potential energy, $k$ the spring stiffness, $|\Delta \textbf{x}|$ is the displacement. Algebraically identifying each term of the Eq.\ref{eq:itttt} compared Eq.\ref{potentialEnergy} leads to a stiffness coefficient of $W_{ij}$ and an equilibrium spring length set to zero.\\
The simplest optimization problem depends on finding the equilibrium positions between two adjacent nodes in the mesh $\mathcal{M}$ based on a network of springs \cite{pedro, firasMS}:
\begin{equation}
    \frac{\partial E}{\partial \textbf{P}}=0  ~~~~~~~ \& ~~~~~~~ \frac{\partial^2 E}{\partial \textbf{P}^2}>0.
\end{equation}
\subsection{Linear system assembly and solution} 
The optimization process of the nodes mesh positions is formulated through the assembly and solution of a linear system, including the following main algorithmic steps:

\begin{enumerate}
    \item The analytic Jacobian is defined as:
    \begin{equation}
        \frac{\partial E_{ij}}{\partial \textbf{P}_\textbf{i}}=-2L_{ij} W_{ij} (\textbf{P}_\textbf{j}-\textbf{P}_\textbf{i})=0.
    \end{equation}
    \item After simplifying the constant and collecting the contributions of each node, we obtain:
    \begin{equation}
        \sum_{j=1}^{n}L_{ij} W_{ij} (\textbf{P}_\textbf{j}-\textbf{P}_\textbf{i})=0.
    \end{equation}
    \item The resulting linear system can be expressed as:
    \begin{equation}
    \label{eq:AP=0}
        \textbf{AP}=0,
    \end{equation}
where
$$
  A_{ij}= \left\{
    \begin{array}{ll}
         -L_{ij} W_{ij},                \mbox {          $~~~~~~~~if  $ $ i\ne j $}\\
         \sum_{j=1}^{n} L_{ij} W_{ij},  \mbox {       $~~if  $ $ i=j$.}
    \end{array}
\right.
$$\\

\item Solving the linear system using an iterative solver, i.e. the Generalized Minimal RESidual (GMRES) algorithm complemented by a parallel Additive Schwartz Preconditioner as provided by the PETSc toolkit \cite{petsc1,petsc2,petsc3,petsc4}.   

\end{enumerate}
When the weight function $W_{ij}$ is a linear combination of the mesh node positions, the optimal solution can be found in a single step. However, in this work, the weight functions depend on both physical and geometrical variables,, thus being non-linear in space. In order to alleviate and overcome the nonlinear effects, we apply the following measures:  
\begin{itemize}
    \item The nodal positions of the mesh $\mathcal{M}$ are computed and updated every $m$ flow field iterations to limit the stiffness of the process and enable the stabilization of the flow field solution.
    \item An under-relaxation factor $\omega$, having an analogous behavior as a mesh velocity, is also added to the mesh adaptation solver to smooth the nodal displacement and to mitigate, for certain cases, the cells overlap. However, since the under-relaxation factor affects negatively the convergence rate, a trade-off between the flow solver convergence and and the pseudo-elastic convergence rate was sought and found in $\omega$ =$\mathcal{O}(10^{-2})$.
    \item $W_{ij} \ge 0$ is imposed in order to preserve the characteristic of a weight function and a stiffness coefficient.
\end{itemize}
As a result, the nodal re-positioning obeys to the following relation:
\begin{equation}
\label{eq:reposition}
\textbf{P}^{k+m}=(1-\omega)\textbf{P}^{k}+\omega \textbf{D},
\end{equation}
where \textbf{D} is the nodal displacement computed from Eq.(\ref{eq:AP=0}).

\subsection{Boundary Conditions}
Two types of the boundary conditions are defined:\\
-Dirichlet (i.e. locked node) where the node position is kept constant: $P_i^m$=$P_i^0$;\\
-Neumann (i.e. moving node in boundary) where only the tangential displacement is allowed, i.e. $\frac{\partial \textbf{P}_\textbf{i} \cdot \textbf{n}_\textbf{i}}{\partial \textbf{x}}=0 $, where $\textbf{n}_\textbf{i}$ is the boundary face normal vector.

\section{Numerical \& Mathematical formulation of the Spring Network}
\label{sec:math}
\subsection{Linear Spring analogy}
The weight function introduced in the Sec.\ref{sec:prob statement} is computed as:
\begin{equation}
\label{eq:k_lin}
    W_{ij}=|U_j-U_i|,
\end{equation}
where $U_i$ is a user-defined flow field state variable related to the node $i$, e.g. density or pressure.
The absolute value ensures the positivity of the weight function and guarantees the minimization of the system's potential energy. $W_{ij}$ in the Eq.\ref{eq:k_lin} is referred as a linear stiffness coefficient between two edge-connected nodes $i$ and $j$, denoted as $k_{ij}^{L}$.\\
During the simulation of extreme conditions, the mesh adaptation creates highly distorted cells due to the large node displacements and high physical gradients. Therefore, the linear spring coefficient needs to be truncated and bounded. The choice of the upper and lower bound values, referred respectively as the minimum percentile (minPer) and the maximum percentile (maxPer), are computed via a $P^2$ algorithm \cite{p2}. This dynamic method estimates the p-percentile as the observations are generated\footnote{for e.g. the median is 0.5-percentile}. The algorithm is independent of the size of the data since the method does not store information about the samples nor data as well as their sizes. Thus, this method requires a confined storage space. The percentile values allow for controlling the stability and the convergence rate of the flow solver.
\subsection{Issues related to the linear spring analogy}
\begin{figure}[H]
\centering
\includegraphics[width=.4\textwidth]{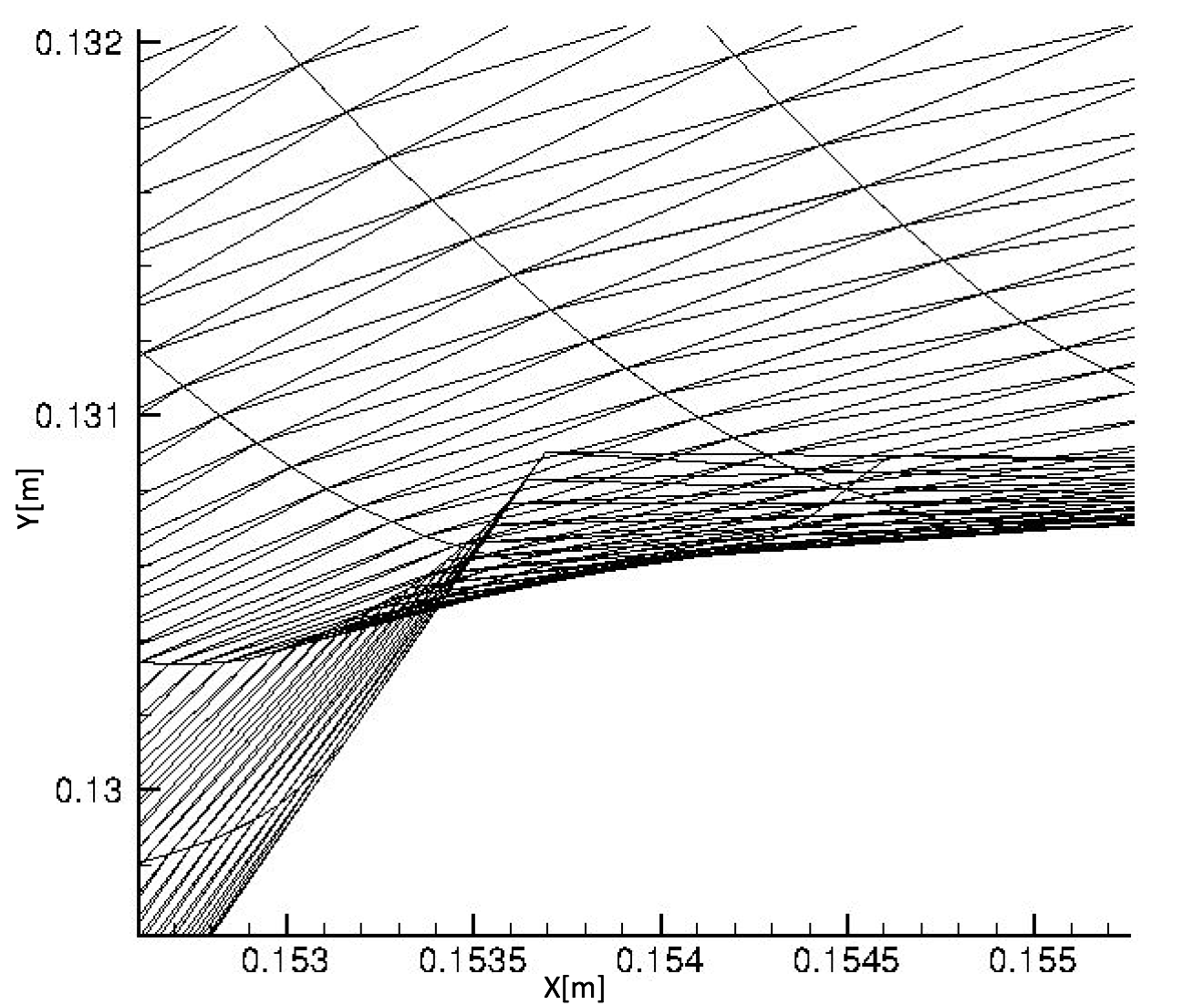}
\caption{Distorted mesh -- Issues related to linear spring analogy}
\label{fig:dist}
\end{figure}

A major drawback of the linear spring analogy appears when the mesh motions and deformations are of large amplitude leading to invalid elements (e.g. negative volumes, areas or grid lines crossovers) \cite{T, ST}, due essentially to the design behavior of a linear spring: the stiffness coefficient $k_{ij}^{L}$ between two neighbor nodes acts only in tension and compression along the connecting edge. Hence, when a mesh cell is experiencing an inversion or a near-inversion state, there is no geometric information about its angles, area (2D) or volume (3D), leading to a free movement of the node, possibly leading to node overlap and edge crossover. According to a solid analogy, we can consider the nodes as articulated ball joints, where there is no blocking momentum at each node. In order to illustrate issues which are related to the linear spring analogy, we consider what happens in the adapted mesh of an axisymmetric double cone test case (see Sec.\ref{sec:DC} for details on the configuration).
As shown in Fig.\ref{fig:dist}, the linear mesh refinement is not well adapted to handle high-aspect ratio meshes, leading to localized edge crossovers close the wall, inside the boundary layer region. 
\subsection{Torsional spring analogy}
\begin{figure}[H]
\centering{\includegraphics[scale=0.5]{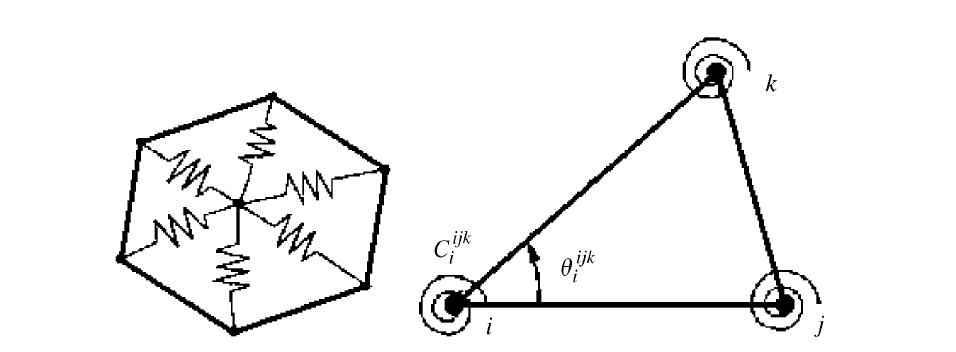}}
\caption{Torsional spring analogy \cite{T}}
\label{fig:triangleTors}
\end{figure}

The linear spring analogy concept can be upgraded by introducing, in the dynamic mesh, a vertex-attached torsional spring in order to add angular momentum. The torsional spring concept will strongly mitigate, by means of local geometrical information, the inversion or near-inversion of the elements \cite{T}.\\
Let $\mathcal{T}_{ijk}$ denotes a triangle and let $\theta_i^{ijk}$ the angle between the edges $ij$ and $ik$ inside $\mathcal{T}_{ijk}$ (see Fig.\ref{fig:triangleTors}). Therefore, the attached $i$-vertex torsional spring coefficient $C_{i}^{ijk}$ is expressed as:
\begin{equation}
    \label{eq:C}
    C_{i}^{ijk} = \frac{1}{sin^2(\theta_i^{ijk})}.
\end{equation}
Eq.\ref{eq:C} conserves the validity of the element, i.e
\begin{equation}
    \text{If} \quad \theta_i^{ijk} \rightarrow \text{0 or} ~ \pi  \Rightarrow \quad C_{i}^{ijk}  \rightarrow \infty 
\end{equation}

Let $N$ denotes the number of the mesh elements attached to the vertex $i$. The torsional spring constant coming from each triangle $\mathcal{T}$ connected to the vertex $i$, contributes to the overall stiffness. Therefore, the torsional spring stiffness $C_{i}$ attached to each vertex $i$ becomes:
\begin{equation}
    \label{eq:Ctot}
    C_{i} = \sum_{m=1}^{N}\frac{1}{sin^2(\theta_i^{m})},
\end{equation}
\cite{ST} shows that this model is expensive regarding memory cost and computational time, especially for 3D simulation. In fact, within this spring concept, the torque system resulting from torsional springs associated to each vertex needs to be transformed into linear forces on nodes in order to compatible with the linear spring analogy and to contribute to the edge global stiffness. In addition, \cite{joliT} shows that the complexity of the torsional spring method, i.e. $\mathcal{O}(n_e^3+n_v^3)$, is mush higher that the linear one, i.e. $\mathcal{O}(n_e^3)$, where $n_e$ and $n_v$ are the number of edges and vertices of the considered mesh. Hence, a simpler model is embraced and introduced in the following.

\subsection{Semi-torsional spring analogy}
\label{sec:semi}
\subsubsection{Mathematical formulation}
This model is based on adding a correction factor to the existing linear spring stiffness coefficient $k^L$ proportional to the area of the triangular mesh element, denoted $k^{ST}$.
The total stiffness of the mesh network related to each edge $ij$ will be \cite{ST}:
\begin{equation}
    k_{ij}=k_{ij}^{L}+k_{ij}^{ST},
\end{equation}
and
\begin{equation}
\label{eq:kST}
    k_{ij}^{ST}= \textsc{p}\sum_{m=1}^{N} \frac{1}{sin^2(\theta_{ij}^m)},
\end{equation}
where \textsc{p} denote a user-defined parameter, $N$ the number of elements attached to the edge $ij$ and $\theta_{ij}$ the angle facing the edge $ij$ as well.\\
\subsubsection{Including the physics}
The mesh r-adaptive algorithms are physics-based. The flow field state variables define the linear stiffness coefficients. Therefore, the formulation of the semi-torsional stiffness must incorporate both physical and geometrical properties. Hence, the factor \textsc{p} will be function of the local physical characteristics.
\subsubsection{2D formulation}
\begin{figure}[H]
\centering{\includegraphics[scale=0.6]{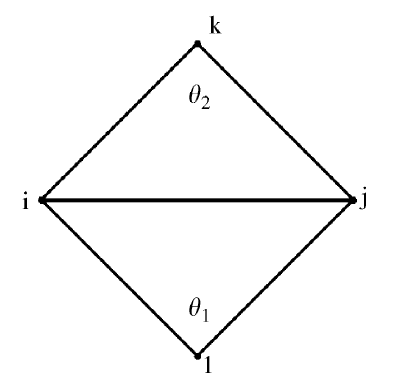}}
\caption{Semi-torsional analogy: 2D triangular case \cite{ST}}
\label{fig:semi}
\end{figure}

For the 2D case, the expression of the semi-torsional spring coefficient becomes:
\begin{equation}
    k_{ij}^{ST}=\textsc{p} \left( \frac{1}{sin^2(\theta_1)}+\frac{1}{sin^2(\theta_2)} \right),
\end{equation}
where $\theta_1$ and $\theta_2$ are the angles defined in Fig.\ref{fig:semi}.
A simpler computation of the $k_{ij}^{ST}$ is based on the following expression:
\begin{equation}
    \label{eq:kSTsimple}
        k_{ij}^{ST}= \textsc{p} \left(\frac{l_{kj}^2 l_{ki}^2}{4 A_{ijk}^2}+\frac{l_{lj}^2 l_{li}^2}{4 A_{ijl}^2}\right),
\end{equation}
where $l_{ij}$ is the distance between nodes $i$ and $j$ and $A_{ijk}$ is the area of the triangular element $ijk$ computed thought the cross product using the formula:
\begin{equation}
A_{ijk} = \frac{1}{2} ||\vec{ki} \times \vec{kj}||.
\end{equation}
\subsubsection{3D formulation}
The probability of creating negative cell volumes increases in the case of the 3D tetrahedral elements since the vertex corner can easily cross the opposite face. The idea was to generalize the semi-torsional spring analogy to be applied to tetrahedral elements \cite{ST}. 
The concept is based on inserting a triangle inside the tetrahedral cell as shown in Fig.\ref{fig:STanalogy3Dtetra1}. This triangle will be the start point of computing the $k^{ST}$. Eq.\ref{eq:kST} is still valid where the angle $\theta_{ij}^m$ is the angle facing the edge as presented in Fig.\ref{fig:STanalogy3Dtetra}:

\begin{figure}[H]
    \captionsetup{justification=centering}
\centering
\begin{minipage}{.42\linewidth}
  \includegraphics[width=\linewidth]{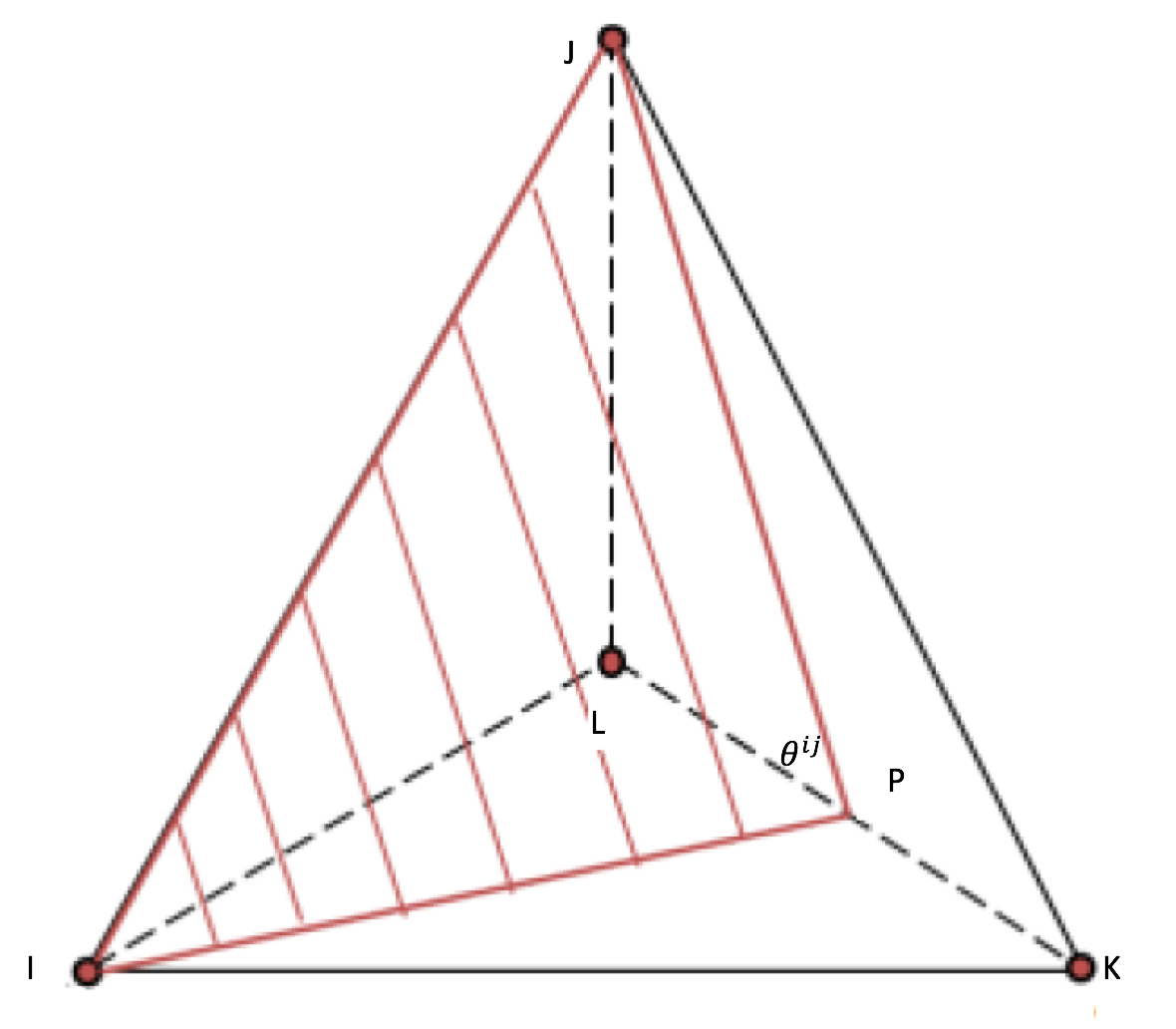}
  \caption{Inserted triangle \cite{joliT}}
  \label{fig:STanalogy3Dtetra1}
\end{minipage}
\hspace{.05\linewidth}
\begin{minipage}{.30\linewidth}
  \includegraphics[width=\linewidth]{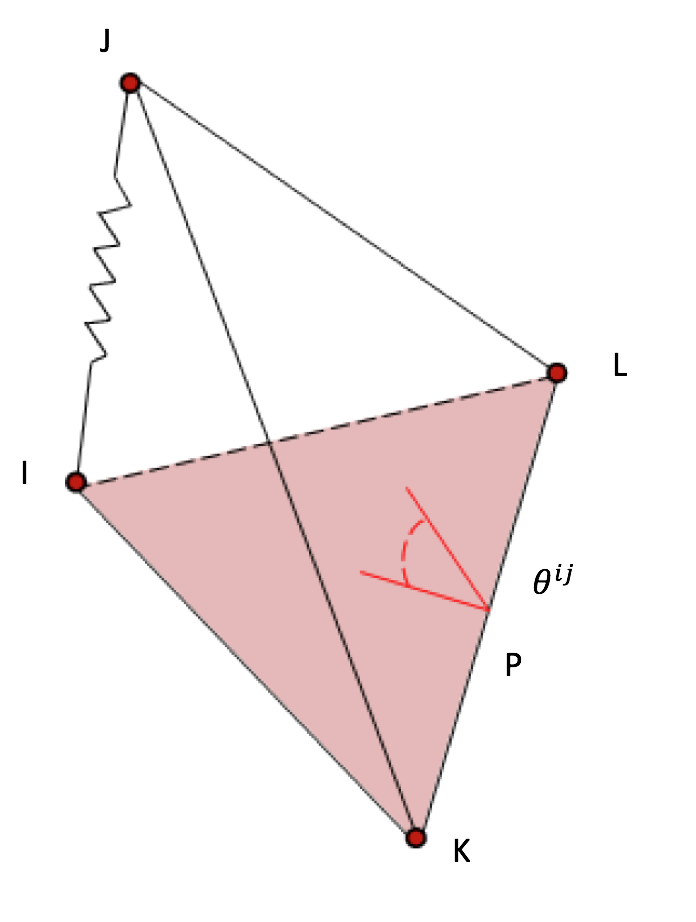}
  \caption{Facing edge angle definition \cite{joliT} }
    \label{fig:STanalogy3Dtetra}
\end{minipage}
\end{figure}

Eq.\ref{eq:3Dksemi} expresses the semi-torsional spring constant within the cell $\mathcal{H}_{m}$ attached to the edge $ij$:
\begin{equation}
\label{eq:3Dksemi}
k^{ST}_{ij}=\textsc{p} \frac{d_{jp}^2 d_{ip}^2}{A_{ijp}^2}.
\end{equation}

\subsection{Ortho-semi-torsional spring analogy}
\begin{figure}[H]
\centering{\includegraphics[scale=0.5]{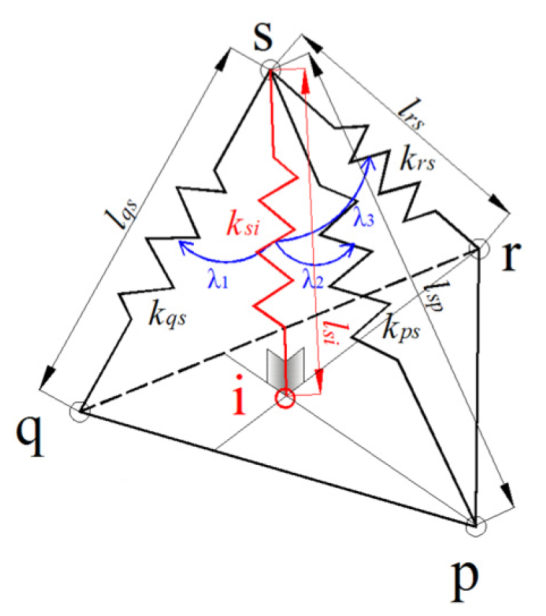}}
\caption{Ortho-semi-torsional spring analogy for 3D tetrahedral mesh \cite{OST}}
\label{fig:OSTtetra}
\end{figure}

For some 3D test cases, the stiffness network provided by the semi-torsional spring coefficients is not sufficient and need to be upgraded. A proposed solution is to use the ortho-semi-torsional spring analogy \cite{OST}. Therefore, the stiffness of an edge $qs$ is described as:
\begin{equation}
\label{eq:TOT}
k_{qs}^{total}=k^{OST}_{qs}+k^{ST}_{qs}+k^{L}_{qs}.
\end{equation}
The goal of this concept is to construct an additional fictitious spring. Therefore,  the mesh stiffness will increase and ensure the validity of the elements. 
Let $i$ be the projection of the vertex corner $s$ on the opposite face (see Fig.\ref{fig:OSTtetra}).
The projection forms geometry based linear springs $k_{si}=\frac{1}{d_{si}}$ and $k_{qi}=\frac{1}{d_{qi}}$, where $d_{\alpha \beta}$ denotes the distance between the point $\alpha$ and $\beta$.\\

The contribution of $k_{si}$ to the edge $qs$ is computed through the following procedure:
\begin{itemize}
    \item compute $d_{tot}=d_{qs}+d_{ps}+d_{rs}$ and $\lambda_{si} = \lambda_{qi} = \frac{d_{qs}}{d_{tot}}$, the linear allocation parameter,
    \item compute $k^{OST}$ according to the following relation:
\end{itemize}
\begin{equation}
\label{eq:OST}
    k^{OST}=\textsc{p}_1 \left(\frac{k_{si}}{\lambda_{si}^\textsc{a}}+\frac{k_{qi}}{\lambda_{qi}^\textsc{a}}\right)^\textsc{b},
\end{equation}
where the constants \textsc{a} and \textsc{b} affect the contribution of $k^{OST}$ to the global stiffness network and $\textsc{p}_1$ will incorporate physical characteristics of the flow field. Choosing $\textsc{a}=\textsc{b}=1$, Eq.\ref{eq:OST} is be transformed into:
\begin{equation}
\label{eq:kostFinal}
k^{OST}=\textsc{p}_1 \left(\frac{k_{si}}{\lambda_{si}}+\frac{k_{qi}}{\lambda_{qi}}\right).
\end{equation}

\subsection{Connectivity information}
We computed and stored the connectivity information, i.e. identifying the edge connected nodes, once and for all within \texttt{std::multimap} during the setup phase of the solver in order to save memory storage and computational time since, in the r-adaptive method, a node's connectivity does not change.
Multimaps are specific containers that can store information so that multiple values can be associated to the same key \cite{multimap}. While providing more flexibility and potentially less memory requirements than corresponding multi-dimensional arrays (with variable row size, as required by our problems), , the major drawback of multimaps is that a binary search algorithm needs to be used for accessing entries instead of constant-time access which could be provided by a multi-dimensional arrays.

\section{Results}
\label{sec:results}
The application of the newly developed physics-based AMR for 2D and 3D cases are presented in this section on the following representative test cases:
\begin{enumerate}
    \item Steady Euler 2D flow: Double Wedge channel flow, triangular mesh.
    \item Steady viscous thermo-chemical non-equilibrium (TCNEQ) 2D flows :
    \begin{itemize}
        \item Double Cone, triangular mesh. 
        \item Hornung Cylinder, quadrilateral mesh.
    \end{itemize}
    \item Steady Euler 3D flow: Hemisphere, tetrahedral mesh.
    \item Magneto Hydro-Dynamics (MHD):
    \begin{itemize}
        \item Unsteady Rotor, 2D triangular mesh.
        \item Steady Solar Wind, 3D tetrahedral mesh.
    \end{itemize}
\end{enumerate}

In this section, three tables are presented for each test case, summarizing:
\begin{enumerate}[label=(\alph*)]
    \item The flow conditions (e.g. free stream, wall temperature in viscous cases);
    \item The mesh characteristics and boundary conditions (BC);\label{pt:2b}
    \item the main settings for the r-adaptation algorithm. 
\end{enumerate}
Moreover, snapshots of the computational domains are also provided. Herein, the numbers on each boundary surface define the  corresponding BC which is applied, as listed in the table \ref{pt:2b}.

\subsection{Wedge channel flow}
\label{sec:DW}
\begin{figure}[H]
\centering{\includegraphics[scale=0.4]{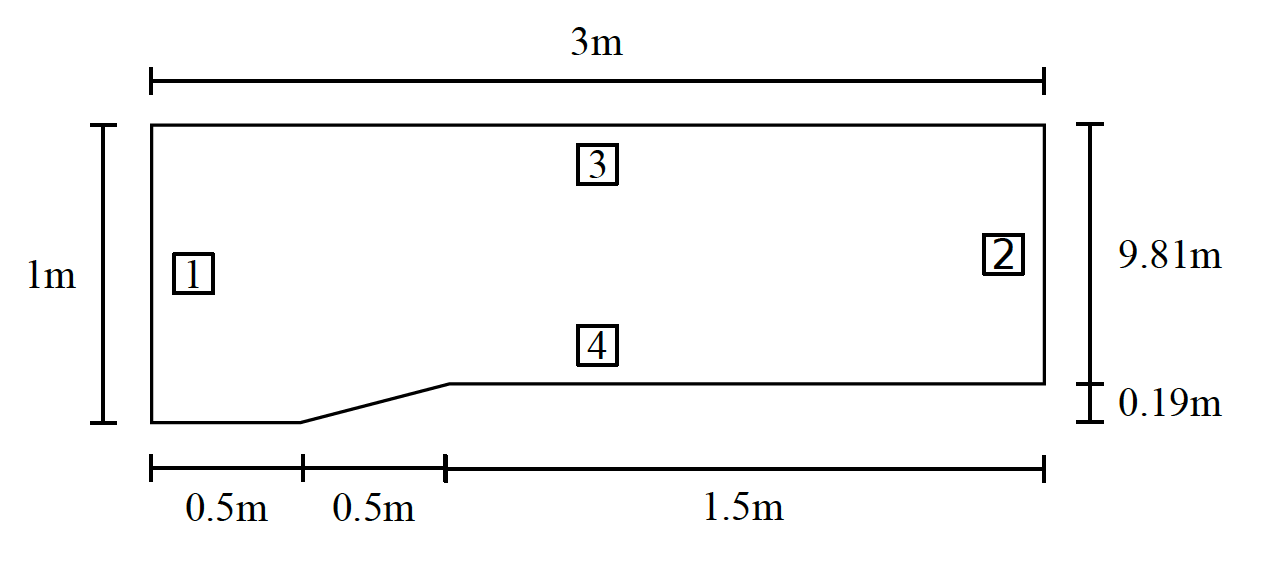}}
\caption{2D double wedge geometry}
\label{fig:DWgeometry}
\end{figure}

\begin{figure}[H]
\centering{\includegraphics[scale=0.25]{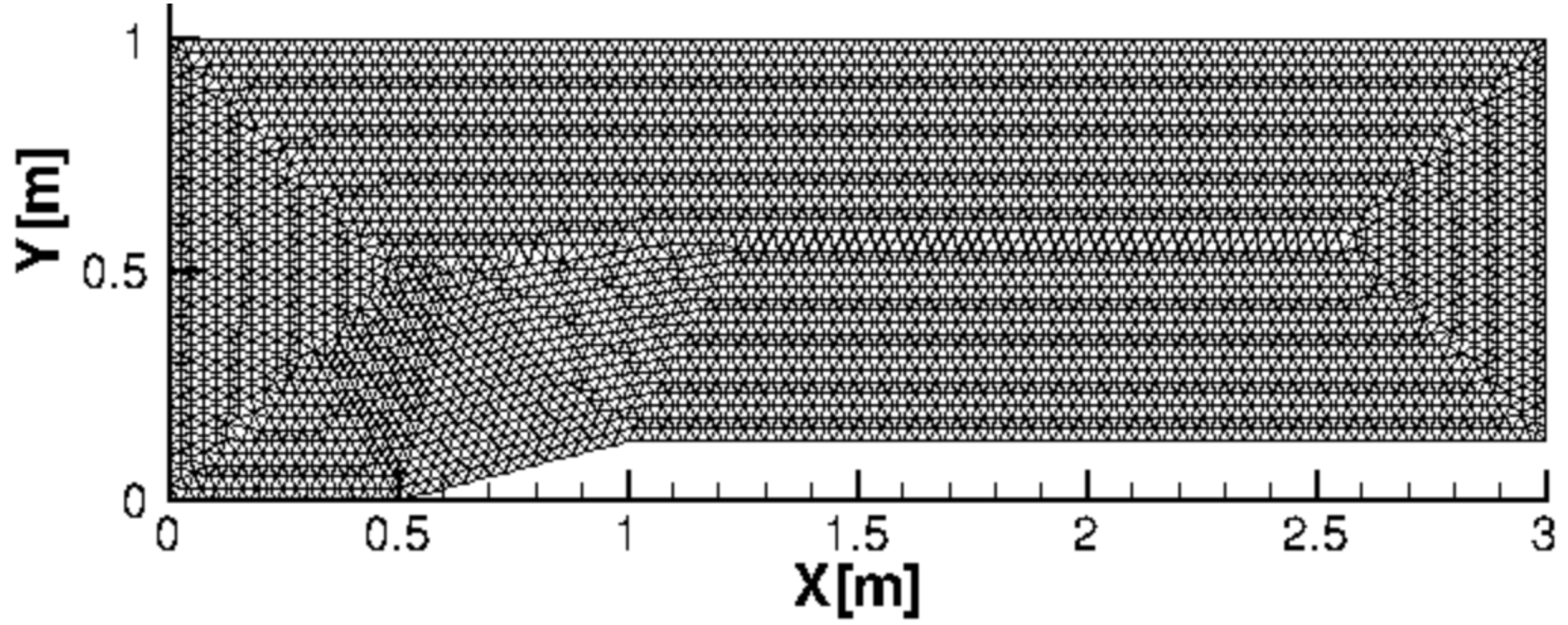}}
\caption{Double wedge  -- initial mesh}
\label{fig:DWinitmesh}
\end{figure}

The 2D supersonic double wedge channel flow test case conditions are presented in Tab.\ref{tab:DWflowchar}, Tab.\ref{tab:DWmeshchar} and Tab.\ref{tab:DWAMR}, while the test case definition and the corresponding unstructured mesh are shown in Fig.\ref{fig:DWgeometry} and Fig.\ref{fig:DWinitmesh} respectively. 

\begin{table}[H]
\centering
\caption{Double wedge -- Flow characteristics}
\label{tab:DWflowchar}
\begin{tabular}{|cccccc|}
\hline
 \footnotesize{Physical Model} & \footnotesize{M} & \footnotesize{$\rho$ [-]} & \footnotesize{$\rho$u [-]} & \footnotesize{$\rho$v [-]} & \footnotesize{$\rho$E [-] }   \\
\footnotesize{Perfect gas} &  \footnotesize{2}    & \footnotesize{1}   & \footnotesize{2.36643}  & \footnotesize{0}   & \footnotesize{5.3}\\
\hline
\end{tabular}
\end{table}

\begin{table}[H]
\centering
\caption{Double wedge -- Mesh characteristics}
\label{tab:DWmeshchar}
\begin{tabular}{|ccccccc|}
\hline
 \footnotesize{Dimensions} & \footnotesize{Type} & \footnotesize{\# Elements} & \footnotesize{BC 1} & \footnotesize{BC 2} & \footnotesize{BC 3} & \footnotesize{BC 4}   \\
\footnotesize{2D} &  \footnotesize{Triangular}    & \footnotesize{6871}   & \footnotesize{Inlet}  & \footnotesize{Outlet}   & \footnotesize{Symmetry} & \footnotesize{no-slip wall}\\
\hline
\end{tabular}
\end{table}

\begin{table}[H]
\centering
\caption{Double wedge -- r-refinement}
\label{tab:DWAMR}
\begin{tabular}{|cccccc|}
\hline
\footnotesize{
 Spring Network} &\footnotesize{Monitor Variable} & \footnotesize{Process Rate} & \footnotesize{Stop AMR Iteration} & \footnotesize{minPer} & \footnotesize{maxPer}    \\
 \footnotesize{Linear}  & \footnotesize{Density} &\footnotesize{20} & \footnotesize{7000} & \footnotesize{0.20}  & \footnotesize{0.65}   \\
\hline
\end{tabular}
\end{table}

\begin{figure}[H]
\centering{\includegraphics[scale=0.25]{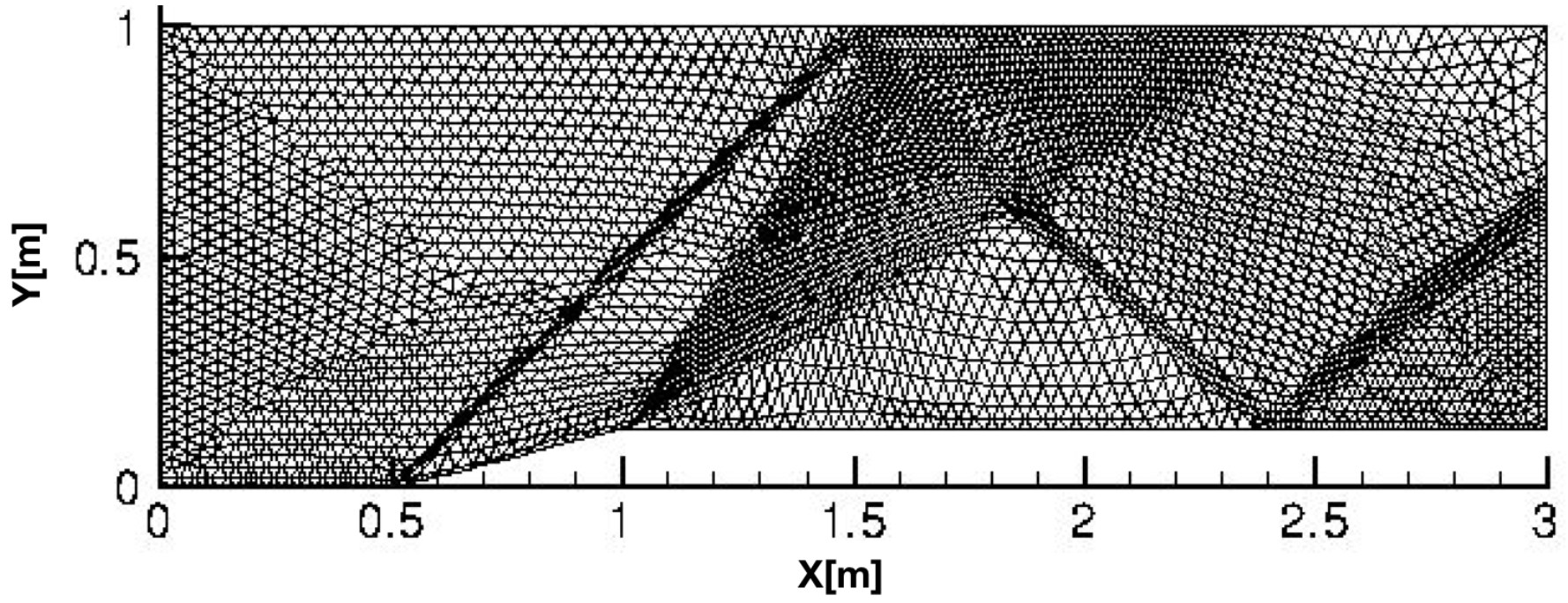}}
\caption{Double wedge  -- final mesh}
\label{fig:DWfinalmesh}
\end{figure}

As shown in Fig.\ref{fig:DWfinalmesh}, in the final adapted solution, the oblique shock, the expansion wave and their reflections are perfectly resolved.

\subsection{Double cone}
\label{sec:DC}

\begin{figure}[H]
    \begin{minipage}[t]{6cm}
        \centering
        \includegraphics[width=5.5cm]{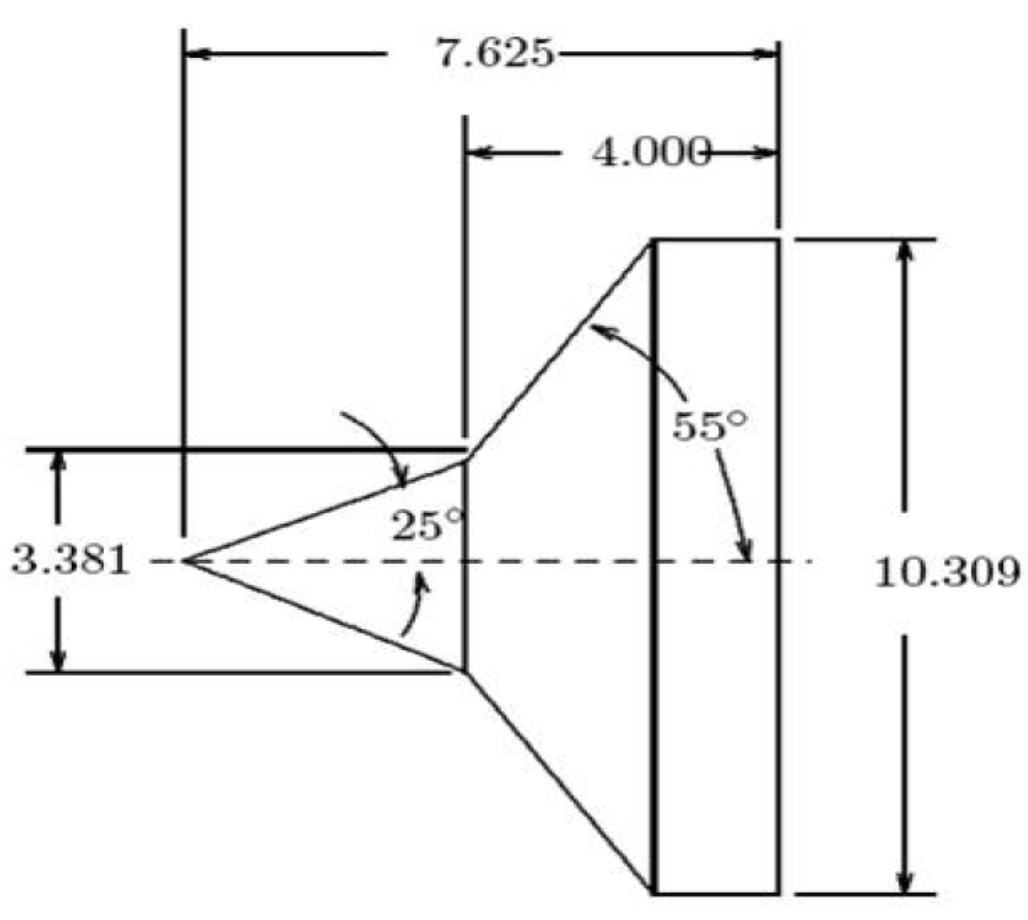}
        \caption{DC geometry - units: 'inches'}
        \label{fig:doublecone}
    \end{minipage}
    \begin{minipage}[t]{6cm}
        \centering
        \includegraphics[width=5.5cm]{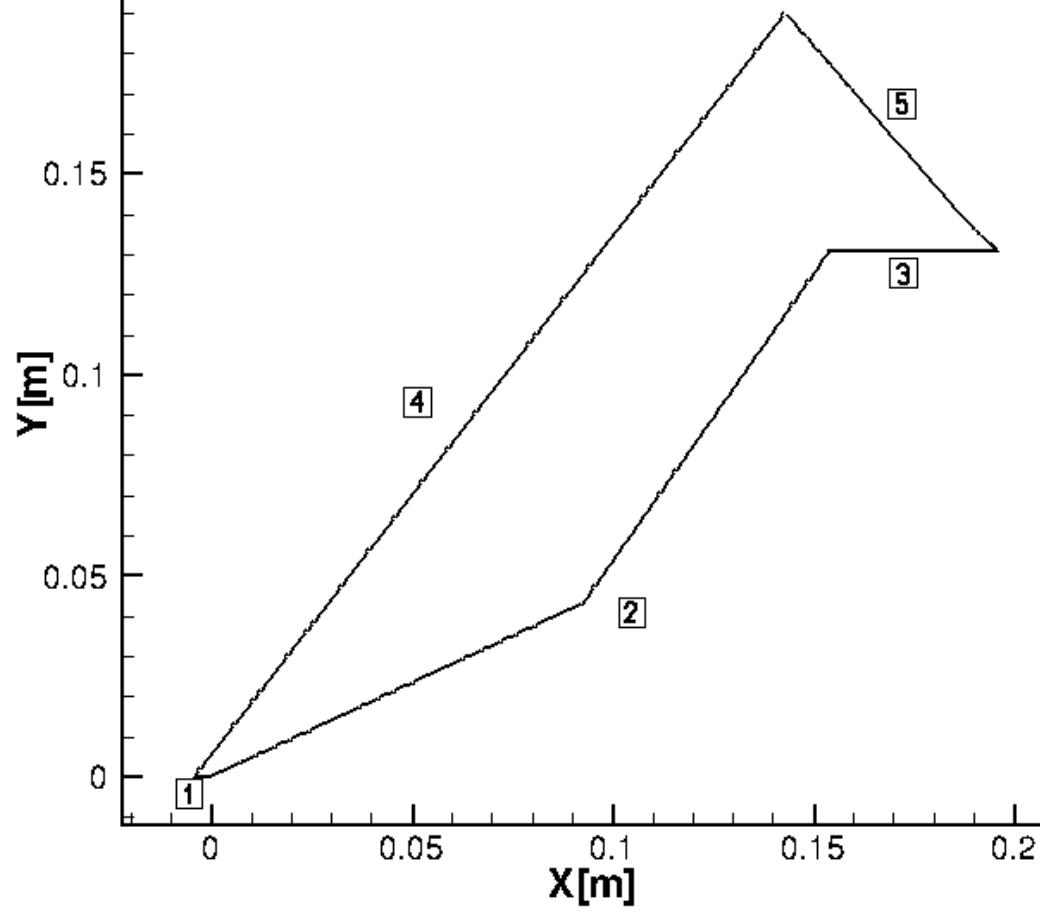}
        \caption{DC - Computational Domain}
        \label{fig:doubleconeComputationalDomain}
    \end{minipage}
\end{figure}

\begin{figure}[H]
    \begin{minipage}[t]{6cm}
        \centering
        \includegraphics[width=5.5cm]{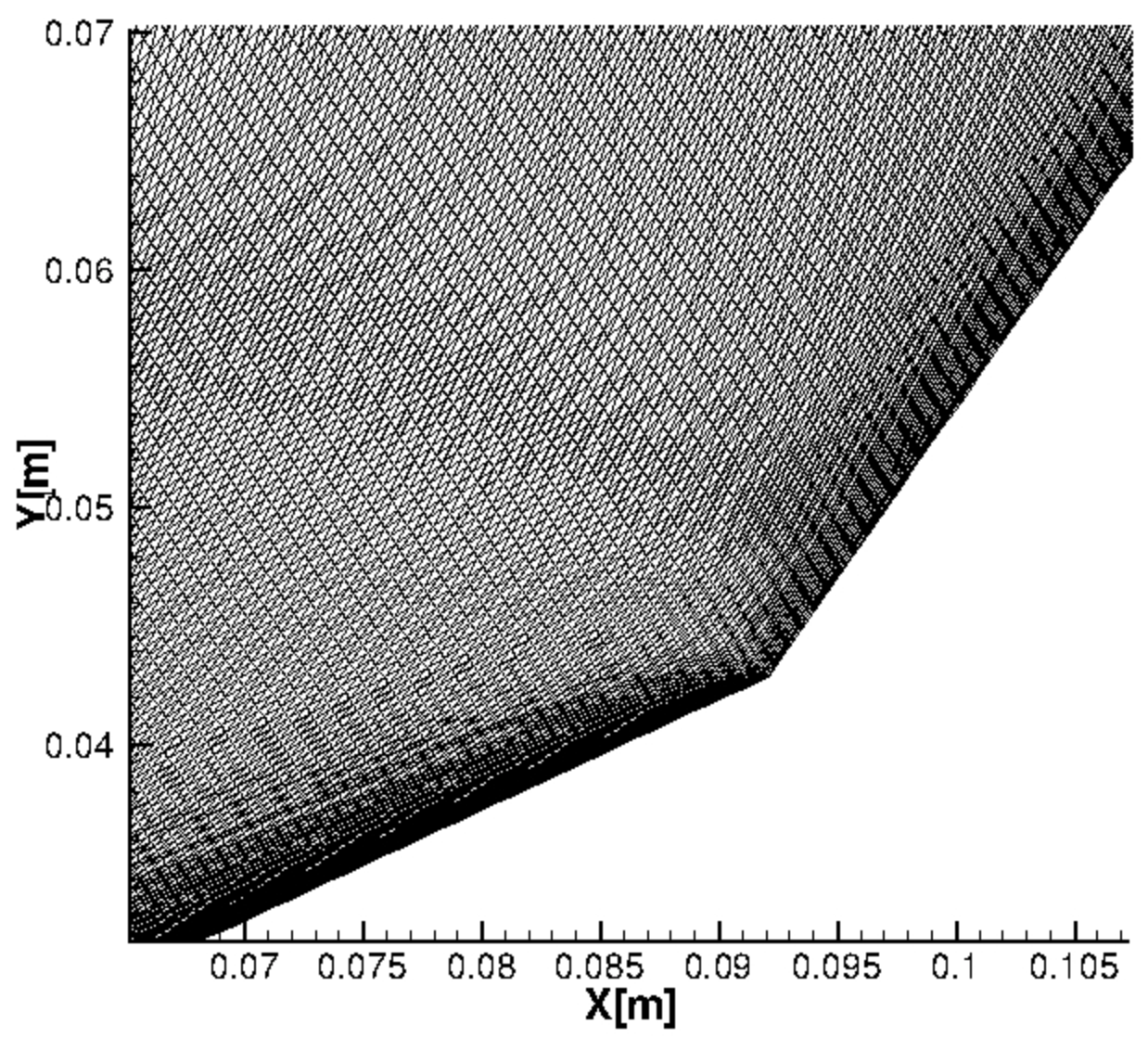}
        \caption{Initial mesh--zoom $1^{st}$ cone}
        \label{fig:init1Cone}
    \end{minipage}
    \begin{minipage}[t]{6cm}
        \centering
        \includegraphics[width=5.5cm]{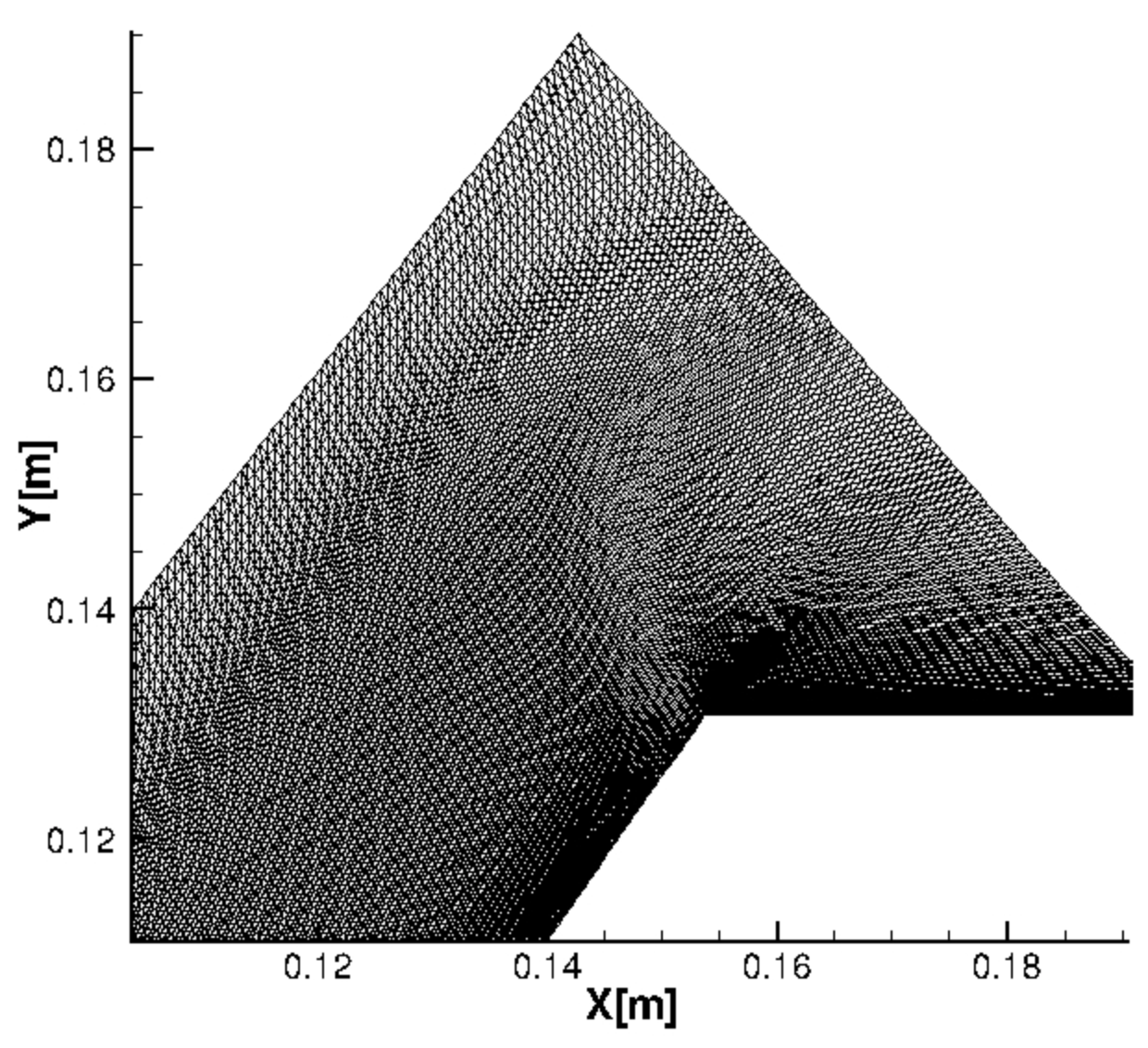}
        \caption{Initial mesh--zoom $2^{nd}$ cone}
        \label{fig:init2Cone}
    \end{minipage}
\end{figure}

The 2D axisymmetric double cone test case conditions are presented in Tab.\ref{tab:DCflowchar}, Tab.\ref{tab:DCmeshchar} and Tab.\ref{tab:DCAMR}, while the test case definition and the corresponding unstructured mesh are shown in Fig.\ref{fig:doublecone}, Fig.\ref{fig:doubleconeComputationalDomain}, Fig.\ref{fig:init1Cone} and Fig.\ref{fig:init2Cone}.

\begin{table}[H]
\centering
\caption{Double cone -- Flow characteristics}
\label{tab:DCflowchar}
\begin{tabular}{|cccccccc|}
\hline
 \footnotesize{Physical Model} & \footnotesize{M}& \footnotesize{$y_{{N}_{2}}$} & \footnotesize{$\rho$ [kg/$m^3$]} & \footnotesize{u [m/s]} & \footnotesize{$T$ [K]} & \footnotesize{$T^{v}$ [K] }  &  \footnotesize{$T^{w}$ [K] }  \\
\footnotesize{TCNEQ ($N-N_{2}$)} &  \footnotesize{11.5 } & \footnotesize{1}   & \footnotesize{0.001468}   & \footnotesize{3849.3}  & \footnotesize{268.7}   & \footnotesize{3160} & \footnotesize{294.7}\\
\hline
\end{tabular}
\end{table}

\begin{table}[H]
\centering
\caption{Double cone -- Mesh characteristics}
\label{tab:DCmeshchar}
\begin{tabular}{|ccccccc|}
\hline
 \footnotesize{Dimensions} & \footnotesize{Type} & \footnotesize{\# Elements} & \footnotesize{BC 1} & \footnotesize{BC 4} & \footnotesize{BC 2 \& 3} & \footnotesize{BC 5}   \\
\footnotesize{2D axisymmetric} &  \footnotesize{Triangular}    & \footnotesize{65280} & \footnotesize{Symmetry}   & \footnotesize{Inlet}  & \footnotesize{Iso-thermal wall}   & \footnotesize{Outlet} \\
\hline
\end{tabular}
\end{table}

\begin{table}[H]
\centering
\caption{Double cone -- r-refinement}
\label{tab:DCAMR}
\begin{tabular}{|cccccc|}
\hline
\footnotesize{Spring Network} &\footnotesize{Monitor Variable} & \footnotesize{Process Rate} & \footnotesize{Stop AMR Iteration} & \footnotesize{minPer} & \footnotesize{maxPer}    \\
 \footnotesize{Semi-torsional}  & \footnotesize{Density} &\footnotesize{10} & \footnotesize{200} & \footnotesize{0.30}  & \footnotesize{0.55}   \\
\hline
\end{tabular}
\end{table}

The semi-torsional spring analogy is applied to the double cone test case. The parameter $\textsc{p}$ is set to be equal to ${k}^{L}$ in order to include the physical characteristics within the adaptation. 
The expression of the global mesh stiffness, between two edge-connected nodes $ij$, is therefore described:
\begin{equation}
\label{eq:doubleCone}
    k_{ij}^{tot}= k_{ij}^{L}\cdot (1+k_{ij}^{ST}).
\end{equation}

\begin{figure}[H]
\captionsetup{justification=centering}
    \begin{minipage}[t]{6cm}
        \centering
        \includegraphics[width=5.5cm]{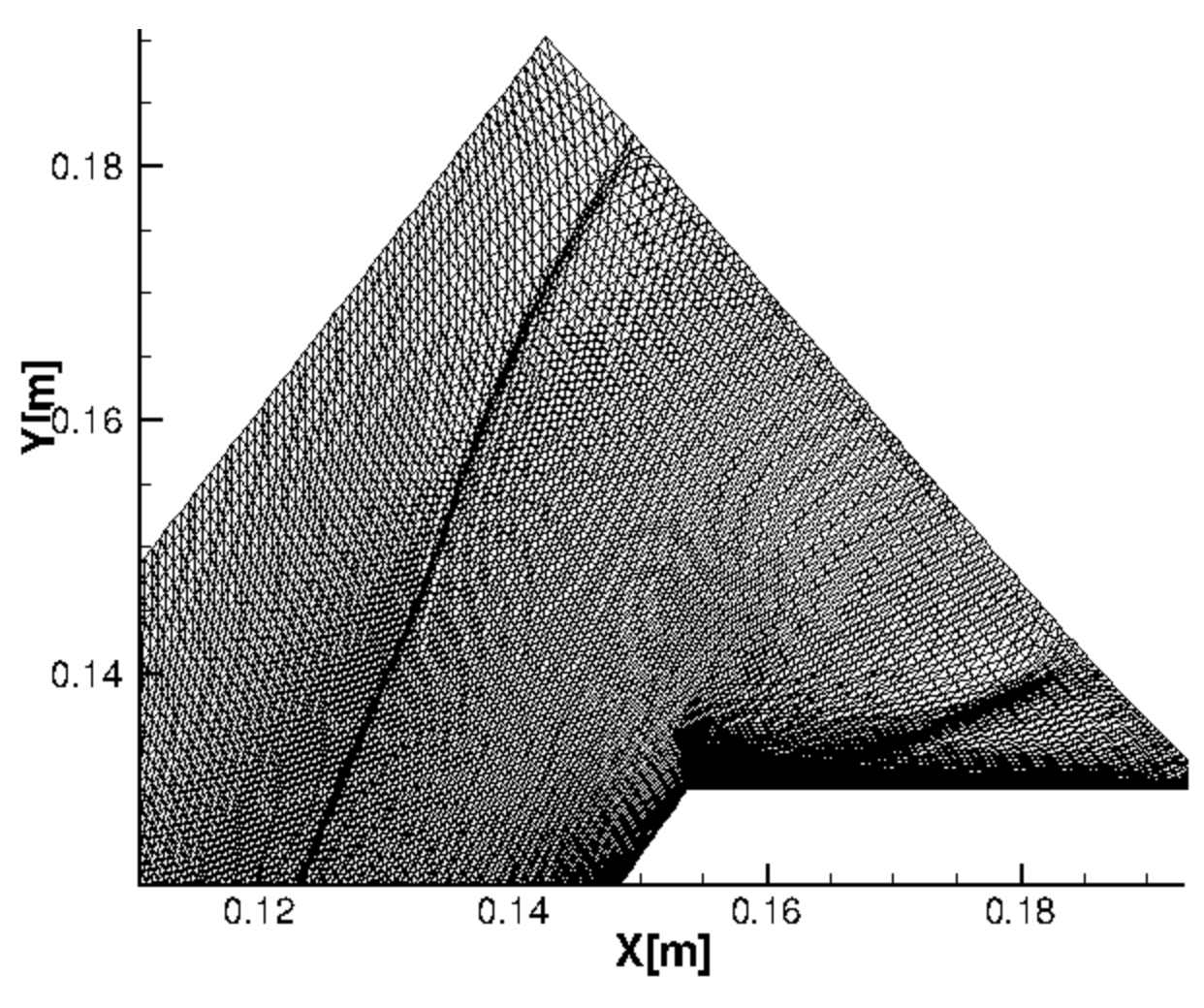}
        \caption{Zoom, $2^{nd}$ cone, as appearing after 200 steps of refinement}
        \label{fig:Zoom 2^{nd} cone}
    \end{minipage}
    \begin{minipage}[t]{6cm}
        \centering
        \captionsetup{justification=centering}
        \includegraphics[width=5.5cm]{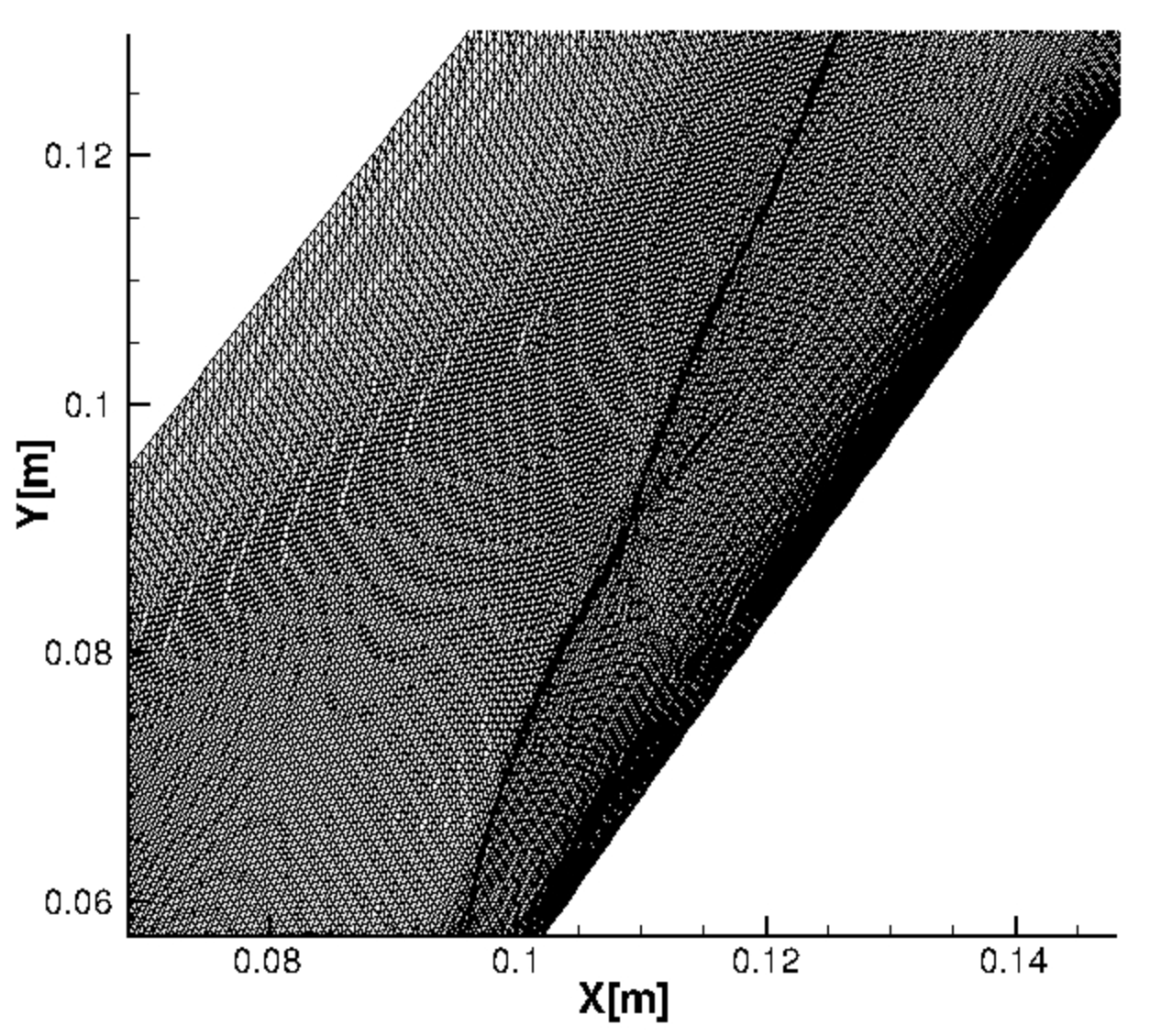}
        \caption{Bow shock as appearing after 200 steps of refinement}
        \label{fig:bow}
    \end{minipage}
\end{figure}

\begin{figure}[H]
    \captionsetup{justification=centering}
    \begin{minipage}[t]{6cm}
        \centering
        \includegraphics[width=5.5cm]{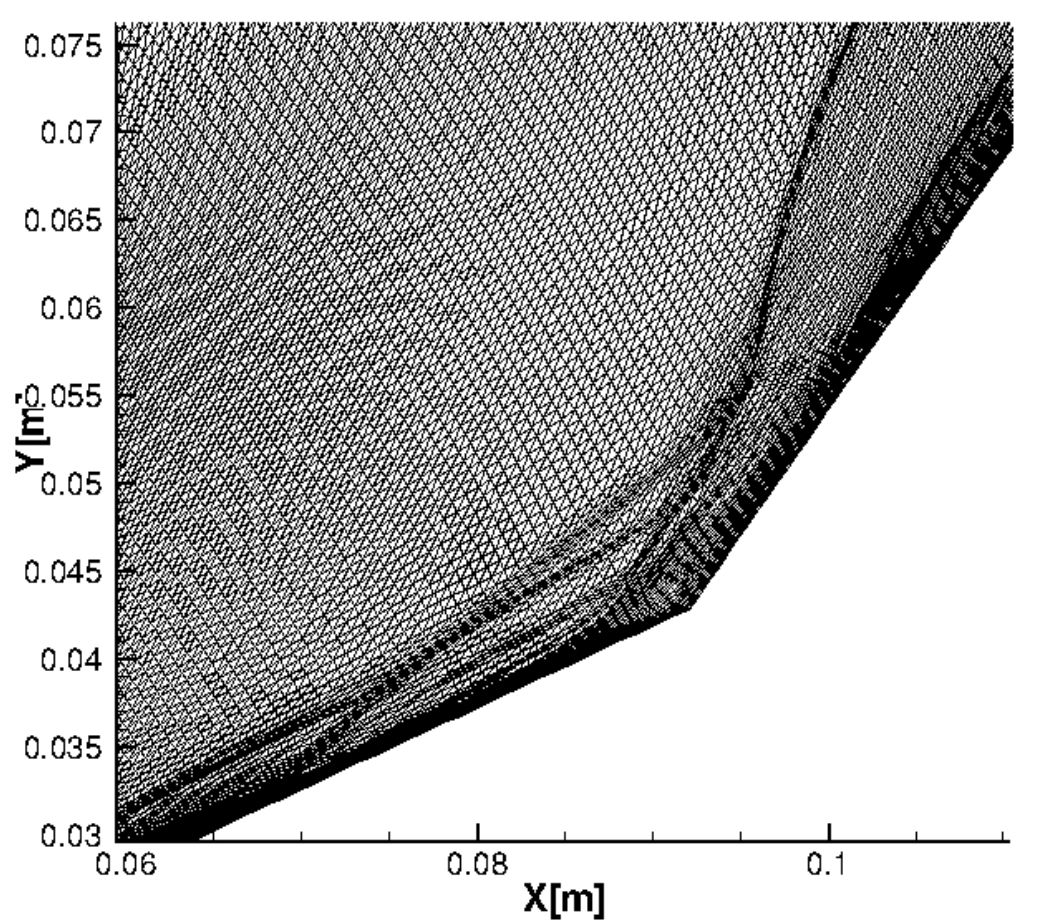}
        \caption{SWBLI as appearing after 200 steps of refinement}
        \label{fig:BL interaction}
    \end{minipage}
    \begin{minipage}[t]{6cm}
        \centering
        \includegraphics[width=5.5cm,height=5cm]{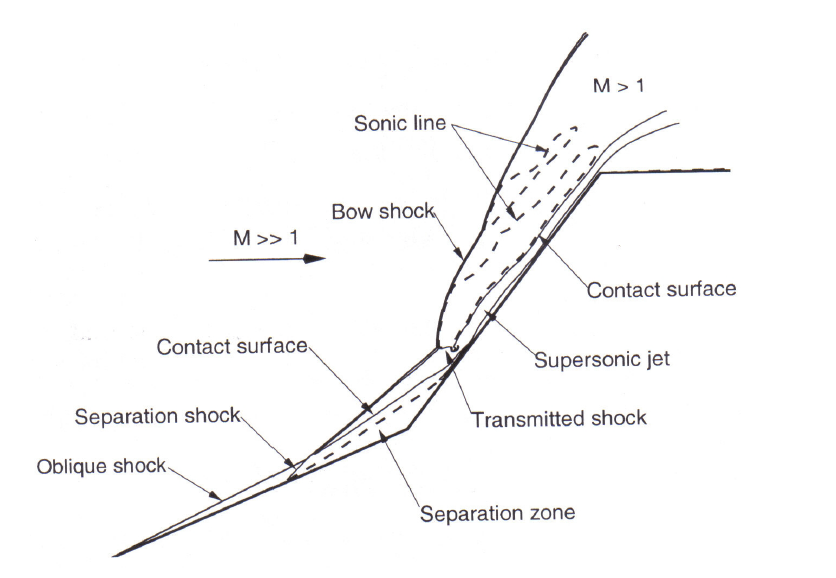}
        \caption{Schematic of the double cone flow field \cite{phd:lani08}}
        \label{fig:coneS}
    \end{minipage}
\end{figure}

Fig.\ref{fig:BL interaction} shows the shock wave boundary layer interactions (SWBLI) occurring near the junction between the first and second cones. The shock structure highlighted by the mesh refinement closely resembles the qualitative solution presented in Fig.\ref{fig:coneS}

\subsection{Hornung}
\label{sec:HC}
\begin{figure}[H]
\centering{\includegraphics[scale=0.3]{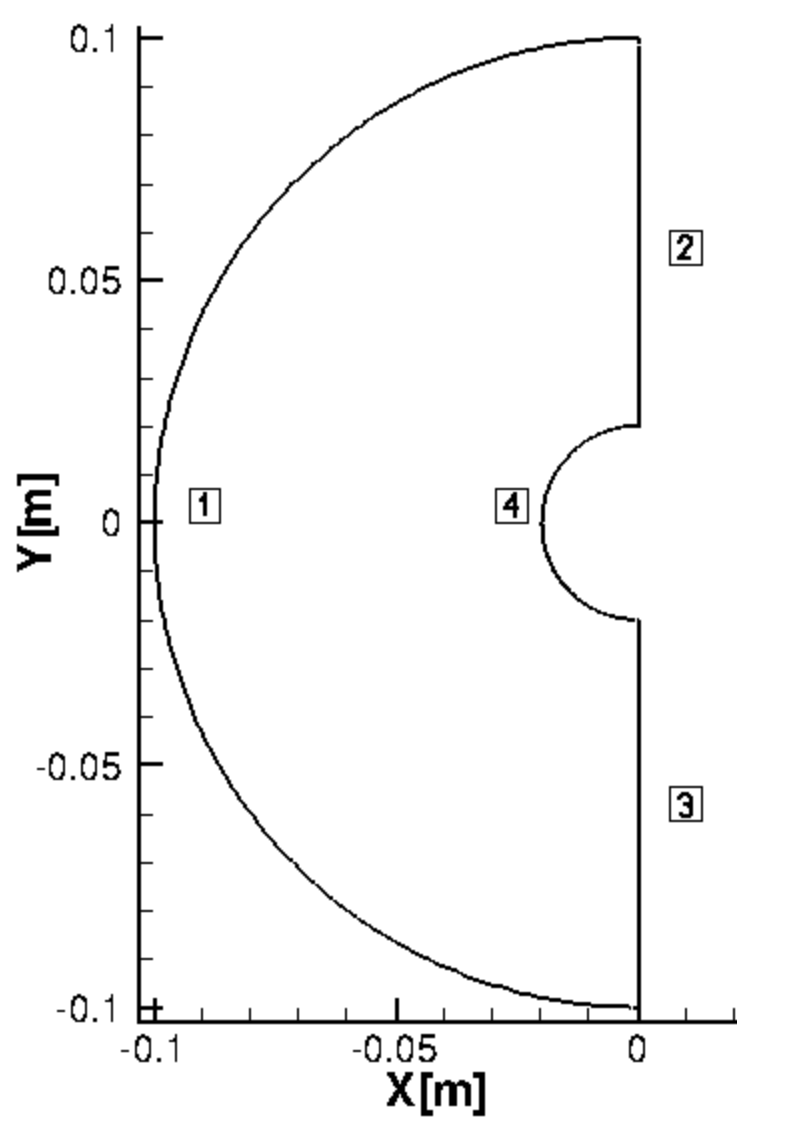}}
\caption{Semi-circle geometry}
\label{fig:geomHornung}
\end{figure}

The 2D semi-cylinder Hornung test case conditions are presented in Tab.\ref{tab:Hornungflowchar}, Tab.\ref{tab:Hornungmeshchar} and Tab.\ref{tab:HornungAMR}, while the test case definition is shown in Fig.\ref{fig:geomHornung}.

\begin{table}[H]
\centering
\caption{Hornung -- Flow characteristics}
\label{tab:Hornungflowchar}
\begin{tabular}{|ccccccc|}
\hline
 \footnotesize{Physical Model} &\footnotesize{M} & $\footnotesize{\rho}_{\footnotesize{{N}}}$ \footnotesize{[kg/$m^3$]} & $\footnotesize{\rho}_{\footnotesize{{{N}_{2}}}}$ \footnotesize{[kg/$m^3$]} & \footnotesize{u [m/s]} & \footnotesize{$T$ [K]} & \footnotesize{$T^{w}$ [K]}  \\
\footnotesize{TCNEQ ($N-N_{2}$)} & 6&\footnotesize{0.0001952}   & \footnotesize{0.004956}   & \footnotesize{5590}  & \footnotesize{1833} & \footnotesize{1000}\\
\hline
\end{tabular}
\end{table}

\begin{table}[H]
\centering
\caption{Hornung -- Mesh characteristics}
\label{tab:Hornungmeshchar}
\begin{tabular}{|cccccc|}
\hline
 \footnotesize{Dimensions} & \footnotesize{Type} & \footnotesize{\# Elements} & \footnotesize{BC 1} & \footnotesize{BC 2 \& 3} & \footnotesize{BC 4}  \\
\footnotesize{2D} &  \footnotesize{Quadrilateral}    & \footnotesize{25000}   & \footnotesize{Inlet}  & \footnotesize{Outlet}   & \footnotesize{Iso-thermal wall} \\
\hline
\end{tabular}
\end{table}

\begin{table}[H]
\centering
\caption{Hornung -- r-refinement}
\label{tab:HornungAMR}
\begin{tabular}{|cccccc|}
\hline
\footnotesize{Spring Network} &\footnotesize{Monitor Variable} & \footnotesize{Process Rate} & \footnotesize{Stop AMR Iteration} & \footnotesize{minPer} & \footnotesize{maxPer}    \\
 \footnotesize{Linear}  & \footnotesize{Flow density} &\footnotesize{10} & \footnotesize{till convergence} & \footnotesize{0.30}  & \footnotesize{0.55}   \\
\hline
\end{tabular}
\end{table}

\begin{figure}[H]
        \captionsetup{justification=centering}

    \begin{minipage}[t]{6cm}
        \centering
        \includegraphics[width=6cm]{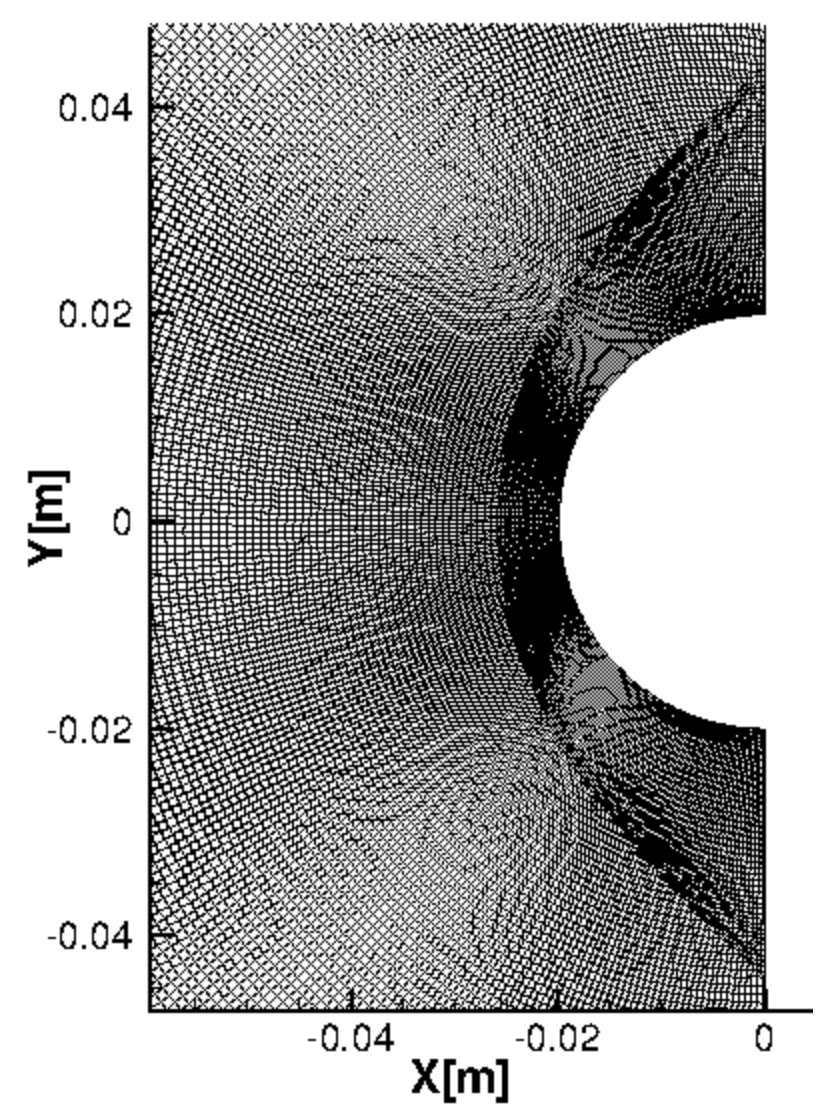}
        \caption{Hornung -- Final mesh}
        \label{fig:final mesh}
    \end{minipage}
    \begin{minipage}[t]{6cm}
        \centering
        \includegraphics[width=6cm,height=8.2cm]{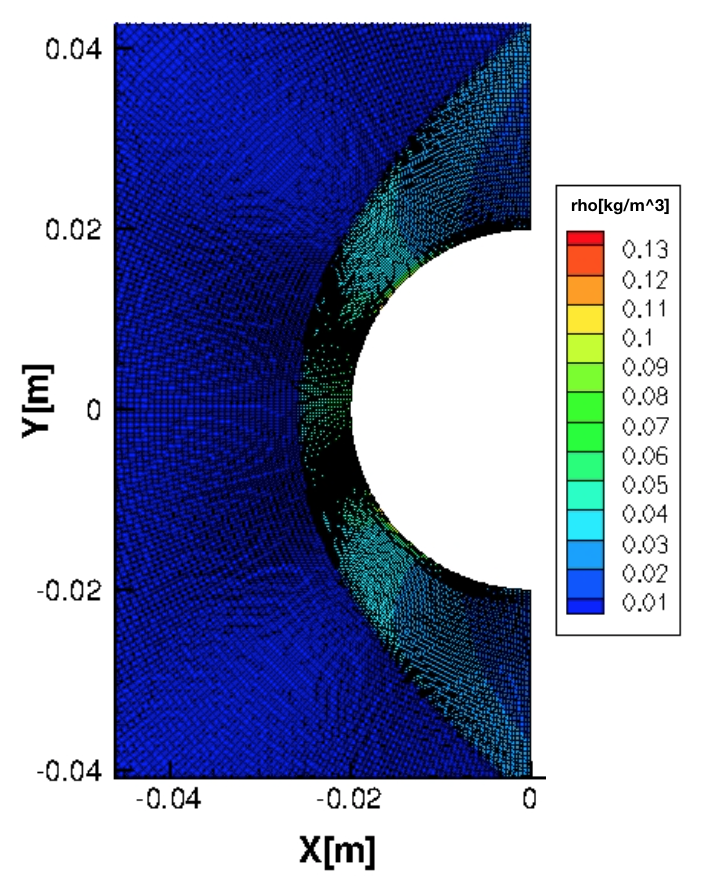}
        \caption{Final mesh and flow field density}
        \label{Final mesh and flow field density}
    \end{minipage}
\end{figure}

\begin{figure}[H]
        \captionsetup{justification=centering}

    \begin{minipage}[t]{6cm}
        \centering
        \includegraphics[width=6cm]{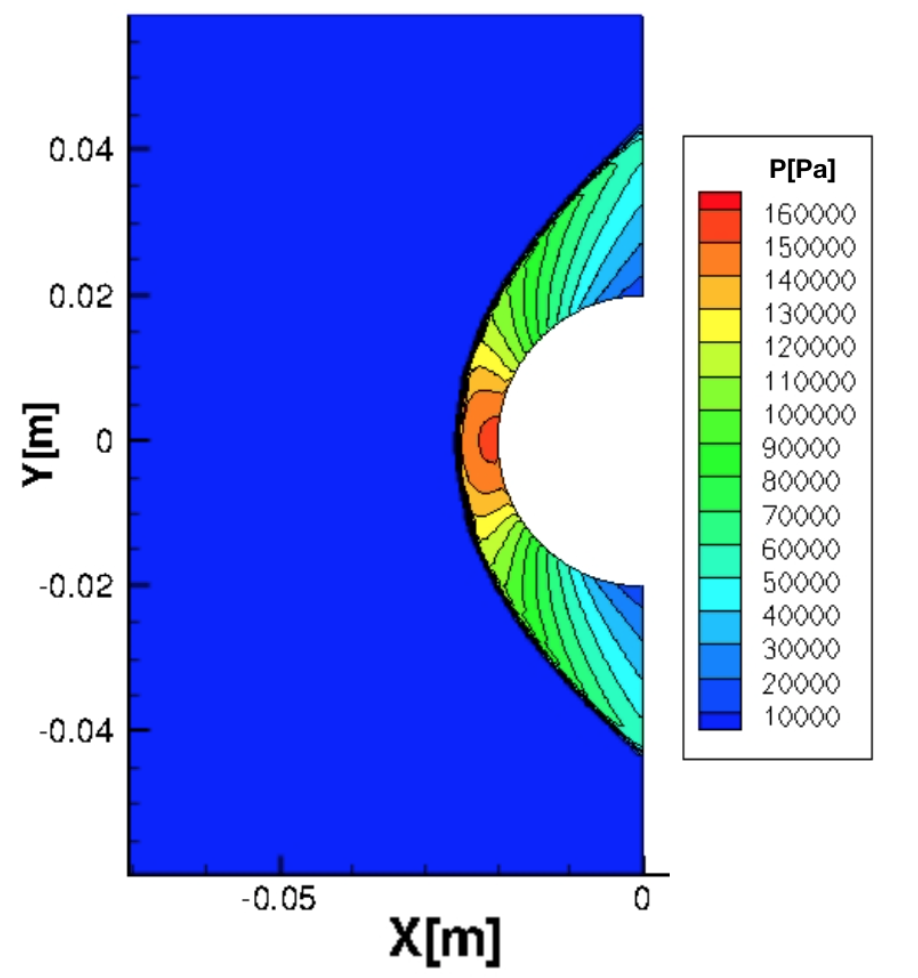}
        \caption{Pressure contours}
        \label{Pressure contours: Converged solution}
    \end{minipage}
    \begin{minipage}[t]{6cm}
        \centering
        \includegraphics[width=6cm,height=6.5cm]{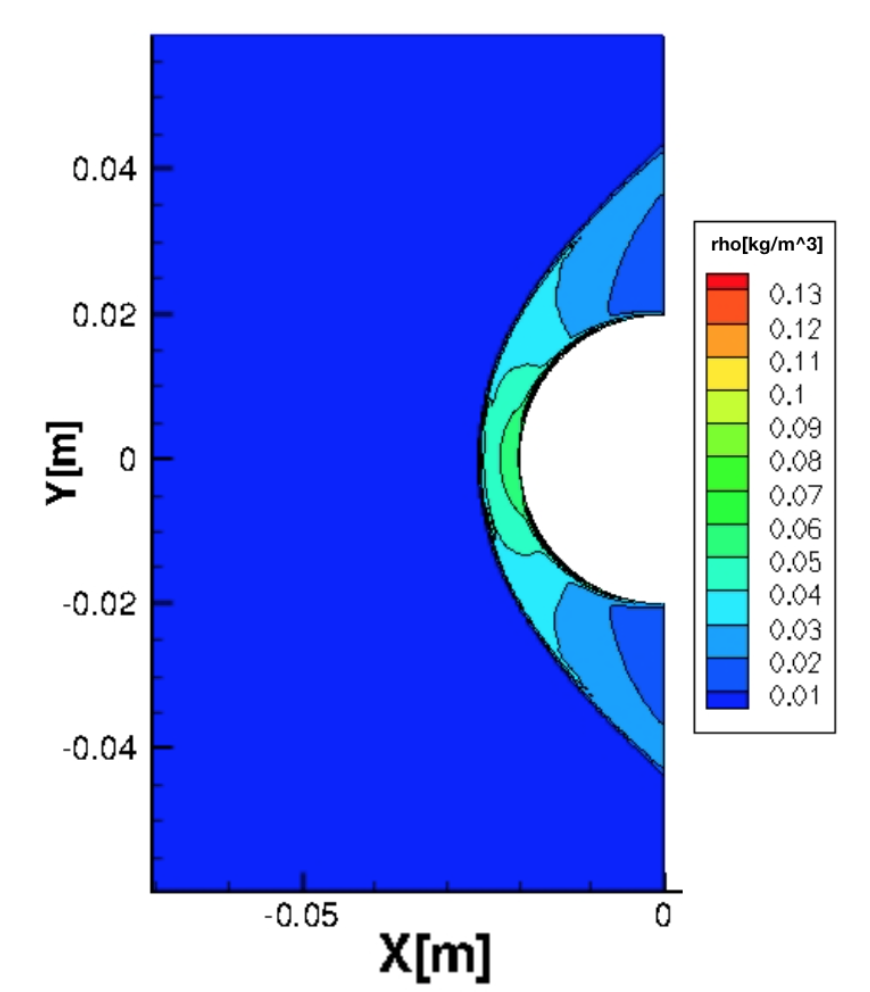}
        \caption{Density contours}
        \label{Density contours: Converged solution}
    \end{minipage}
\end{figure}

The simulation uses the linear spring analogy.
The mesh refinement result is presented in Fig.\ref{fig:final mesh} showing a perfect bow shock adaptation.
The flow field pressure and density contours, presented in Fig.\ref{Pressure contours: Converged solution} and \ref{Density contours: Converged solution}, show a symmetrical solution. The refined shock, based on the flow field density, and the density contours match properly as shown in Fig.\ref{Final mesh and flow field density}.

\subsection{Hemisphere}
\label{sec:Hemisphere}
\begin{figure}[H]
\centering
\begin{minipage}{.45\linewidth}
  \includegraphics[width=\linewidth]{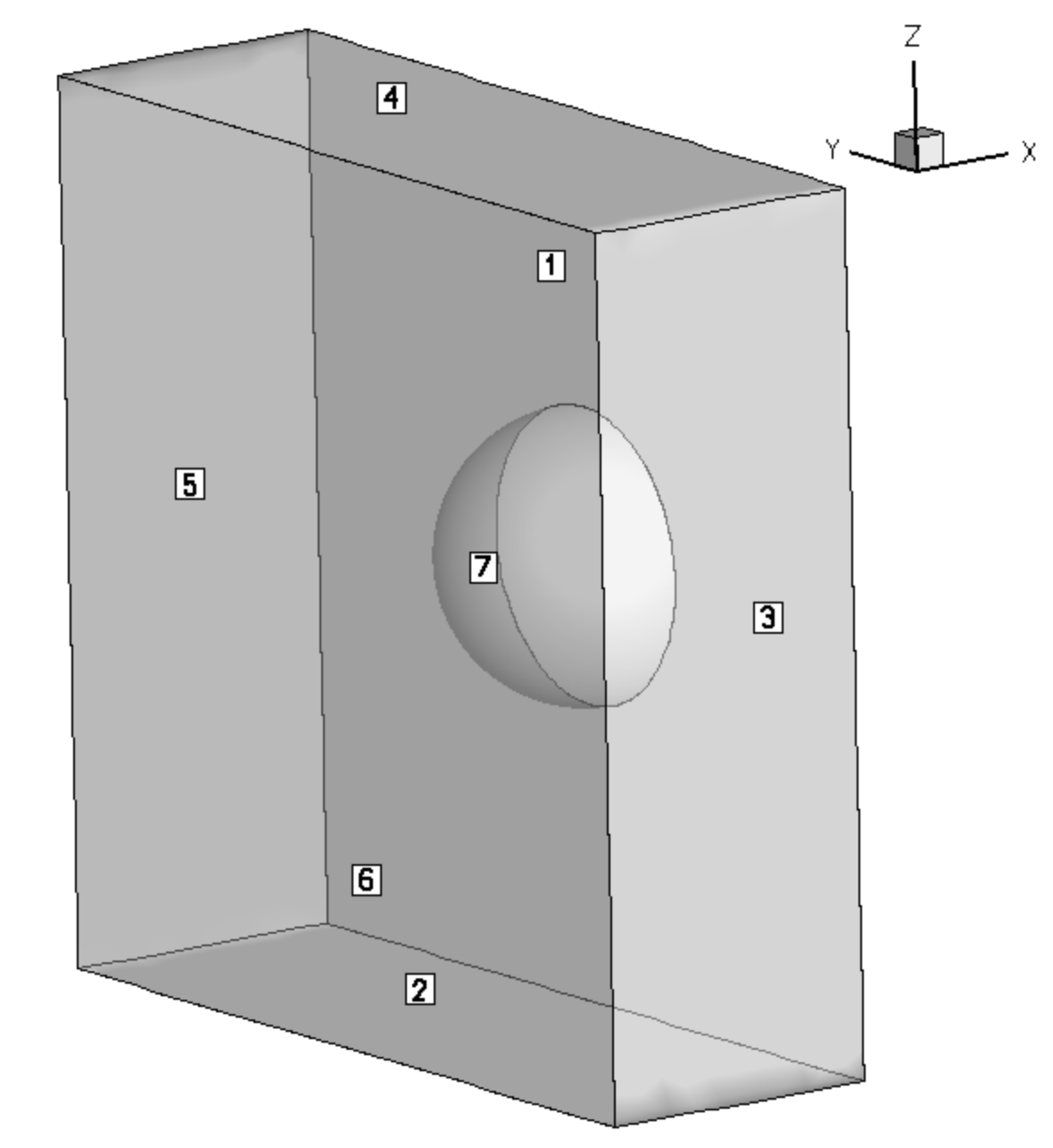}
  \caption{Hemisphere geometry}
  \label{fig:hemisphereGeom}
\end{minipage}
\hspace{.05\linewidth}
\begin{minipage}{.45\linewidth}
  \includegraphics[width=0.55\linewidth]{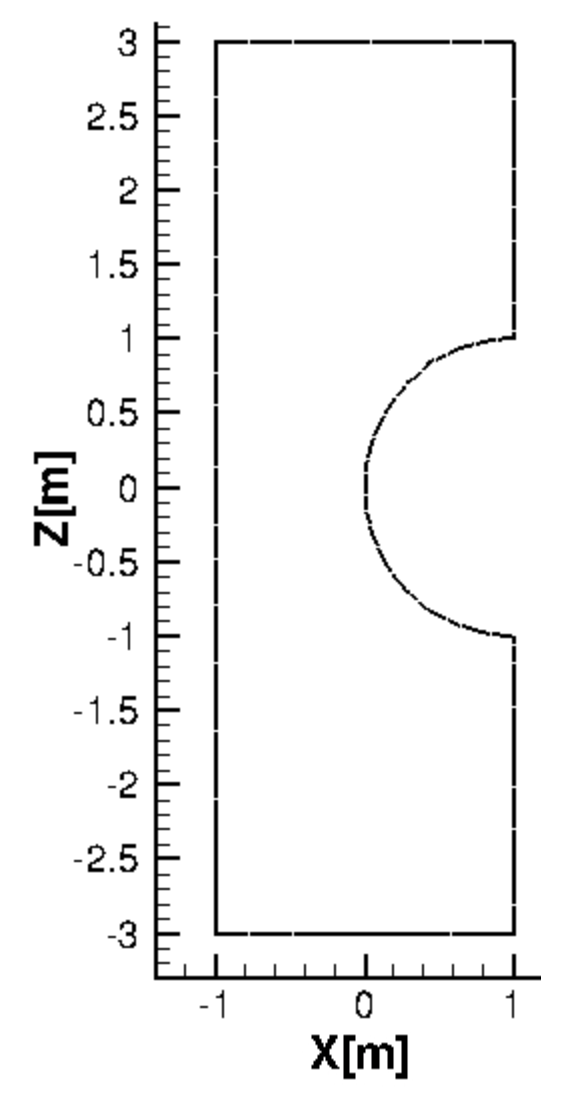}
  \caption{2D section}
  \label{fig:hemisphereGeom2D}
\end{minipage}
\end{figure}

The 3D hemisphere test case conditions are presented in Tab.\ref{tab:HemisphereFC}, Tab.\ref{tab:Hemispheremeshchar} and Tab.\ref{tab:HemisphereAMR}, while the computational domain and a 2D section are shown in Fig.\ref{fig:hemisphereGeom} and Fig.\ref{fig:hemisphereGeom2D} respectively.

\begin{table}[H]
\centering
\caption{Hemisphere -- Flow characteristics}
\label{tab:HemisphereFC}
\begin{tabular}{|cccccccc|}
\hline
 \footnotesize{Physical Model} & \footnotesize{M} &\footnotesize{P [Pa]} & \footnotesize{u [m/s]} & \footnotesize{v [m/s]} & \footnotesize{w [m/s]} &\footnotesize{T [K]} &  \footnotesize{$\rho$} \footnotesize{[kg/$m^3$]}  \\
 
\footnotesize{Perfect gas} & 10 &\footnotesize{1000}   & \footnotesize{3413.8}   & \footnotesize{0}  & \footnotesize{0}  & \footnotesize{290}  & \footnotesize{0.0120129}  \\
\hline
\end{tabular}
\end{table}

\begin{table}[H]
\centering
\caption{Hemisphere -- Mesh characteristics}
\label{tab:Hemispheremeshchar}
\begin{tabular}{|cccccc|}
\hline
 \footnotesize{Dimensions} & \footnotesize{Type} & \footnotesize{\# Elements} & \footnotesize{BC 1 .. 5} & \footnotesize{BC 6} & \footnotesize{BC 7}  \\
\footnotesize{3D} &  \footnotesize{Tetrahedral}    & \footnotesize{190485}   & \footnotesize{Inlet}  & \footnotesize{Outlet}   & \footnotesize{no-slip wall} \\
\hline
\end{tabular}
\end{table}

\begin{table}[H]
\centering
\caption{Hornung -- r-refinement}
\label{tab:HemisphereAMR}
\begin{tabular}{|cccccc|}
\hline
\footnotesize{Spring Network} &\footnotesize{Monitor Variable} & \footnotesize{Process Rate} & \footnotesize{Stop AMR Iteration} & \footnotesize{minPer} & \footnotesize{maxPer}    \\
 \footnotesize{orth-semi-torsional}  & \footnotesize{Pressure} &\footnotesize{20} & \footnotesize{300} & \footnotesize{0.30}  & \footnotesize{0.55}   \\
\hline
\end{tabular}
\end{table}

The ortho-semi-torsional spring analogy, coupled with the linear and semi-torsional spring analogy, is used within this test case. The global mesh stiffness obeys to Eq.\ref{eq:TOT} where the ortho-semi-torsional spring analogy in Eq.\ref{eq:kostFinal} is transformed into:
\begin{equation}
    k^{OST}_{qs}= \frac{k^{L}_{qs}}{2}\left(\frac{k_{si}}{\lambda_{si}}+\frac{k_{qi}}{\lambda_{qi}}\right),
\end{equation}
while the semi-torsional spring analogy in Eq.\ref{eq:3Dksemi} will be transformed into:
\begin{equation}
    k^{ST}_{qs}=k^{L}_{qs} \frac{d_{ql}^2 d_{sl}^2}{A_{qsl}^2},
\end{equation}
where $l$ has the same geometrical signification as the point $p$ in Eq.\ref{eq:3Dksemi}.
The mesh is adequately refined around the shock. In fact, the mesh nodes density increases around the zone of pressure variation.

\begin{figure}[H]
\centering{\includegraphics[scale=0.35]{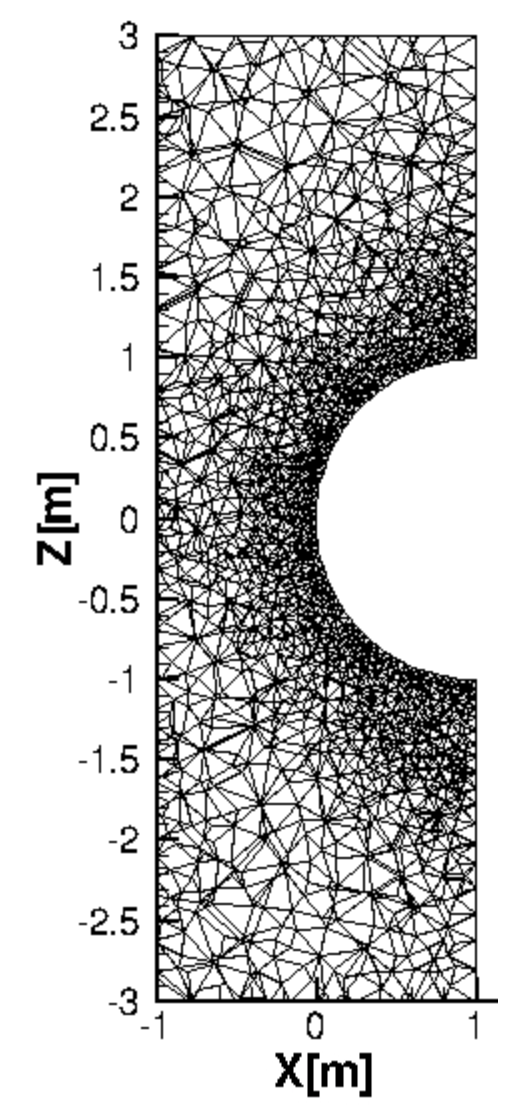}}
\caption{Initial mesh: section Y=0 }
\label{fig:bow1}
\end{figure}

\begin{figure}[H]
\centering
\begin{minipage}{.45\linewidth}
  \includegraphics[scale=0.35]{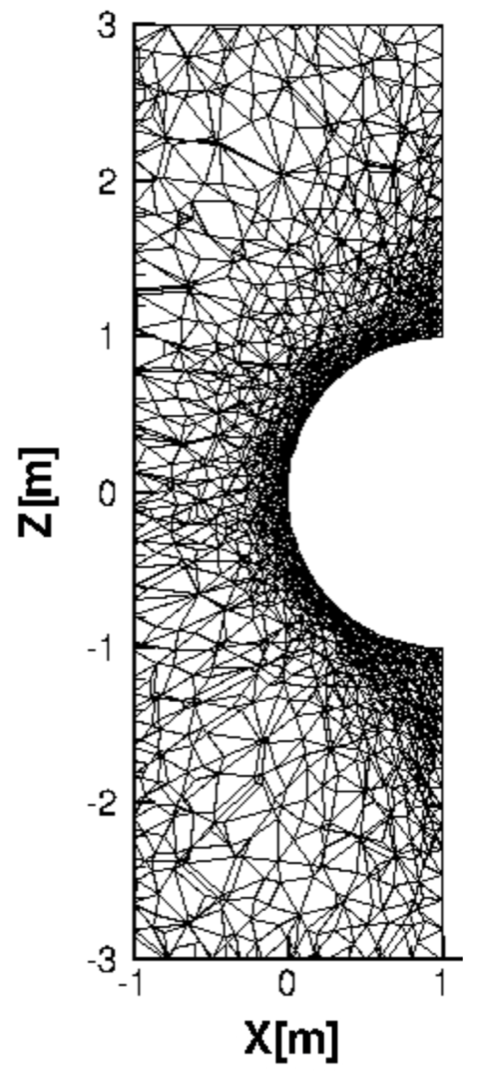}
  \caption{Final mesh: section Y= 0}
  \label{fig:Mesh000}
\end{minipage}
\hspace{.05\linewidth}
\begin{minipage}{.45\linewidth}
  \includegraphics[scale=0.35]{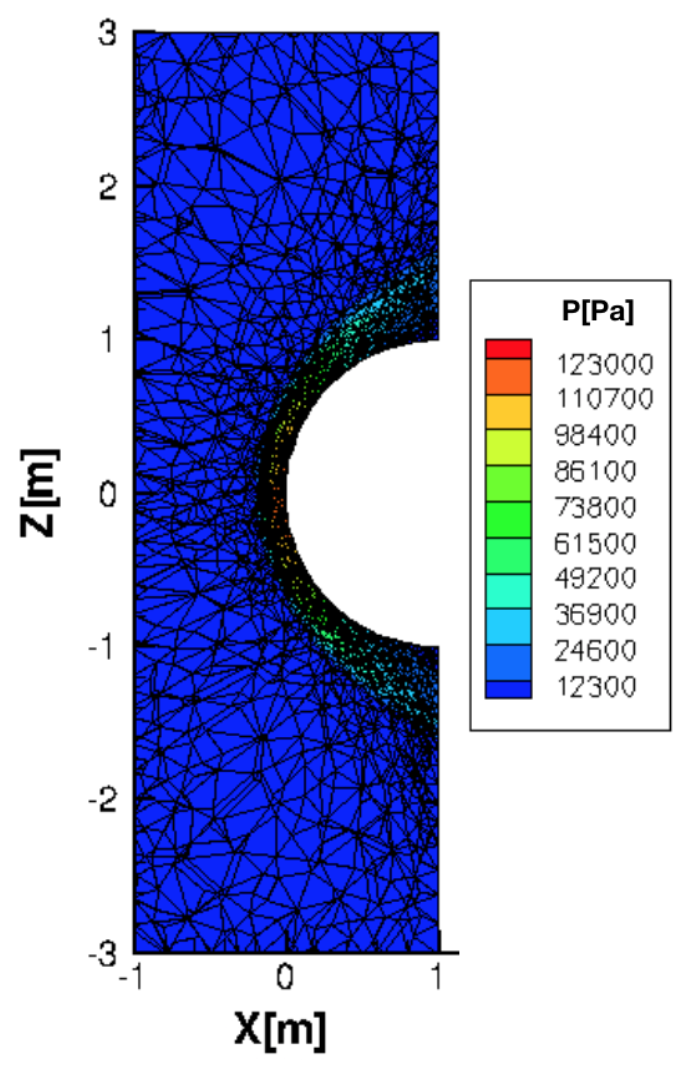}
  \caption{Mesh and pressure contours}
  \label{fig:pressureMesh}
\end{minipage}
\end{figure}

\begin{figure}[H]
\centering
        \captionsetup{justification=centering}

\begin{minipage}{.45\linewidth}
  \includegraphics[width=\linewidth]{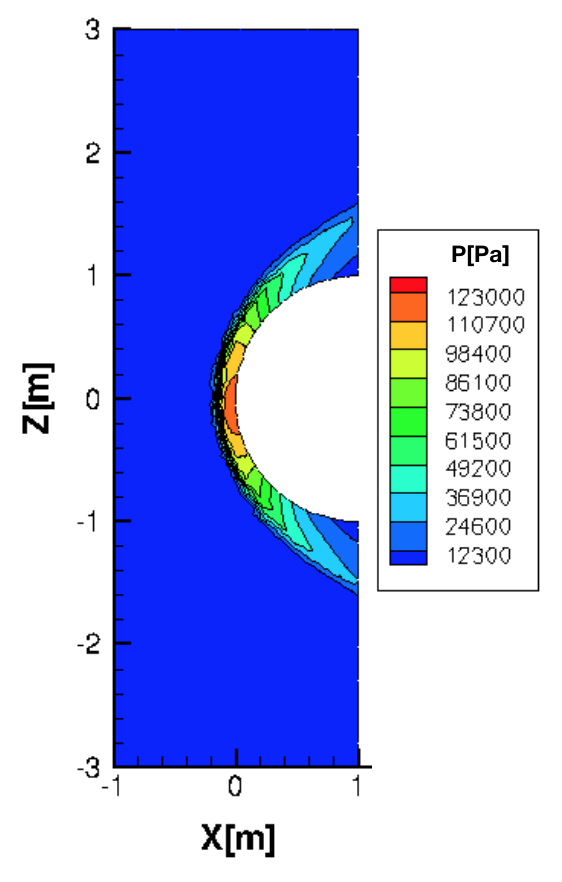}
  \caption{Pressure Contours: section Y=0}
  \label{fig:pressureContours0}
\end{minipage}
\hspace{.05\linewidth}
\begin{minipage}{.45\linewidth}
  \includegraphics[width=\linewidth]{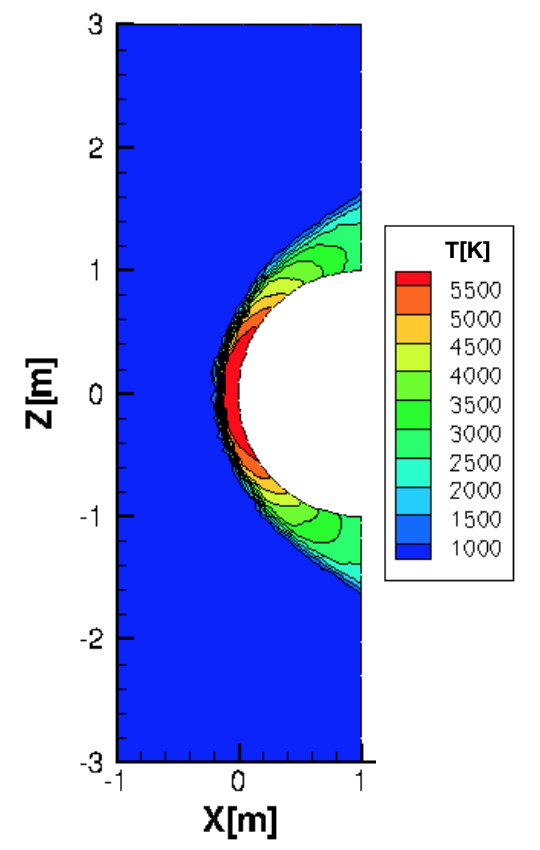}
  \caption{Temperature contours: section Y=0}
  \label{fig:temperatureContours0}
\end{minipage}
\end{figure}

\subsection{Solar wind/Earth's magnetosphere interaction}
\label{sec:SW}
This test case simulates a Solar Wind/Earth's Magnetosphere Interaction that occurred during a magnetic storm on April the $6^{th}$, $2000$. The inlet conditions correspond to real data which were recorded by the NASA's Advanced Composition Explorer (ACE) satellite at the Lagrangian point L1 \cite{solarwindA}. The test case conditions (in adimensional form, as explained in \cite{solarwindA}) are presented in Tab.\ref{tab:SolarWindFC} and Tab.\ref{tab:SolarWindAMR}.

\begin{table}[H]
\centering
\caption{Solar wind -- Flow characteristics}
\label{tab:SolarWindFC}
\begin{tabular}{|ccccc|}
\hline
 \footnotesize{Physical Model} & \footnotesize{$\rho$ [-]} & \footnotesize{u [-]} & \footnotesize{v [-]} & \footnotesize{w [-] } \\
\footnotesize{MHD} & \footnotesize{1.26020}   & \footnotesize{-10.8434}   & \footnotesize{-0.859678}  & \footnotesize{0.0146937}  \\

 \hline
 \footnotesize{$B_x$} \footnotesize{[-]}  & \footnotesize{$B_y$} \footnotesize{[-]} & \footnotesize{$B_z$} \footnotesize{[-]} & \footnotesize{p [-]}&\\
 \footnotesize{0.591792} & \footnotesize{-2.13282}  & \footnotesize{-0.602181} & \footnotesize{0.565198 } &\\
\hline
\end{tabular}
\end{table}

\begin{table}[H]
\centering
\caption{Solar wind -- r-refinement}
\label{tab:SolarWindAMR}
\begin{tabular}{|cccccc|}
\hline
\footnotesize{Spring Network} &\footnotesize{Monitor Variable} & \footnotesize{Process Rate} & \footnotesize{Stop AMR Iteration} & \footnotesize{minPer} & \footnotesize{maxPer}    \\
 \footnotesize{semi-torsional}  & \footnotesize{Flow density} &\footnotesize{20} & \footnotesize{1045} & \footnotesize{0.30}  & \footnotesize{0.55}   \\
\hline
\end{tabular}
\end{table}

The computational domain is a rectangular box and a sphere (modelling the earth)  centered at the origin, as explicitly defined in \cite{solarwindA} and shown in Fig.\ref{fig:SWgeom}:

\begin{figure}[H]
\centering{\includegraphics[scale=0.3]{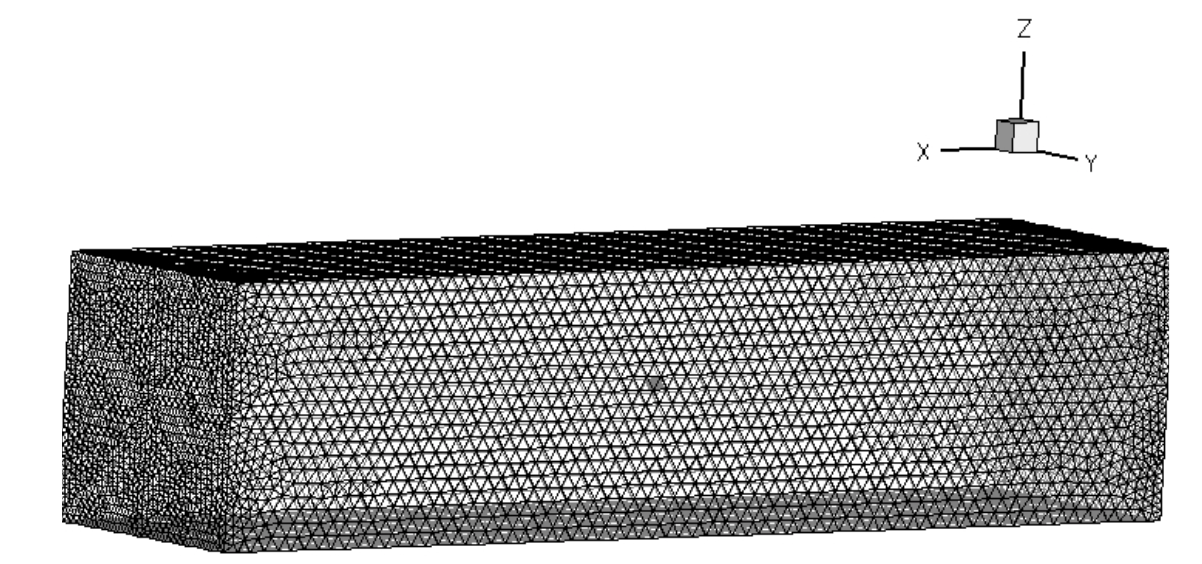}}
\caption{Computational domain, -200$\le$x$\le$235, -50$\le$y,z$\le$50, radius of the sphere $r=2.5$}
\label{fig:SWgeom}
\end{figure}

The semi-torsional spring analogy is applied to the solar wind test case:
\begin{equation}
    k^{ST}_{qs}=k^{L}_{qs} \frac{d_{ql}^2 d_{sl}^2}{A_{qsl}^2},
\end{equation}

\begin{figure}[H]
\centering
\begin{minipage}{.45\linewidth}
  \includegraphics[width=\linewidth]{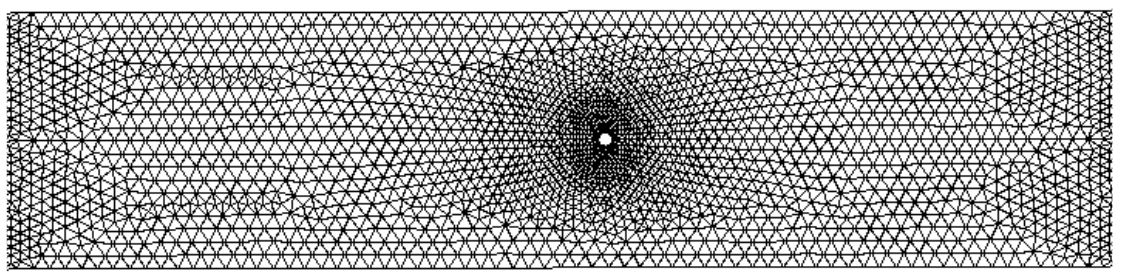}
  \caption{Initial mesh, section Y=0}
  \label{fig:fullView}
\end{minipage}
\hspace{.05\linewidth}
\begin{minipage}{.45\linewidth}
  \includegraphics[width=\linewidth]{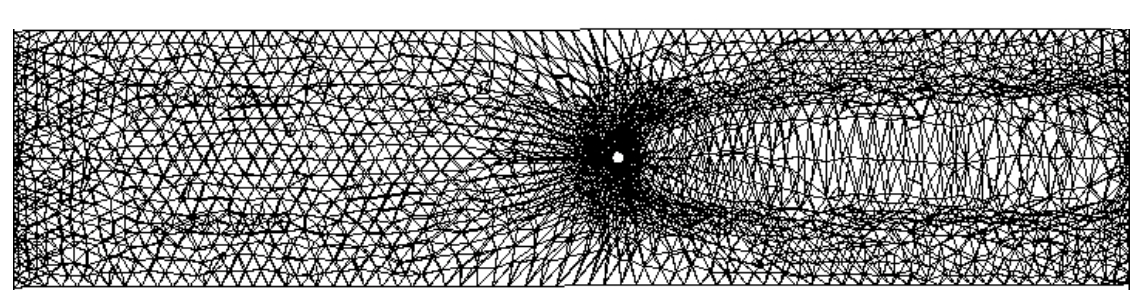} 
  \caption{Final mesh, section Y=0}
  \label{fig:FinalMeshView}
\end{minipage}
\end{figure}

\begin{figure}[H]
\centering
\begin{minipage}{.45\linewidth}
  \includegraphics[width=\linewidth]{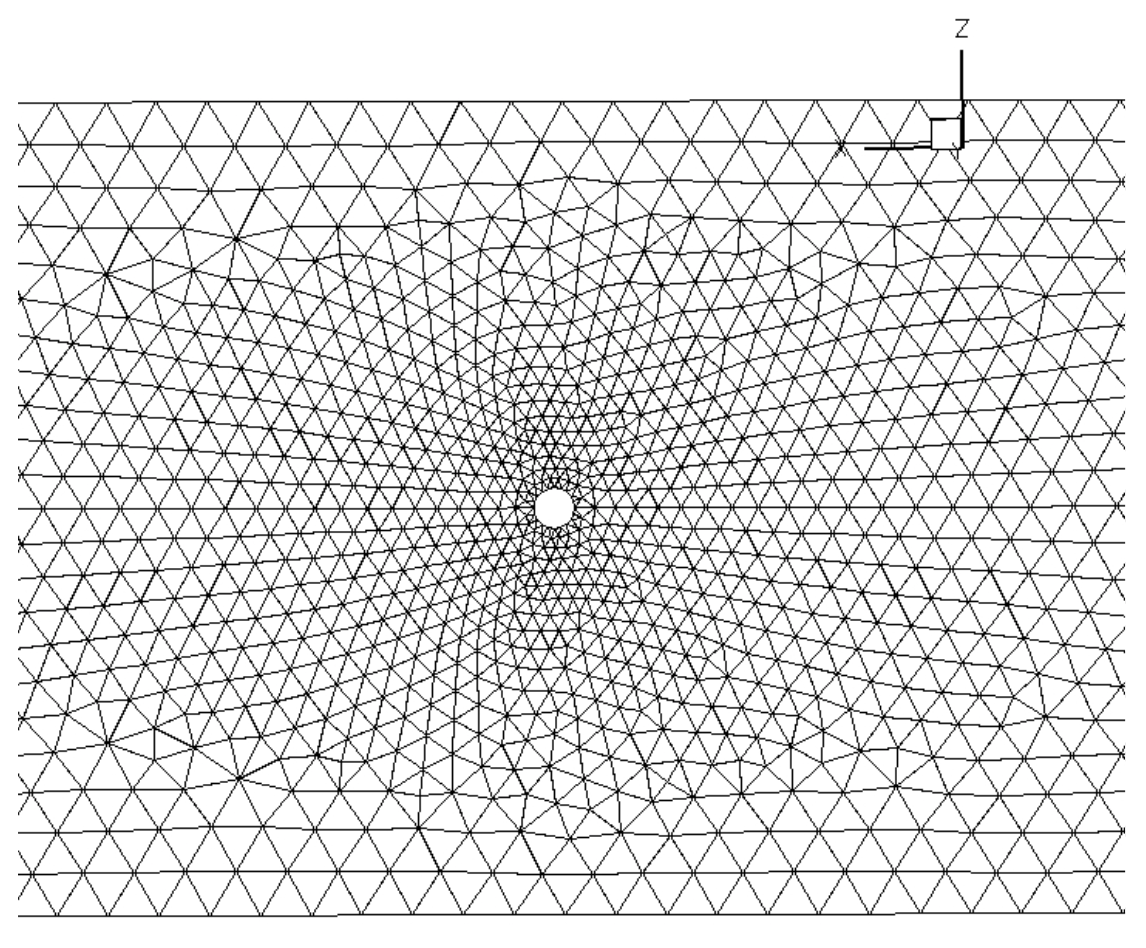} 
  \caption{Initial mesh-zoom, section Y=0}
  \label{fig:zoomArroundEarth}
\end{minipage}
\hspace{.05\linewidth}
\begin{minipage}{.45\linewidth}
  \includegraphics[width=\linewidth]{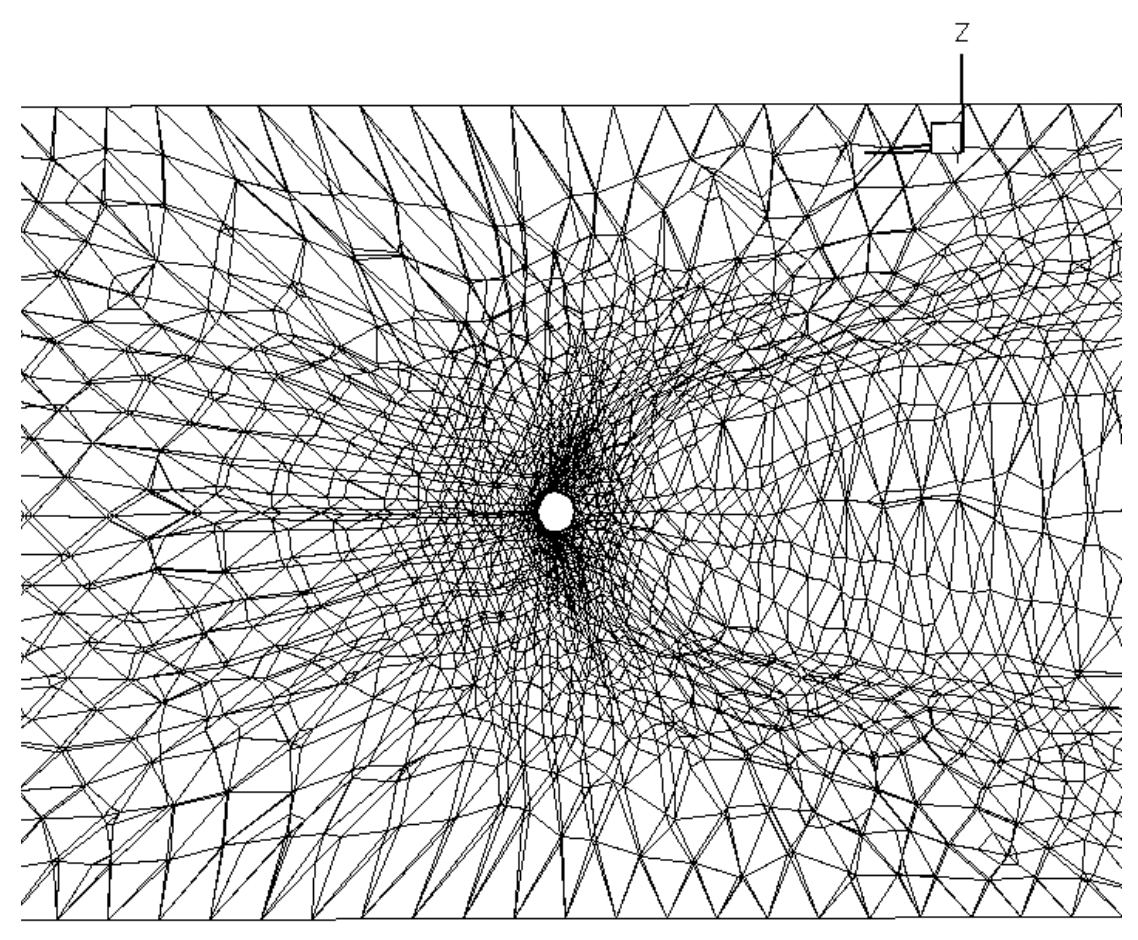}
  \caption{Final mesh- zoom, section Y=0}
  \label{fig:FinalMeshZoomView}
\end{minipage}
\end{figure}

\begin{figure}[H]
\centering{\includegraphics[scale=0.55]{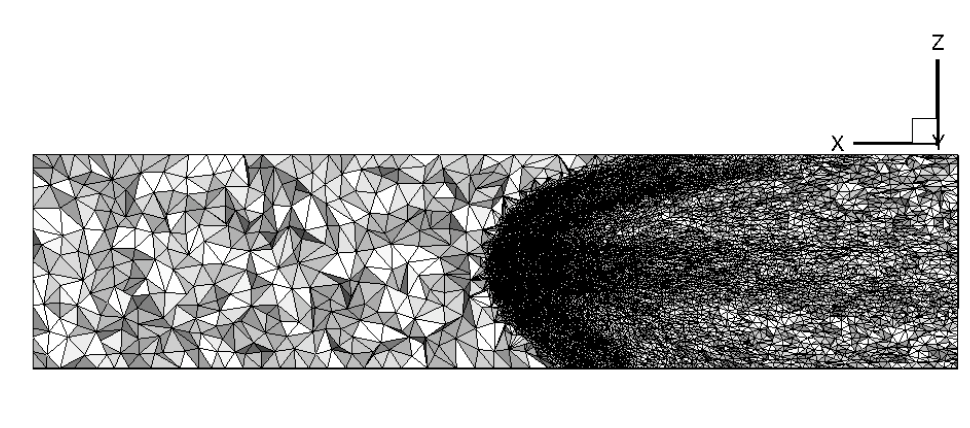}}
\caption{Adapted mesh from \cite{solarwindA}, section Y=0 }
\label{fig:SWadapted}
\end{figure}

The initial mesh is shown in Fig.\ref{fig:fullView} (full view) and Fig.\ref{fig:zoomArroundEarth} (zoom around the Earth), while the final adapted mesh corresponding to the converged steady state solution is presented in Fig.\ref{fig:FinalMeshView} (full view) and Fig.\ref{fig:FinalMeshZoomView} (zoom around the Earth). The reference solution for this case was computed on a mesh with $2773426$ tetrahedral elements (see.Fig.\ref{fig:SWadapted}), while this work shows promising results (at least qualitatively) even for those kind of complex applications using only $197060$ tetrahedral elements. 

\begin{figure}[H]
        \captionsetup{justification=centering}

    \begin{minipage}[t]{6cm}
        \centering
        \includegraphics[width=6cm]{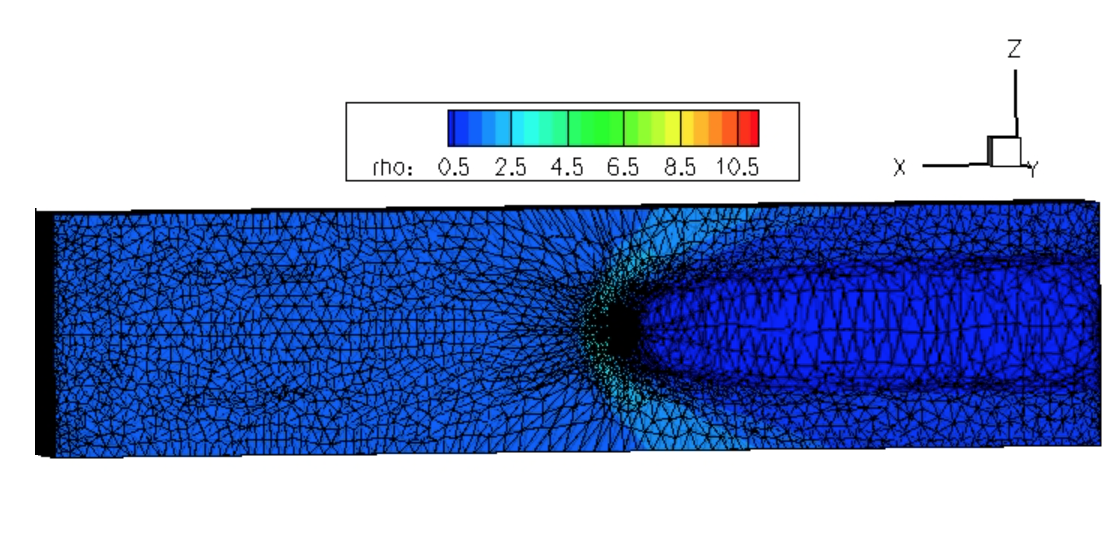}
        \caption{Mesh and density, section Y=0}
    \end{minipage}
    \begin{minipage}[t]{6cm}
        \centering
        \includegraphics[width=6cm]{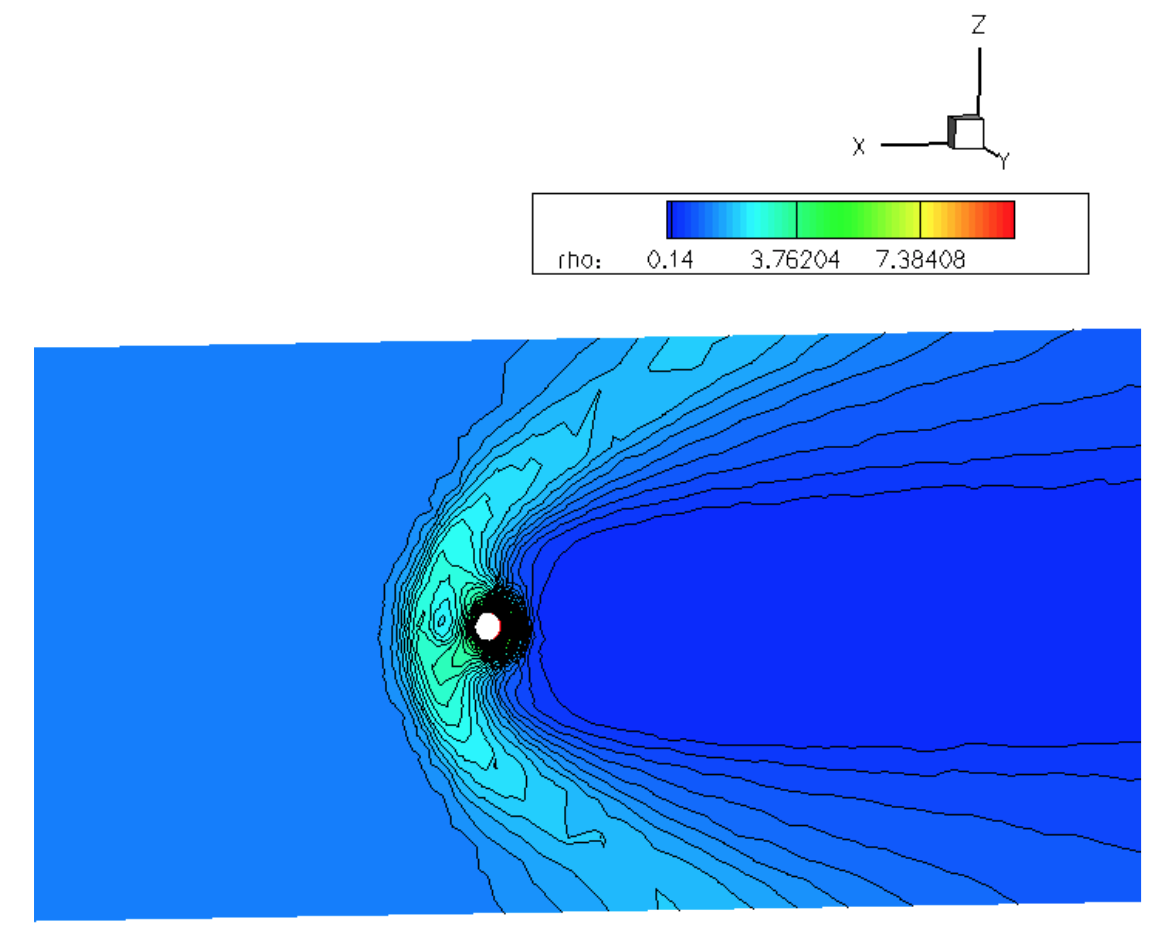}
        \caption{Density contours, section Y=0}
    \end{minipage}
\end{figure}

\begin{figure}[H]
\centering{\includegraphics[scale=0.4]{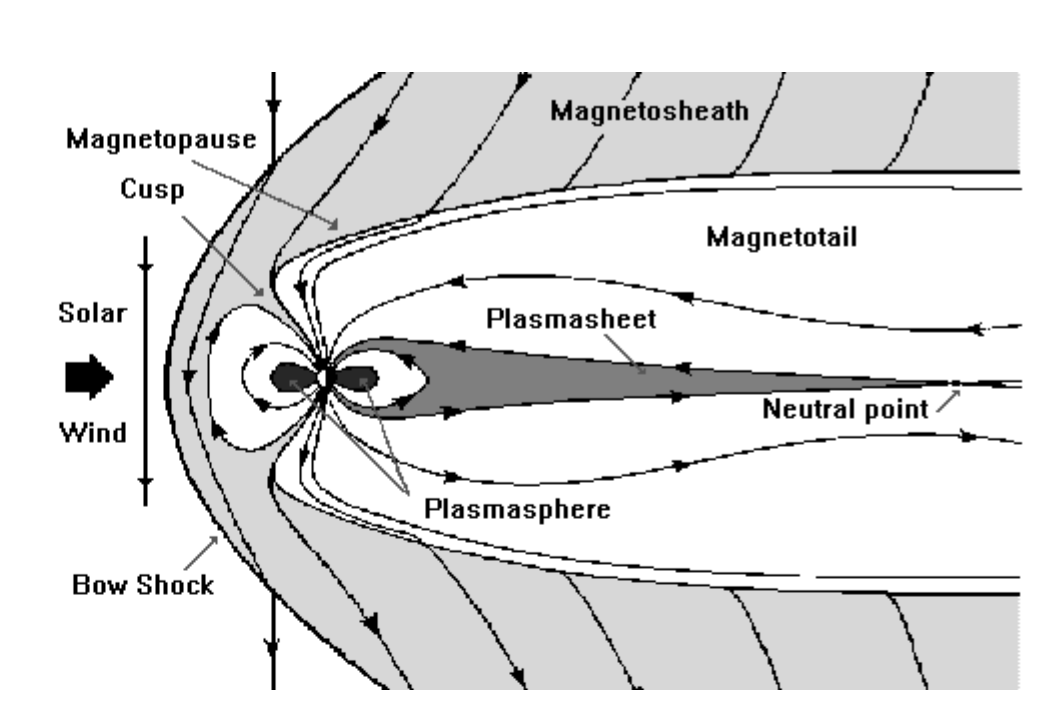}}
\caption{General flow features of the solar wind/Earth's magnetosphere interaction \cite{solarwindtheory}}
\label{fig:solarwindt}
\end{figure}

\begin{figure}[H]
\centering{\includegraphics[scale=0.5]{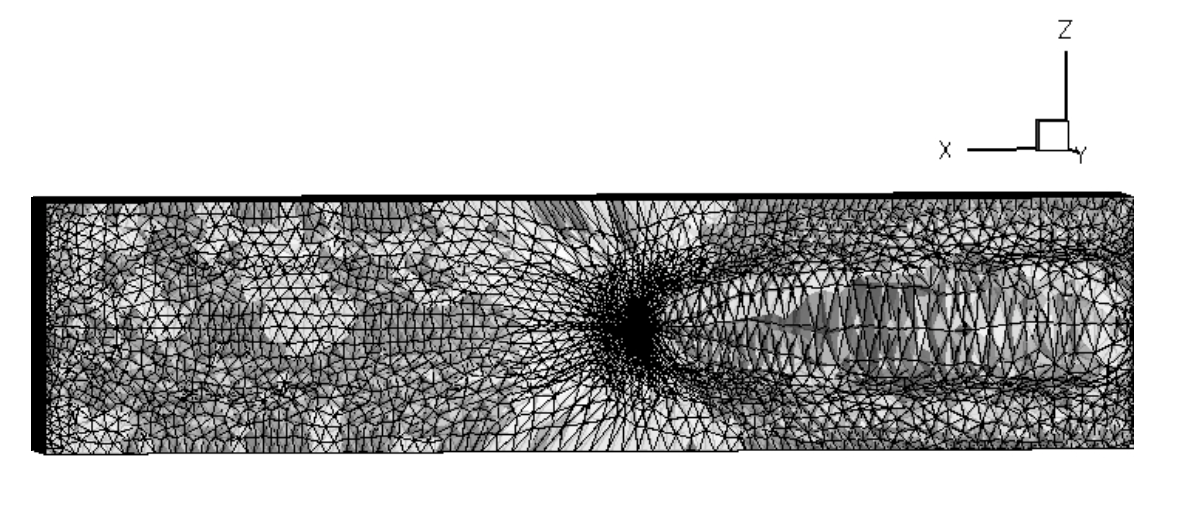}}
\caption{Final mesh, section Y=0}
\label{fig:SW3D}
\end{figure}

The main features of the plasma field in the Earth magnetosphere are detected by the r-adaptation as compared to the sketch in Fig.\ref{fig:SW3D}. In particular, the bow shock and the magnetopause are well resolved as shown in Fig.\ref{fig:solarwindt}.

\subsection{MHD Rotor}
\begin{figure}[H]
        \captionsetup{justification=centering}

    \begin{minipage}[t]{6cm}
        \centering
        \includegraphics[width=6cm]{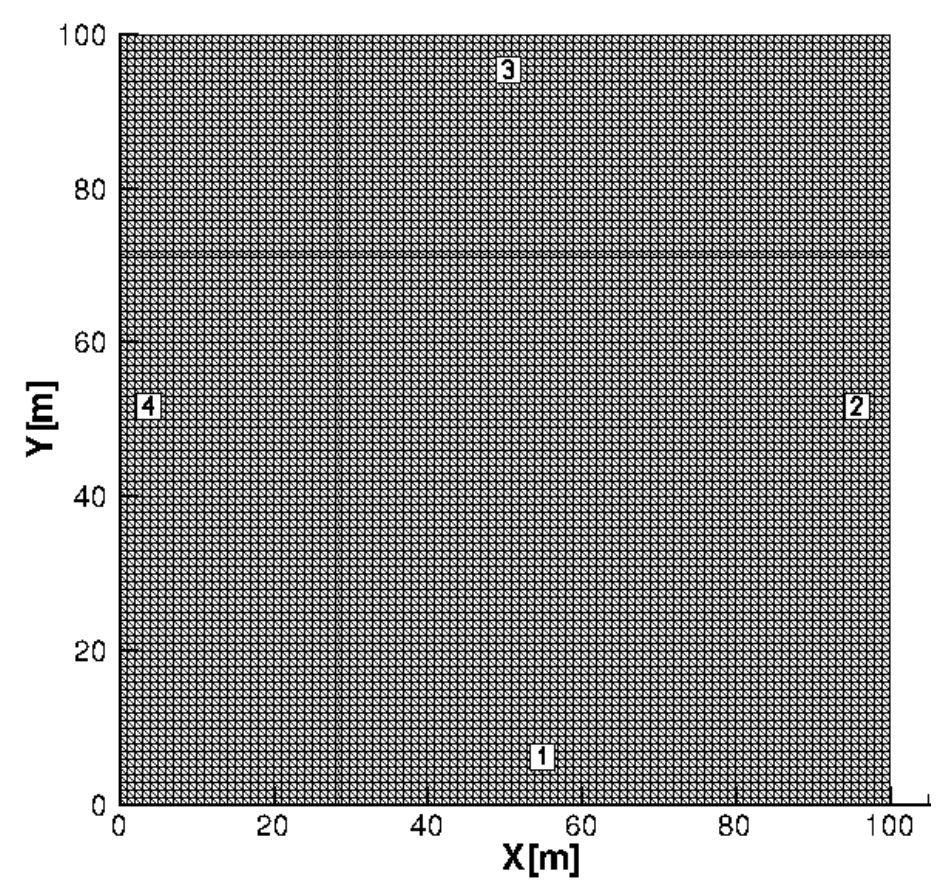}
        \caption{Computational Domain -- Rotor}
        \label{fig:RotorCD}
    \end{minipage}
    \begin{minipage}[t]{6cm}
        \centering
        \includegraphics[width=6cm]{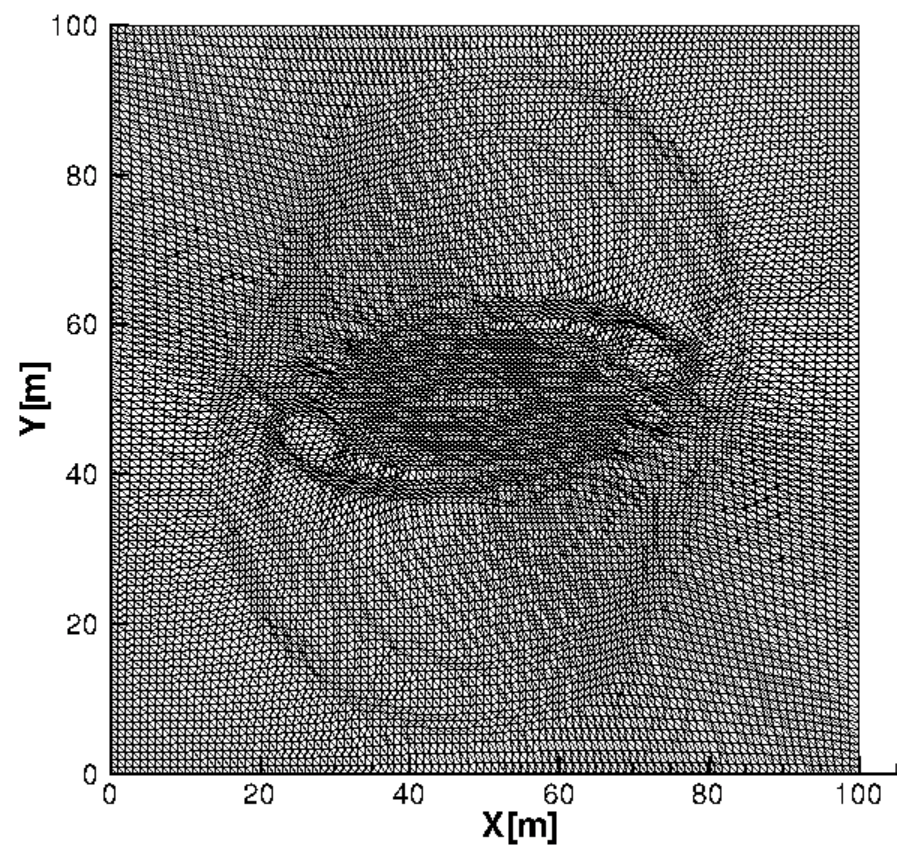}
        \caption{Rotor -- Adapted Mesh}
        \label{fig:AdaptedRotor}
    \end{minipage}
\end{figure}

The test case studies the evolution of strong torsional Alfv\'en waves in ideal MHD. More details about this case can be found in \cite{ALVAREZLAGUNA}.
The ideal 2D MHD Rotor test case conditions are presented in Tab.\ref{tab:RotorFC}, Tab.\ref{tab:RotorMesh} and Tab.\ref{tab:RotorAMR}, while the corresponding unstructured mesh is shown in Fig.\ref{fig:RotorCD}. 

\begin{table}[H]
\centering
\caption{Rotor -- Flow characteristics at $t=0$}
\label{tab:RotorFC}
\begin{tabular}{|ccccccc|}
\hline
 \footnotesize{Physical Model} & \footnotesize{$\textbf{B}$} & \footnotesize{$\textbf{E}$} &  \multicolumn{3}{c|}{\footnotesize{$\rho$}} \\
\footnotesize{MHD} & \footnotesize{$(2.5/ \sqrt{4\pi}, 0, 0)$}   & \footnotesize{(0, 0, $B_x$ $u_y$)}   &  \multicolumn{3}{c|}{\footnotesize{1+9f(t)}}  \\
\hline
\footnotesize{$u_x$}  & \footnotesize{$u_y$} & \footnotesize{T} & \multicolumn{3}{c|}{\footnotesize{$f(r)$}} \\
\footnotesize{-2$f(r)$y/10; r<10}  & \footnotesize{2$f(r)$x/10; r<10} & \footnotesize{0.5/(1+9$f(t)$)}  & \multicolumn{3}{c|}{\footnotesize{1; r<10 -- 0; r>11.5 }}\\
\footnotesize{-2$f(r)$y/r; r$\ge$10 }  & \footnotesize{2$f(r)$x/r; r$\ge$10} &\footnotesize{} &\multicolumn{3}{c|}{\footnotesize{$\frac{200}{3}(11.5-r)$; 10$\le$ r $\le$ 11.5}} \\
\hline
\end{tabular}
\end{table}

\begin{table}[H]
\centering
\caption{Rotor -- Mesh characteristics}
\label{tab:RotorMesh}
\begin{tabular}{|cccc|}
\hline
 \footnotesize{Dimensions} & \footnotesize{Type} & \footnotesize{\# Elements} & \footnotesize{BC 1 .. 4}   \\
\footnotesize{2D} &  \footnotesize{Triangle}    & \footnotesize{20000}   & \footnotesize{Outlet} \\
\hline
\end{tabular}
\end{table}

\begin{figure}[H]
        \captionsetup{justification=centering}

    \begin{minipage}[t]{6cm}
        \centering
        \includegraphics[width=6cm]{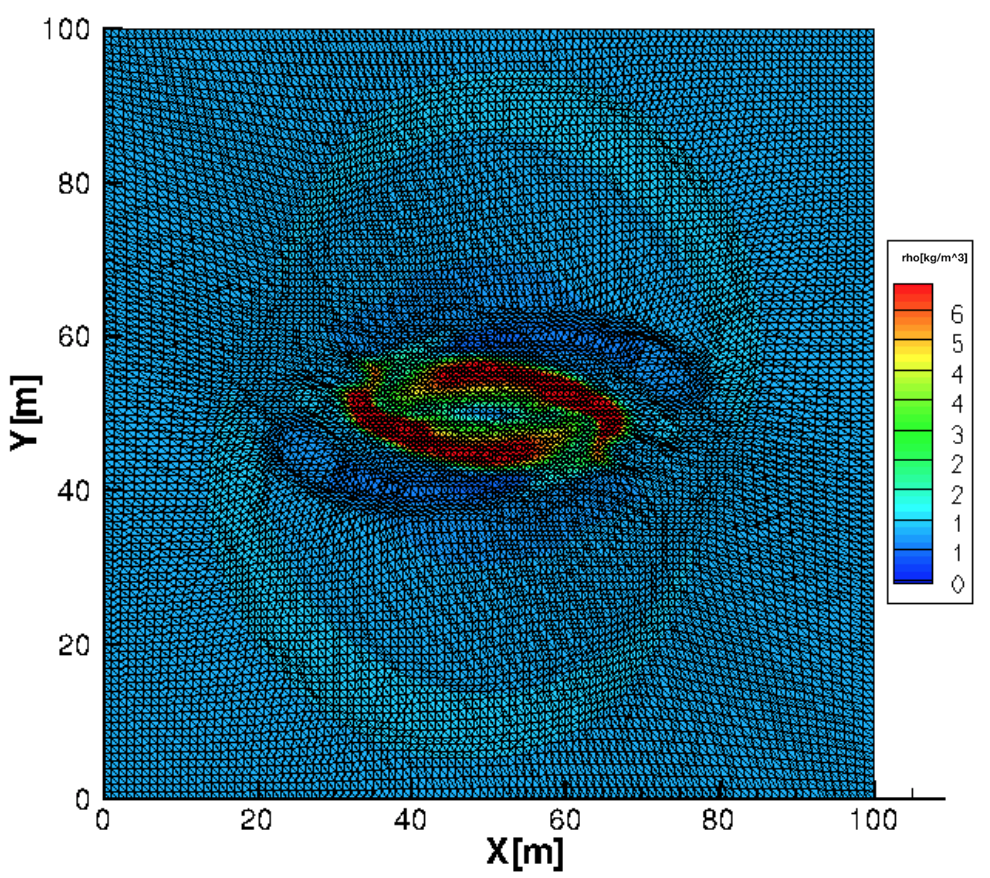}
        \caption{Rotor -- Density field}
        \label{fig:MHDRotorAMRdensity}
    \end{minipage}
    \begin{minipage}[t]{6cm}
        \centering
        \includegraphics[width=6cm]{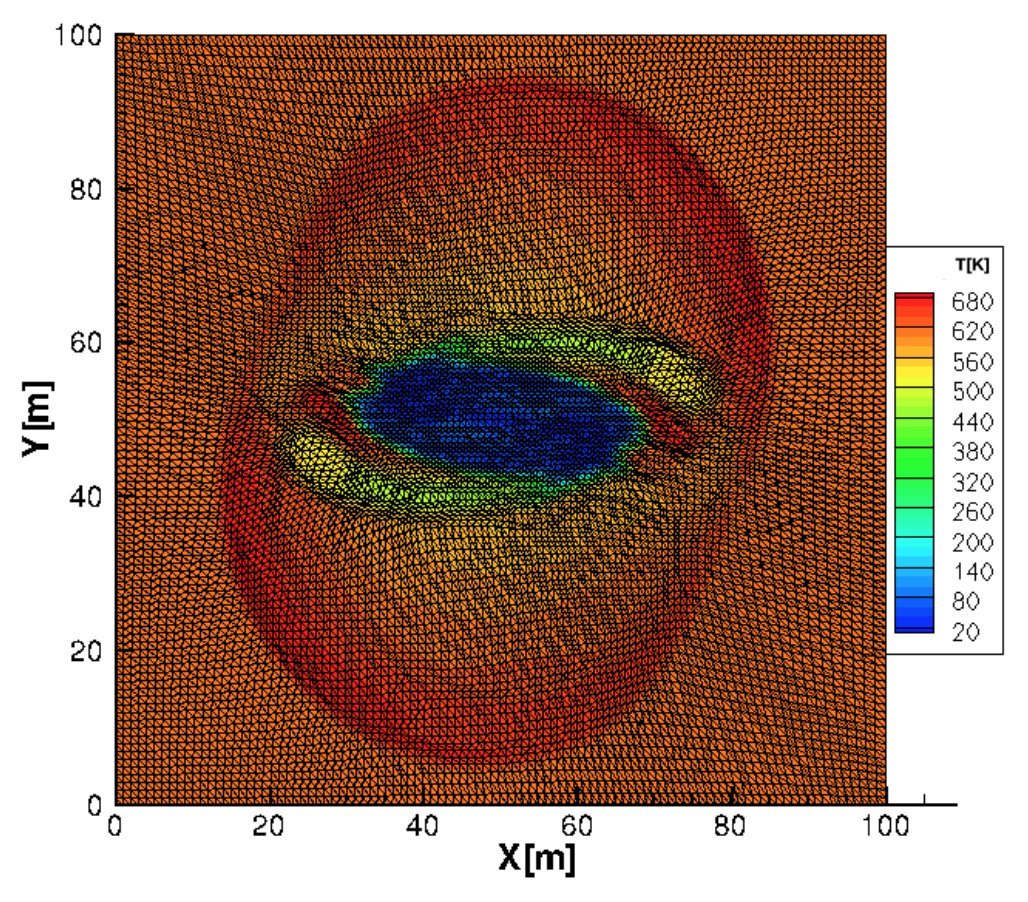}
        \caption{Rotor -- Temperature field }
        \label{fig:MHDRotorAMRT}
    \end{minipage}
\end{figure}

\begin{table}[H]
\centering
\caption{ Rotor -- r-refinement}
\label{tab:RotorAMR}
\begin{tabular}{|cccccc|}
\hline
\footnotesize{Spring Network} &\footnotesize{Monitor Variable} & \footnotesize{Process Rate} & \footnotesize{Stop AMR time} & \footnotesize{minPer} & \footnotesize{maxPer}    \\
 \footnotesize{Linear}  & \footnotesize{Flow density} &\footnotesize{1} & \footnotesize{t=0.2962} & \footnotesize{0.30}  & \footnotesize{0.55}   \\
\hline
\end{tabular}
\end{table}
The initial mesh in Fig.\ref{fig:RotorCD} is unstructured and obtained by splitting a uniform structured mesh. The refined mesh in Fig.\ref{fig:AdaptedRotor} appears to follow closely the main flow features, as highlighted in Fig.\ref{fig:MHDRotorAMRdensity} (density) and Fig.\ref{fig:MHDRotorAMRT} (temperature).

\section{Mesh Quality Indicator}
\label{sec:MQI}
\subsection{Motivation}
The following section will present a new method to grade an adapted mesh qualitatively. The mesh r-adaptive algorithm re-positions grid nodes according to a certain monitor flow field variable. For instance, if one monitors the density of the flow field, the nodes will migrate and the local concentration of the mesh node will increase at discontinuities. Hence, for an adequate refinement, the cells around the discontinuity result to be highly distorted. The Author's key idea is based on defining a certain cell distortion criteria and coupling it to the local physical properties of the monitored flow field variable. Let $\mathcal{D}_{init}$ be the measure of a cell distortion on the initial un-modified mesh and $\mathcal{D}_{final}$ at the end of the refinement, both extrapolated to nodal values.\\
In order to reflect the physics of the problem, the function $f(\mathcal{D}_{init},\mathcal{D}_{final})$ is multiplied by the ratio of the monitored flow state variable. Let $\mathcal{S}_{init}$ the initial monitored nodal state and $\mathcal{S}_{final}$ at the end of the refinement.\\
The proposed mesh quality indicator ($\mathcal{MQI}$) is expressed as:
\begin{equation}
    \label{eq:quality}
    \mathcal{MQI} = f(\mathcal{D}_{init},\mathcal{D}_{final})~ \frac{\mathcal{S}_{final}}{\mathcal{S}_{init}}.
\end{equation}
\subsection{Analysis of MQI}
\label{point:analysis}
 \begin{itemize}
    \item For the free-stream flow, the ratio $\frac{\mathcal{S}_{final}}{\mathcal{S}_{init}}$ should be equal to 1. Since the AMR is physic driven, the mesh nodes within the free stream do not move. Therefore, $f(\mathcal{D}_{init},\mathcal{D}_{final})=\mathcal{C}$, where $\mathcal{C}$ is a constant yielding to  $\mathcal{MQI}=\mathcal{C}$.
    \item If both the ratio $\frac{\mathcal{S}_{final}}{\mathcal{S}_{init}}$ and the distortion function measurement $f(\mathcal{D}_{init},\mathcal{D}_{final})$ increase (resp. decrease), then, $MQI >> \mathcal{C}$ (resp. $\mathcal{MQI} << \mathcal{C}$). As a result, the mesh fitting is inadequate.
    \item If the ratio $\frac{\mathcal{S}_{final}}{\mathcal{S}_{init}}$ increases, the local refinement is needed. Therefore, the function $f(\mathcal{D}_{init},\mathcal{D}_{final})$ must incorporate the philosophy of the distortion criteria and reflect the increase of the local mesh nodes density.
 \end{itemize}

\subsection{MQI applied to a 2D mesh}

\subsubsection{Triangular mesh}
\begin{figure}[H]
\centering
\begin{minipage}{.44\linewidth}
  \includegraphics[width=\linewidth]{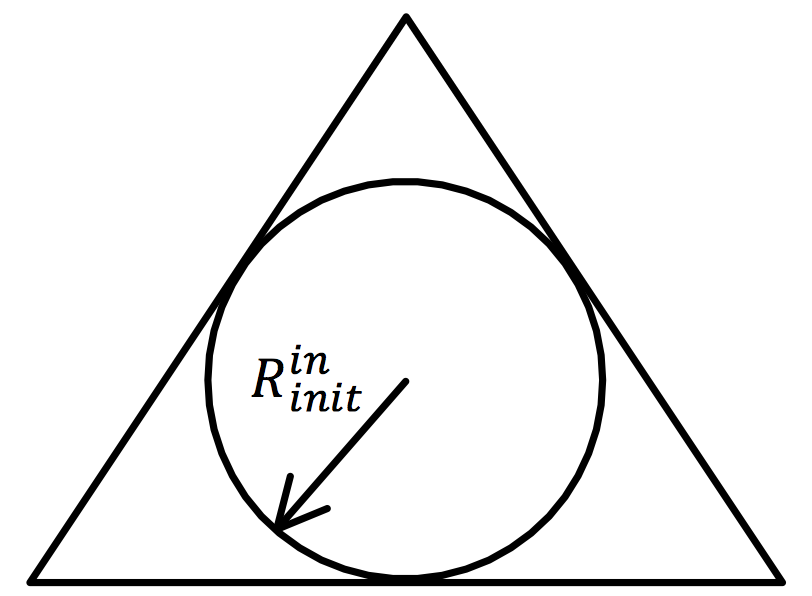}
  \caption{Initial element}
  \label{fig:incircle}
\end{minipage}
\hspace{.05\linewidth}
\begin{minipage}{.41\linewidth}
  \includegraphics[width=\linewidth]{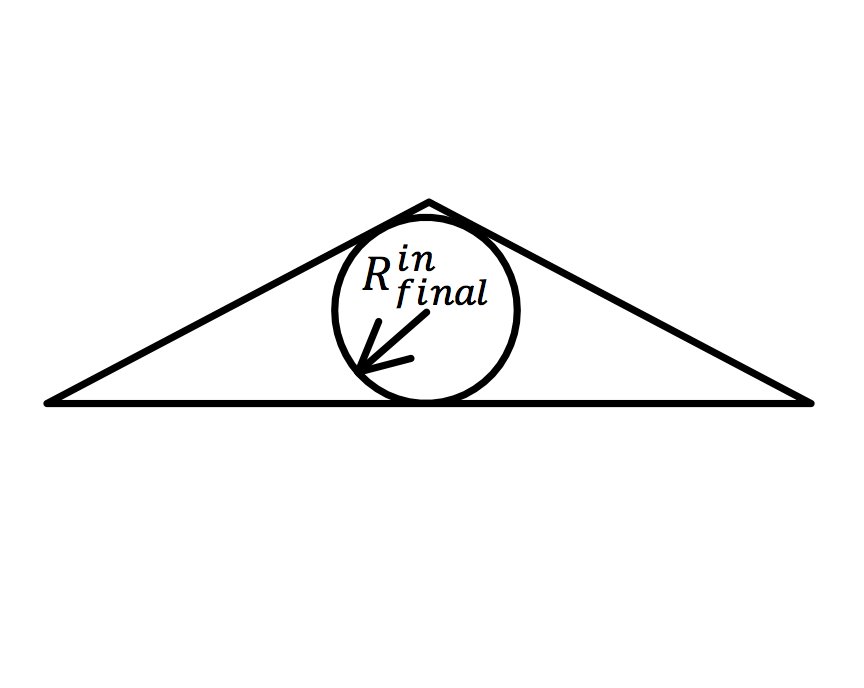}
  \caption{Distorted element}
  \label{fig:incircle2}
\end{minipage}
\end{figure}

The cell distortion criteria $\mathcal{D}$ is defined as the radius of the inscribed circle of the mesh triangular element and denoted as $\mathcal{R}^{in}$. The in-circle radius formulation gives a direct information about the triangle distortion as Fig.\ref{fig:incircle} and Fig.\ref{fig:incircle2} show.

The Eq.\ref{eq:quality} is transformed into:
\begin{equation}
\label{eq:MQI_radius}
    \mathcal{MQI} = \frac{\mathcal{R}_{final}^{in}}{\mathcal{R}_{init}^{in}} ~ \frac{\mathcal{S}_{final}}{\mathcal{S}_{init}}.
\end{equation}\\
\begin{itemize}
    \item \textit{Discussion: Choice Of $f(\mathcal{D}_{init},\mathcal{D}_{final})$ }
\end{itemize}

\begin{figure}[H]
\centering{\includegraphics[scale=0.5]{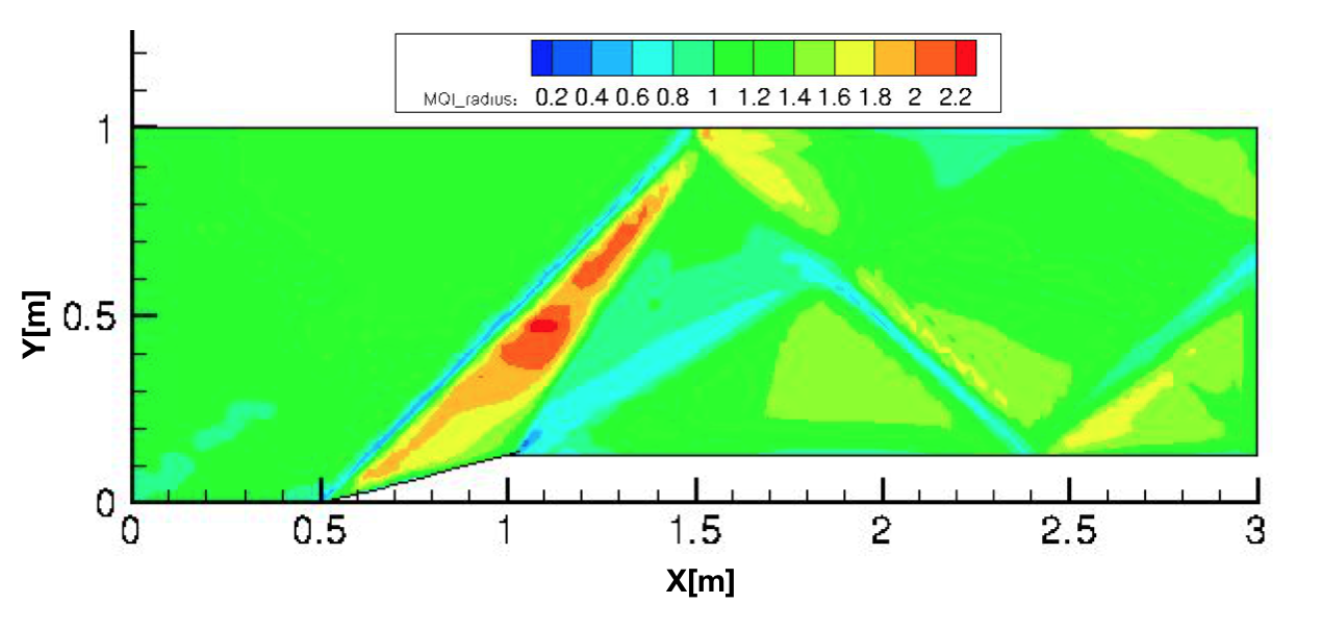}}
\caption{$\mathcal{MQI}$ applied to the 2D triangular double wedge}
\label{fig:MQI_radius}
\end{figure}

First, the ratio $\frac{\mathcal{R}_{final}}{\mathcal{R}_{init}}$  is further investigated:
$$
   \frac{\mathcal{R}_{final}}{\mathcal{R}_{init}} \left\{
    \begin{array}{ll}
         =1,  \mbox {      if the cell keeps the same shape; }\\
         <1,  \mbox {      if the cell becomes narrow;}\\
        >1,  \mbox {      if the cell becomes extended.}
    \end{array}
\right.
$$
\begin{itemize}
    \item \textit{Computation of the in-circle radius}
\end{itemize}
\cite{Rin} expresses, for a triangle $ijk$, the in-radius $\mathcal{R}^{in}$ formulation based on Eq.\ref{eq:inRadius}:
\begin{equation}
    \label{eq:inRadius}
    \mathcal{R}^{in}=\frac{2 A_{ijk}}{d_{ij}+d_{ik}+d_{jk}},
\end{equation}
where $A_{ijk}$ denotes the area of the triangle $ijk$ and $d_{ij}$ denotes the distance between $i$ and $j$ vertices. The extrapolation to a nodal value is done by averaging all the in-circle radius of the $N$ triangles attached to the considered vertex $i$.
\begin{equation}
\label{Extrapolation1}
    \mathcal{R}^{in}_{i}=\frac{1}{N}\sum_{m=1}^{N}\frac{2 A_{ijk}^{m}}{d_{1}^{m}+d_{2}^{m}+d_{3}^{m}}.
\end{equation}
\begin{itemize}
    \item \textit{Results}
\end{itemize}
\underline{2D Wedge}\\
The results of computing the $\mathcal{MQI}$, defined by Eq.\ref{eq:MQI_radius}, are presented in Fig.\ref{fig:MQI_radius}.

\begin{figure}[H]
\centering{\includegraphics[scale=0.38]{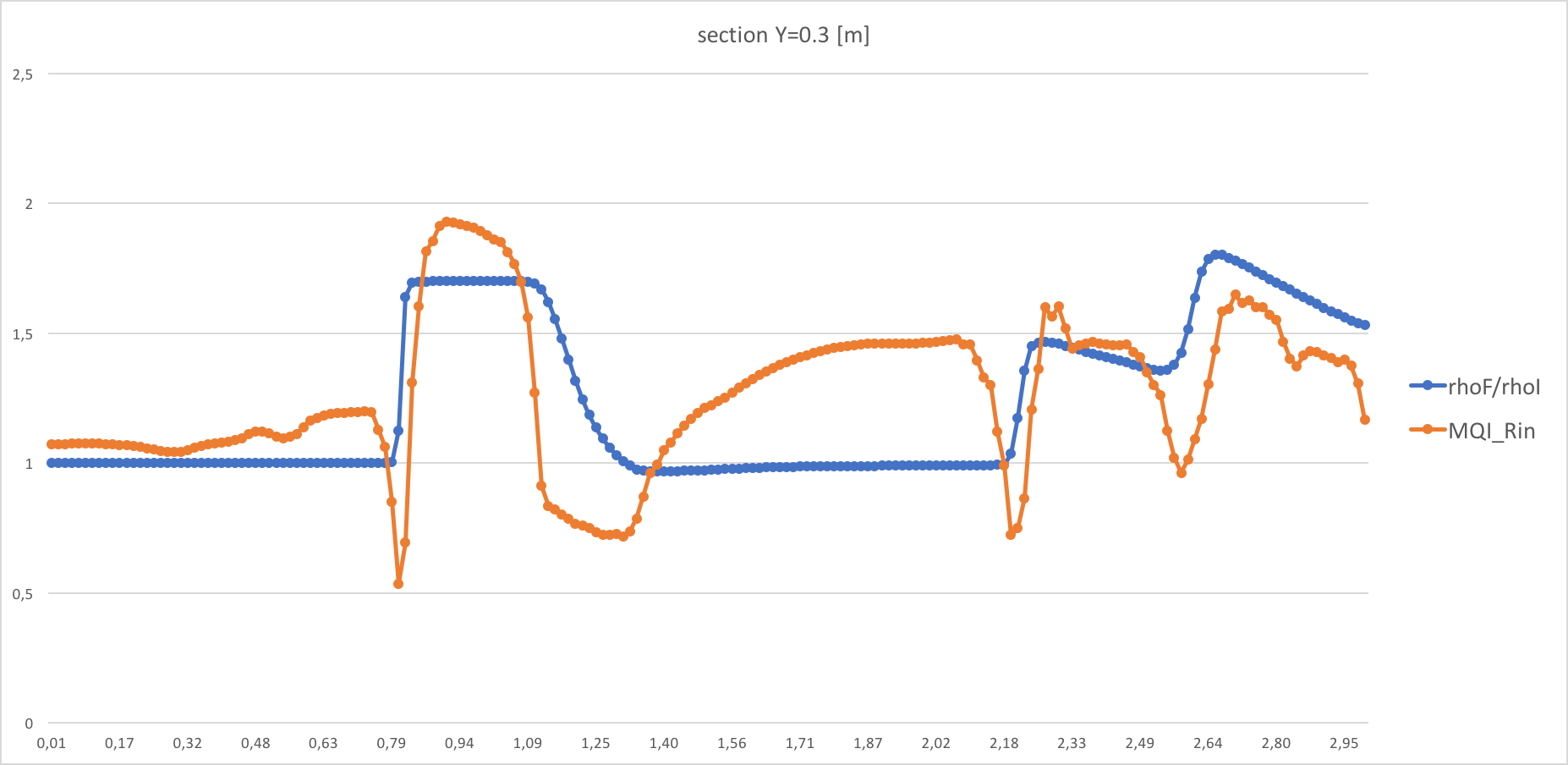}}
\caption{$\mathcal{MQI}$ for double wedge triangular test case at a line section $Y=0.3[m]$}
\label{fig:secWedgeT0.3}
\end{figure}

\begin{figure}[H]
\centering{\includegraphics[scale=0.4]{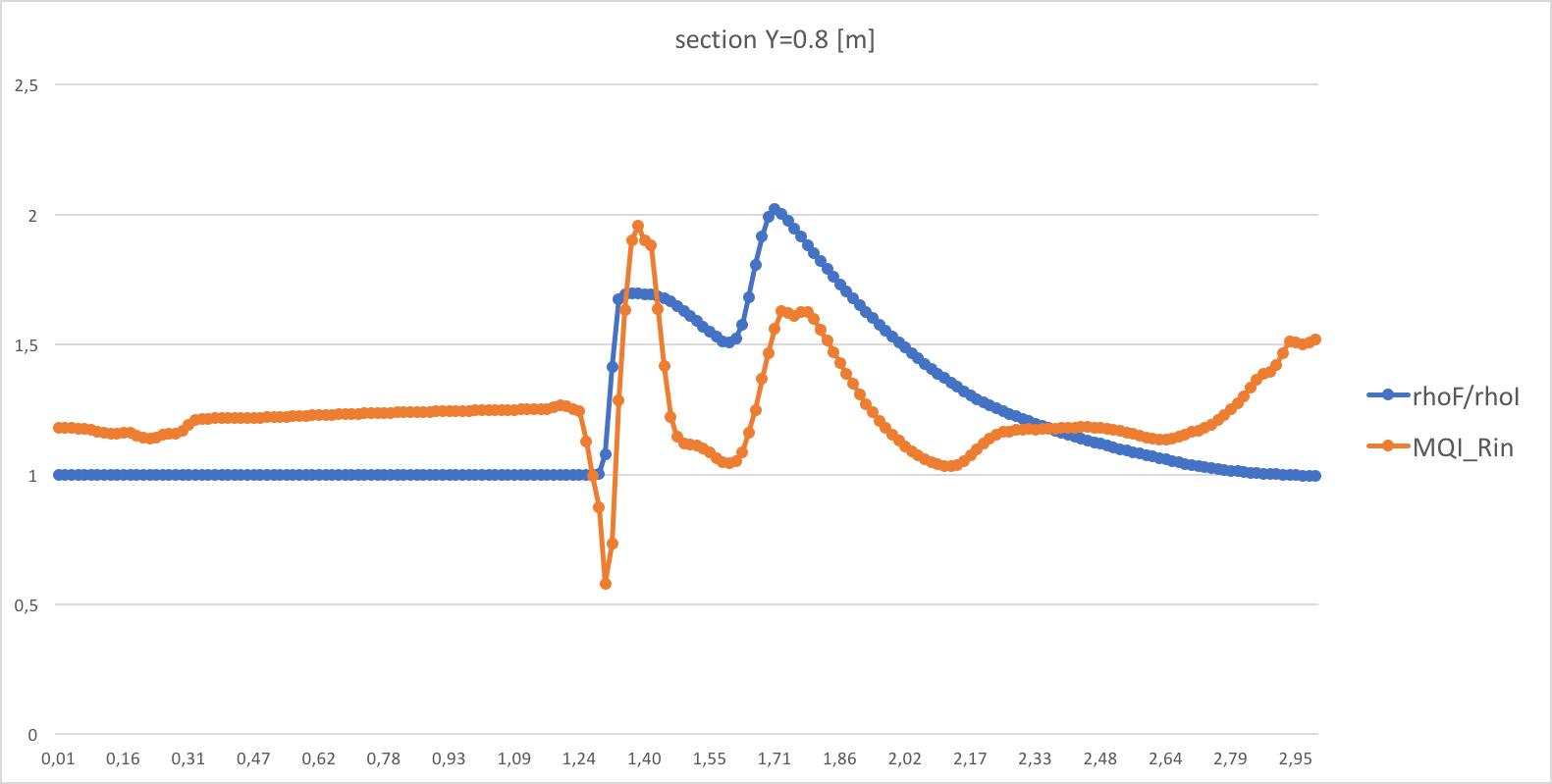}}
\caption{$\mathcal{MQI}$= for double wedge triangular test case at a line section $Y=0.8[m]$}
\label{fig:secWedgeT0.8}
\end{figure}

The free-stream presents a value of $\mathcal{MQI}$ equal 1. The increase of the $\mathcal{MQI}$ value after the discontinuities (red zone after the first oblique shock and yellow zone after the first reflection of the oblique shock) is explained by the increase of the ratio $\frac{\mathcal{R}_{final}}{\mathcal{R}_{init}}$. Since the nodes adjacent to a discontinuity will contribute to the growth of the local grid resolution and the r-adaptive technique does not either add nodes nor change connectivity, then, the cells size next to a discontinuity will increase. Further analysis are presented in Fig.\ref{fig:secWedgeT0.3} and Fig.\ref{fig:secWedgeT0.8}.\\

Fig.\ref{fig:secWedgeT0.3} shows a $\mathcal{MQI}$ decrease at discontinuities (i.e $\frac{\rho_{Final}}{\rho_{Init}}$ increases at the oblique shock and its reflections). \\
Let $\mathcal{S}_\infty$ be set of nodes with $ X$ $\in$ [0, 1.18].\\
Let $\mathcal{S}_1$ be set of nodes with $X$ $\in$ [1.40, 2.1].\\
For nodes $\in$ $\mathcal{S}_\infty$, the $\mathcal{MQI}$ $\ne$ 1. This is due to the mesh relaxation and the equilibrium node position after the refinement. The goal is to better refine the main oblique shock. As a consequence, $\mathcal{MQI}$ $\approx$ 1. Yet, this increase is not too large and can be accepted since the oblique shock is better refined and nothing of interest happens in the free stream.\\
The jumps in the density ratio reflects the existence of shocks. At those positions, the $\mathcal{MQI}$ shows a strong peak with respect to the state jump. Therefore, this can be explained by the fact that the cells are becoming smaller and smaller implying a good mesh refinement. Hence, $\mathcal{MQI}$ peaks (e.g. peak I, peak II and peak III in Fig.\ref{fig:secWedgeT0.3} at the positions $X=0.75[m]$, $X=2.18[m]$ and $X=2.64[m]$ respectively indicating the position of the $1^{st}$ oblique shock and its reflections) reflect partially the ability of a cell to deform and show the intensity of the aforementioned shocks.  
The $\mathcal{MQI}$'s overshoots with respect to the density ratio indicate cells enlargement. For example, for nodes $\in$ $\mathcal{S}_1$, the $\mathcal{MQI}$ $\ne$ 1. This overshoot was expected since the grid nodes in $\mathcal{S}_1$ are pulled to contribute to both the main oblique shock and its first reflection.

Fig.\ref{fig:secWedgeT0.8} shows the same conclusions as Fig.\ref{fig:secWedgeT0.3} for the free stream flow, main oblique shock and the $\mathcal{MQI}$ overshoot.\\
Let $\mathcal{S}_2$ be set of nodes with $X$ $\in$ [1.6, 2.4].\\
The mesh nodes $\in$ $\mathcal{S}_2$ are subject to the expansion wave and reflection of the oblique shock interaction. Since $\frac{\rho_{Final}}{\rho_{Init}}$ $>$ $\mathcal{MQI}$ $\Rightarrow$ 
$\frac{R_{Final}}{R_{Init}}$ $< 1$, the refinement is applied consistently.\\
The $\mathcal{MQI}$ value at the outlet section of the double wedge mesh is greater than $\frac{\rho_{Final}}{\rho_{Init}}$. Hence, the cells are becoming enlarged. In fact, those cells, not subject to any shocks, are pulled and contribute to the refinement of the third reflection of the oblique shock.\\

\begin{figure}[H]
\centering
        \captionsetup{justification=centering}

\begin{minipage}{.44\linewidth}
  \includegraphics[width=\linewidth]{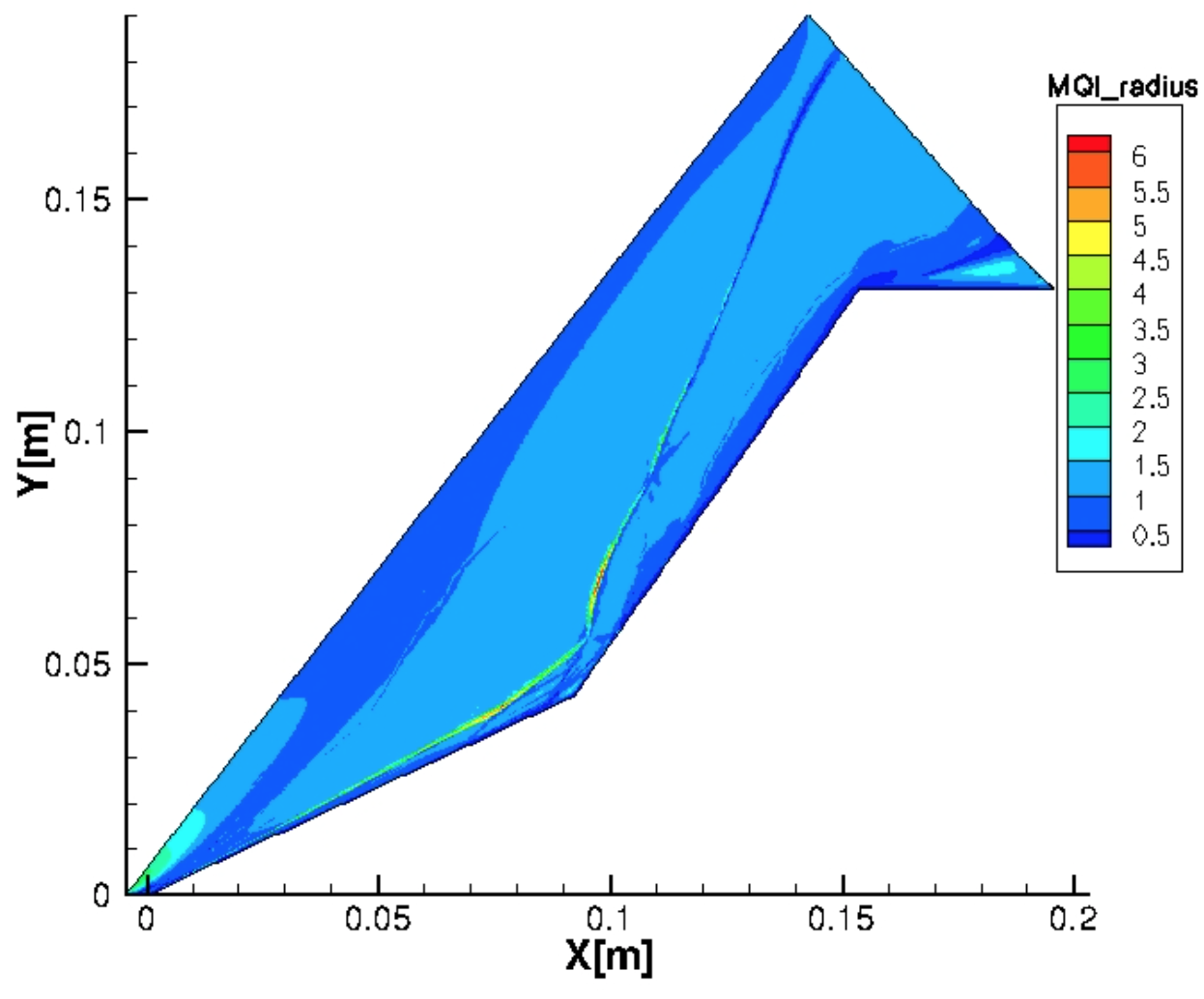}
  \caption{$\mathcal{MQI}$ applied to 2D double cone test case}
  \label{fig:MQIdoublecone}
\end{minipage}
\hspace{.05\linewidth}
\begin{minipage}{.44\linewidth}
  \includegraphics[width=\linewidth]{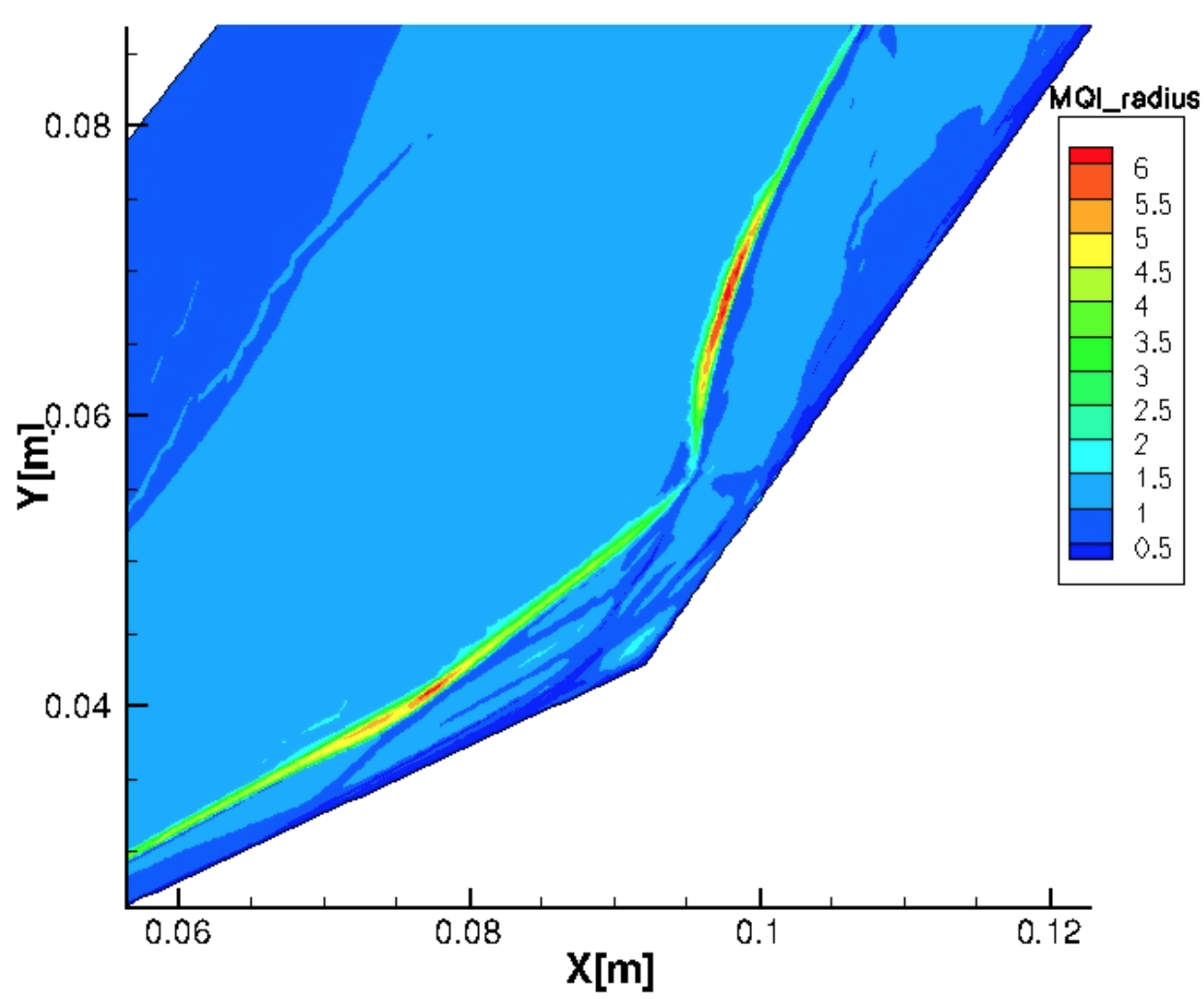}
  \caption{$\mathcal{MQI}$ applied to 2D double cone test case--zoom}
  \label{fig:MQIdoubleconezoom}
\end{minipage}
\end{figure}

\begin{figure}[H]
\centering
\begin{minipage}{.44\linewidth}
  \includegraphics[width=\linewidth]{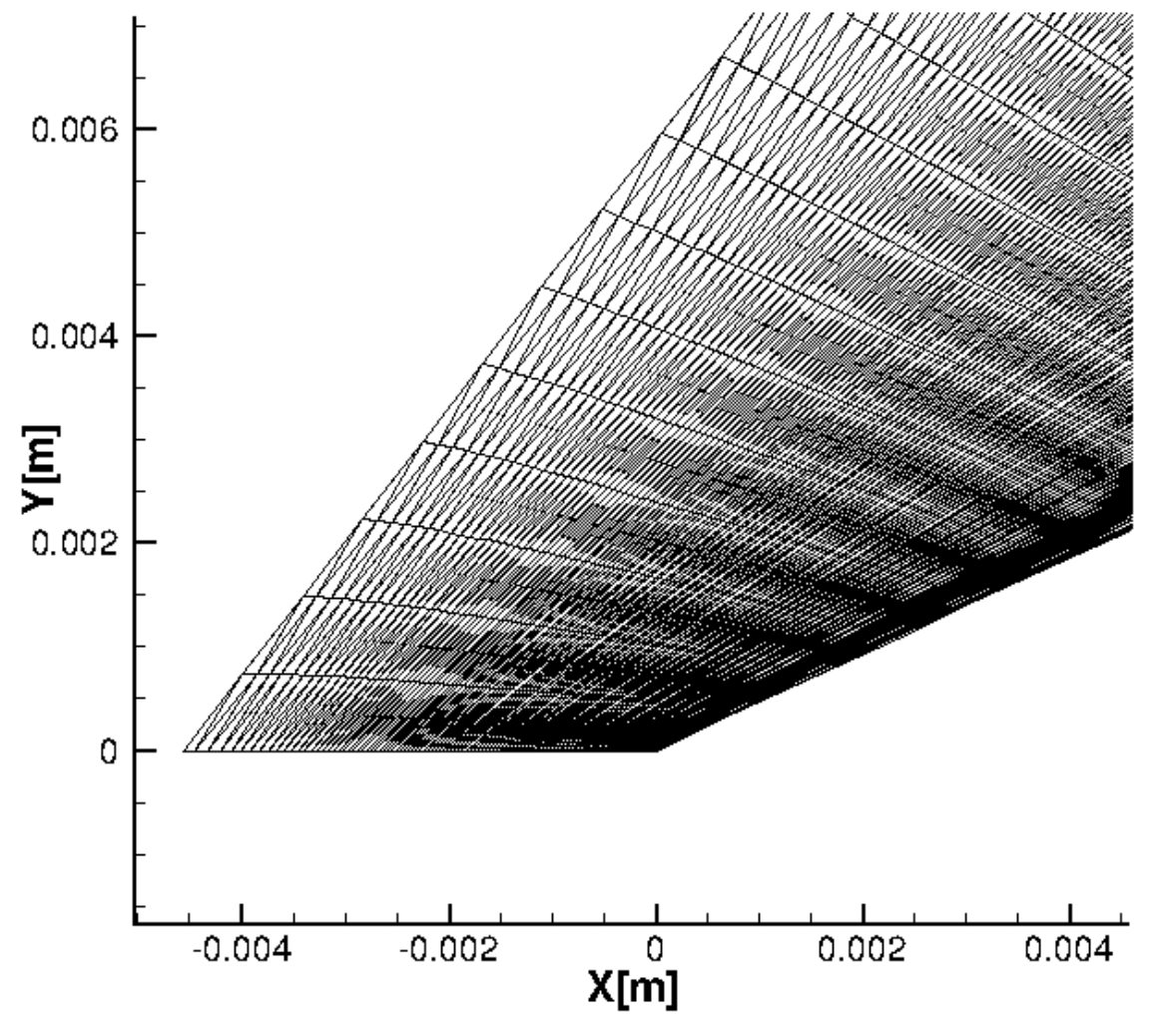}
  \caption{Initial mesh-zoom inlet}
  \label{fig:InletInitial}
\end{minipage}
\hspace{.05\linewidth}
\begin{minipage}{.44\linewidth}
  \includegraphics[width=\linewidth]{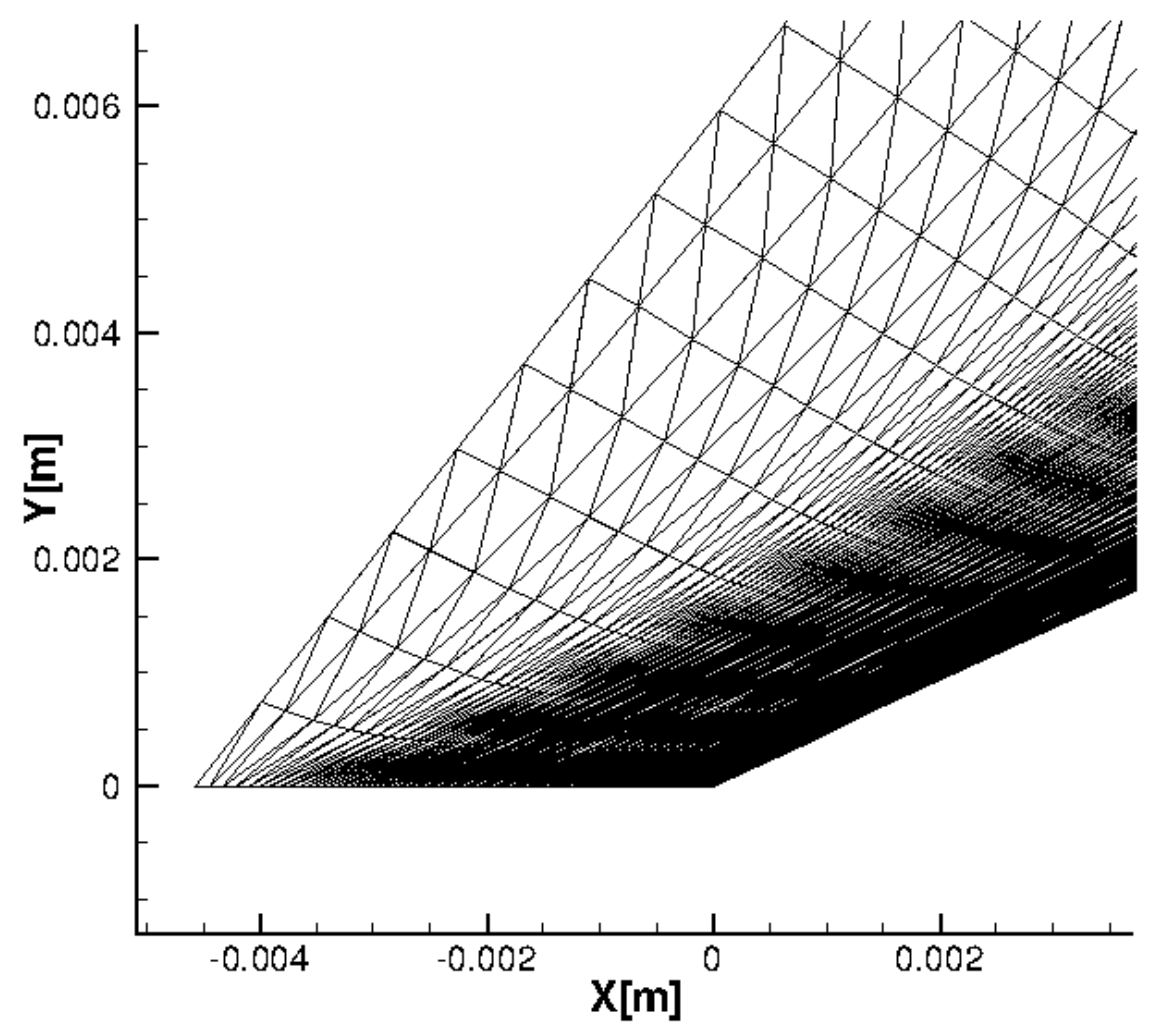}
  \caption{Final mesh-zoom inlet}
  \label{fig:InletFinal}
\end{minipage}
\end{figure}

\underline{Double cone}\\
Fig.\ref{fig:MQIdoublecone} and Fig.\ref{fig:MQIdoubleconezoom} show the mesh quality indicator for the double cone test case, especially the distribution of the $\mathcal{MQI}$ arround the SWBLI. The nodes located within the red zones contribute to the refinement of the adjacent shocks. Therefore, the triangles are enlarged and the radius of the in-circle increases leading to an increase of the $\mathcal{MQI}$ value.\\ 
Fig.\ref{fig:MQIdoublecone} presents a light blue-turquoise zone at the inlet of the double cone due to the nodal contribution to the oblique shock as shown in Fig.\ref{fig:InletInitial} and Fig.\ref{fig:InletFinal}. 

The simulation of the test case is crashing when applying an AMR technique based on the density. Thanks to $\mathcal{MQI}$, we observe a blue-turquoise zone at the level of the second cone that indicates an enlargement of the cells. Those cells are located within the boundary layer and since they become too big and they create a zone of negative pressure. Hence, in order to be able to converge the double cone test case, we would need to add more points in the original mesh or follow another monitor flow field variable. 

\subsection{Quadrilateral mesh}
The cell distortion criteria $\mathcal{D}$ definition for 2D quadrilateral meshes is more complex compared to the triangular mesh. Depending on the test case, $\mathcal{D}$ will be based on the aspect ratio or the skewness of the quadrilateral element. Hence, for the Hornung test case in Sec.\ref{sec:HC}, the aspect ratio $\mathcal{AR}$ will be used as a distortion criterion, whereas, for the quadrilateral double wedge in Sec.\ref{sec:DW}, the skewness $\Theta$ of the element will be adopted.
\begin{figure}[H]
\centering
\begin{minipage}{.40\linewidth}
  \includegraphics[width=\linewidth]{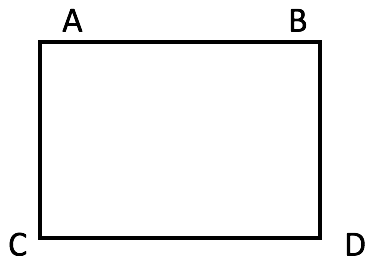}
  \caption{Initial cell}
  \label{fig:AR}
\end{minipage}
\hspace{.05\linewidth}
\begin{minipage}{.47\linewidth}
  \includegraphics[width=\linewidth]{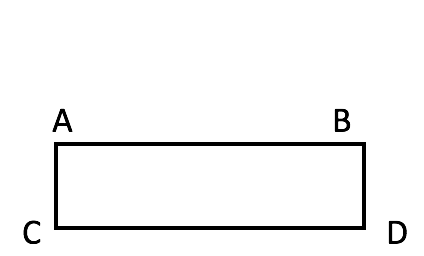}
  \caption{Distorted cell}
  \label{fig:AR2}
\end{minipage}
\end{figure}
Eq.\ref{eq:quality} is transformed into:
\begin{equation}
\label{eq:MQI_AR}
    \mathcal{MQI} = \frac{\mathcal{AR}_{initial}}{\mathcal{AR}_{final}} ~ \frac{\mathcal{S}_{final}}{\mathcal{S}_{init}}.
\end{equation}

First, the ratio $\frac{\mathcal{AR}_{init}}{\mathcal{AR}_{final}}$ will be further investigated.
$$
   \frac{\mathcal{AR}_{init}}{\mathcal{AR}_{final}} \left\{
    \begin{array}{ll}
         =1,  \mbox {      if the cell keeps the same shape; }\\
         <1,  \mbox {      if the cell becomes narrow;}\\
        >1,  \mbox {      if the cell becomes extended.}
    \end{array}
\right.
$$\\
For a quadrilateral $ABDC$, the aspect ratio $\mathcal{AR}$ is determined through the following relation:
\begin{equation}
    \label{eq:AR}
    \mathcal{AR}=\frac{d_{AB}}{d_{AC}},
\end{equation}
where $d_{AC}$ denotes the distance between the nodes $A$ and $C$.
The extrapolation to a nodal value is done by averaging all the aspect ratio of the $N$ elements attached to the considered vertex $i$, according to:
\begin{equation}
\label{Extrapolation}
    \mathcal{AR}_{i}=\frac{1}{N}\sum_{m=1}^{N}\frac{d_{AB}^{m}}{d_{AC}^{m}}.
\end{equation}

\begin{figure}[H]
\centering{\includegraphics[scale=0.4]{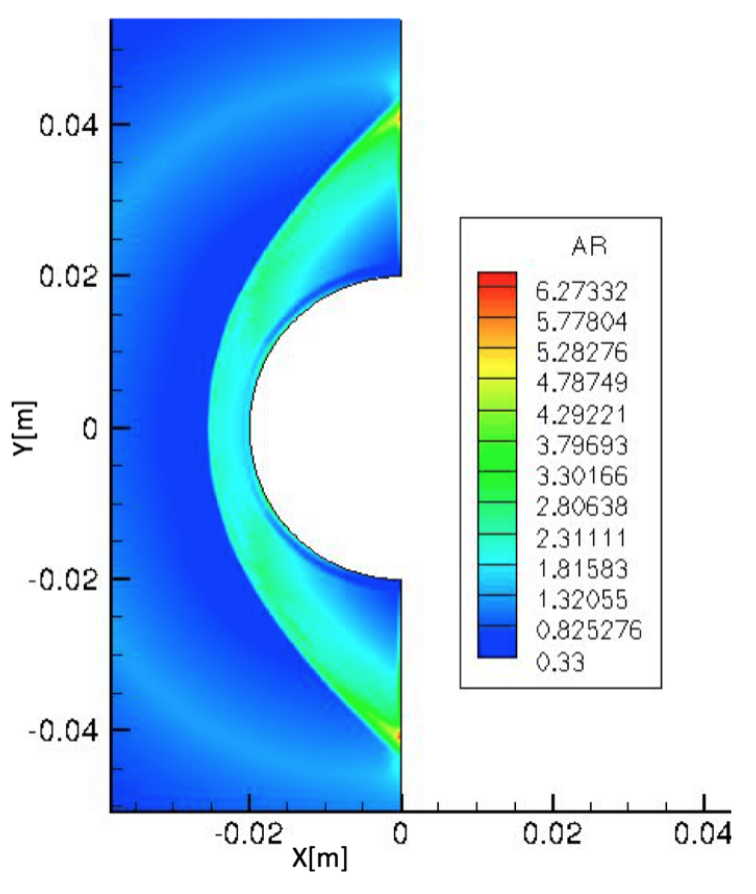}}
\caption{$\mathcal{MQI}$ applied to the Hornung test case-- based on the Aspect Ratio}
\label{fig:AR_start}
\end{figure}

The results of $\mathcal{MQI}$ are shown in Fig.\ref{fig:AR_start}.

\begin{figure}[H]
\centering{\includegraphics[scale=0.4]{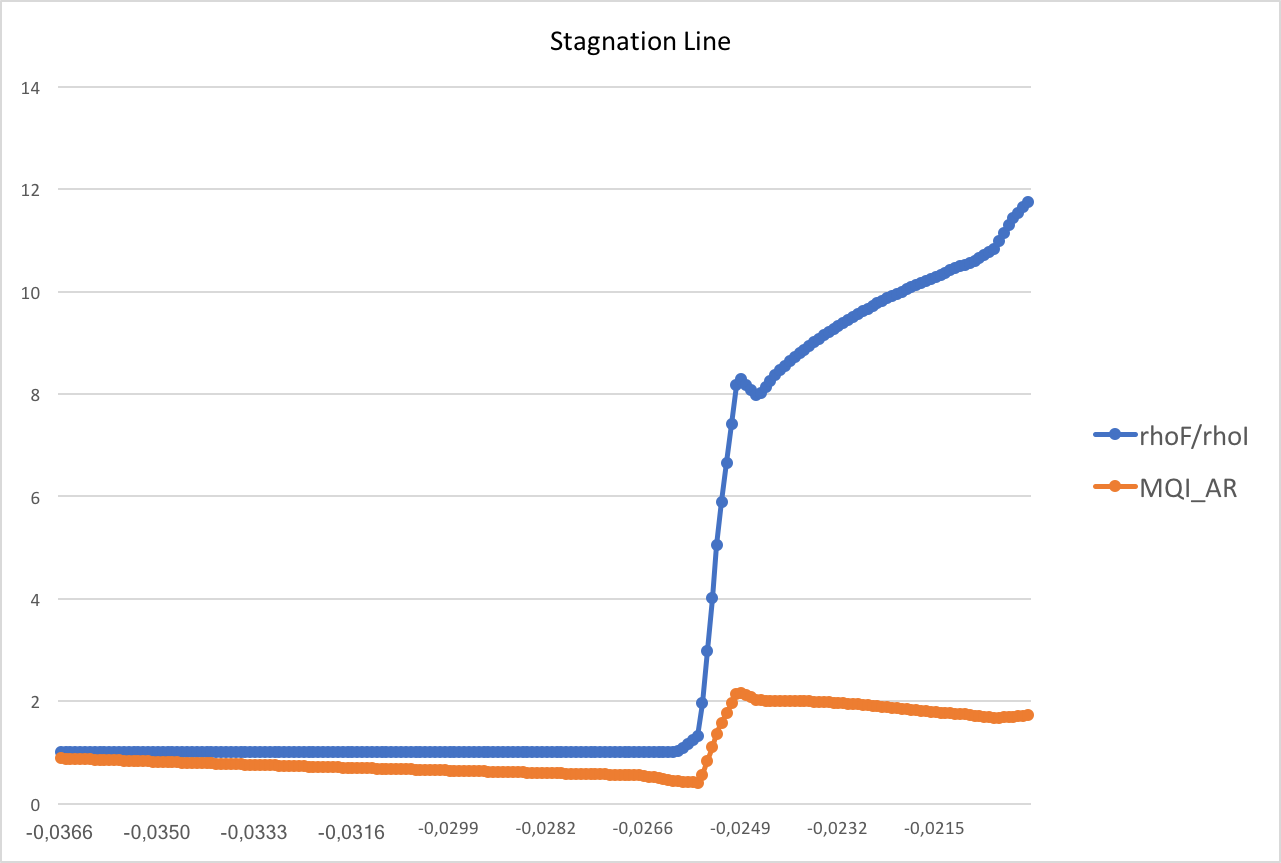}}
\caption{$\mathcal{MQI}$ applied to the Hornung test case -- stagnation line}
\label{fig:AR_0}
\end{figure}

\begin{figure}[H]
\centering{\includegraphics[scale=0.4]{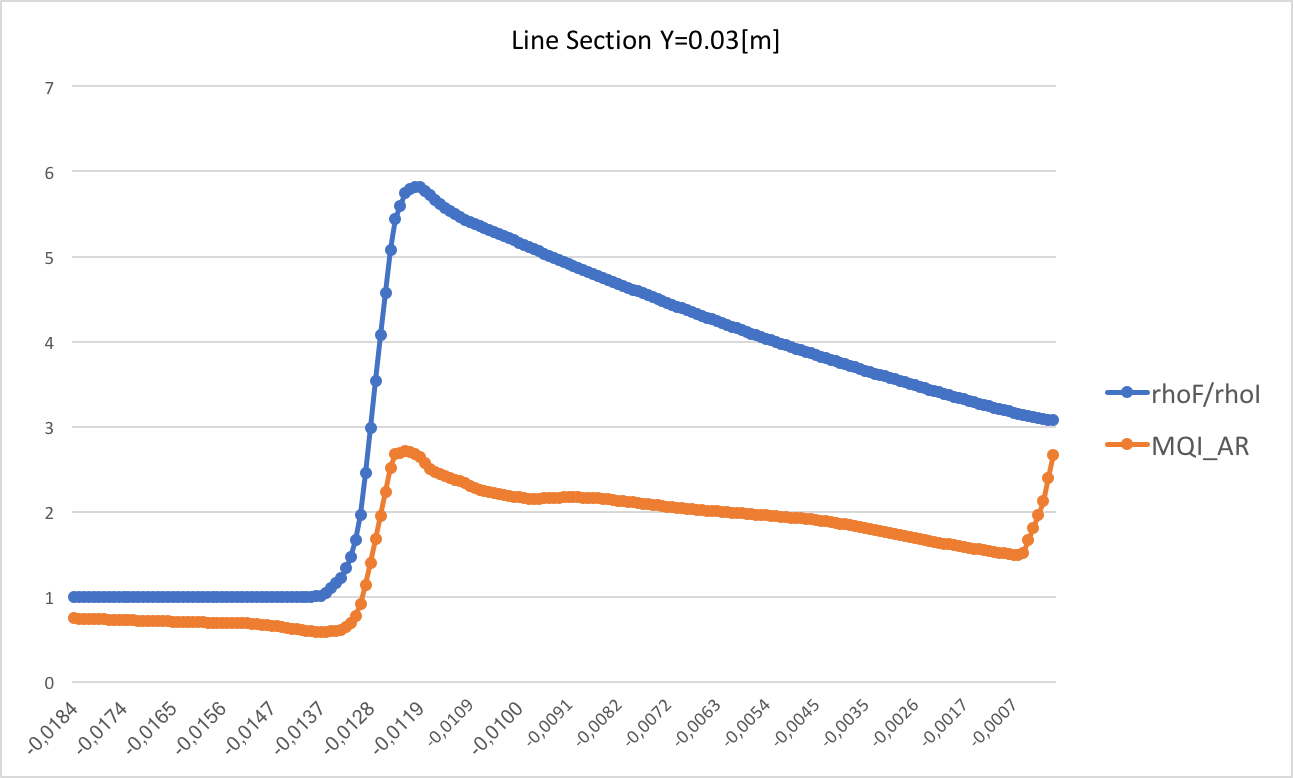}}
\caption{$\mathcal{MQI}$ applied to the Hornung test case -- section Y=0.03[m]}
\label{fig:AR_003}
\end{figure}

The free-stream flow presents, as expected, a value of $\mathcal{MQI}$ $\approx$ 1. One can observe that the value of $\mathcal{MQI}$, at the two tip-end of the bow shock, is too high and presents a maximum red spot. In that region, the mesh elements present a decrease in $\mathcal{AR}$, implying an increase of the ratio $\frac{\mathcal{AR}_{init}}{\mathcal{AR}_{final}}$. Fig.\ref{fig:AR_003} shows both an increase in $\mathcal{MQI}$ at the first jump of the density and close to the outlet. The choice of the distortion criterion at section $Y=0.03[m]$ does not respond to our first hypothesis (see Analysis of MQI in Sec.\ref{point:analysis}), therefore, the quality assessment of the refinement is performed in two steps:
\begin{enumerate}
    \item a mesh quality indicator based on the aspect ratio is adopted to estimate the mesh quality close to the stagnation line where the quadrilateral mesh elements are compressed or enlarged due to the AMR process;
    \item a mesh quality indicator based on the skewness distortion criteria $\Theta$ is adopted, since, the quadrilateral mesh element at the bow shock's tips are skewed.
\end{enumerate}
Fig.\ref{fig:AR_0} shows an expected behaviour whereas Fig.\ref{fig:AR_003} present an increase of the $\mathcal{MQI}$ values instead of a steady behaviour close to the outlet.

\begin{figure}[H]
\centering{\includegraphics[scale=0.4]{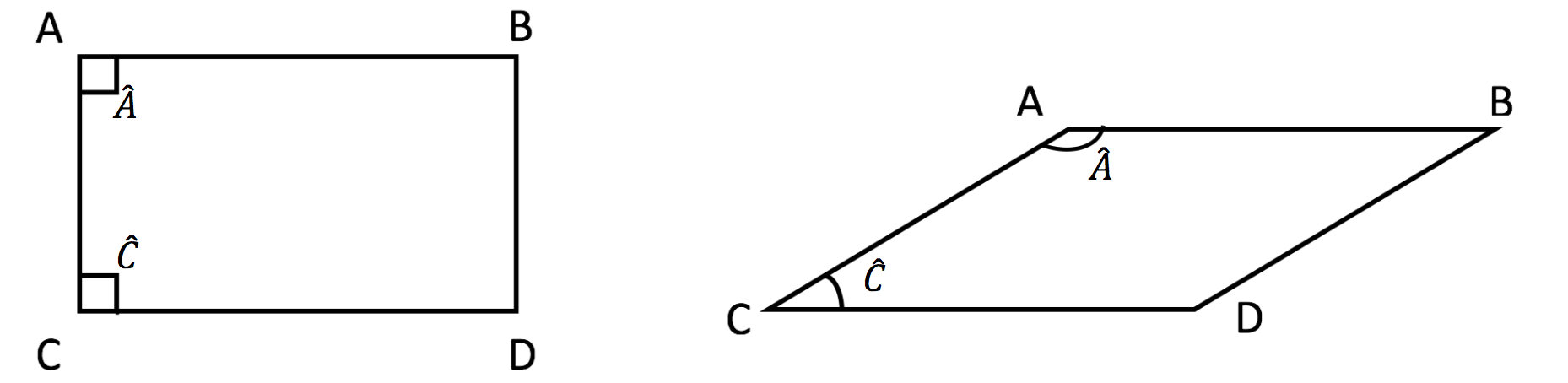}}
\caption{Left: Initial cell-- Right: Distorted cell}
\label{fig:skew}
\end{figure}

Eq.\ref{eq:quality} is transformed into
\begin{equation}
     \mathcal{MQI}= \Delta \Theta ~ \frac{\mathcal{S}_{final}}{\mathcal{S}_{init}}, \quad \text{where}
\end{equation}

$$
    \Delta \Theta \left\{
   \begin{array}{ll}
         =0,  \mbox {       if the cell keeps the same shape. }\\
         else,  \mbox {      if the cell becomes narrow.}\\
    \end{array}
\right.
$$\\
For a quadrilateral element $ABDC$, the skewness of an element is computed through the following formula \cite{skew}:
\begin{equation}
     \Theta = max[\frac{\alpha_{max}-\alpha_{ref}}{180^\circ-\alpha_{ref}},\frac{\alpha_{ref}-\alpha_{max}}{\alpha_{ref}} ],
\end{equation}
where $\alpha_{ref}$ =90$^\circ$ for a quadrilateral element, $\alpha_{max}$ and $\alpha_{min}$ are respectively the maximum and minimum angle in the quadrilateral element.
The extrapolation to a nodal value is done by averaging all the element's skewness of the $N$ elements attached to the considered vertex $i$:
\begin{equation}
     \Theta_i = \frac{1}{N}\sum_{m=1}^{N} \Theta^m.
\end{equation}
\underline{Hornung}

\begin{figure}[H]
\centering{\includegraphics[scale=0.35]{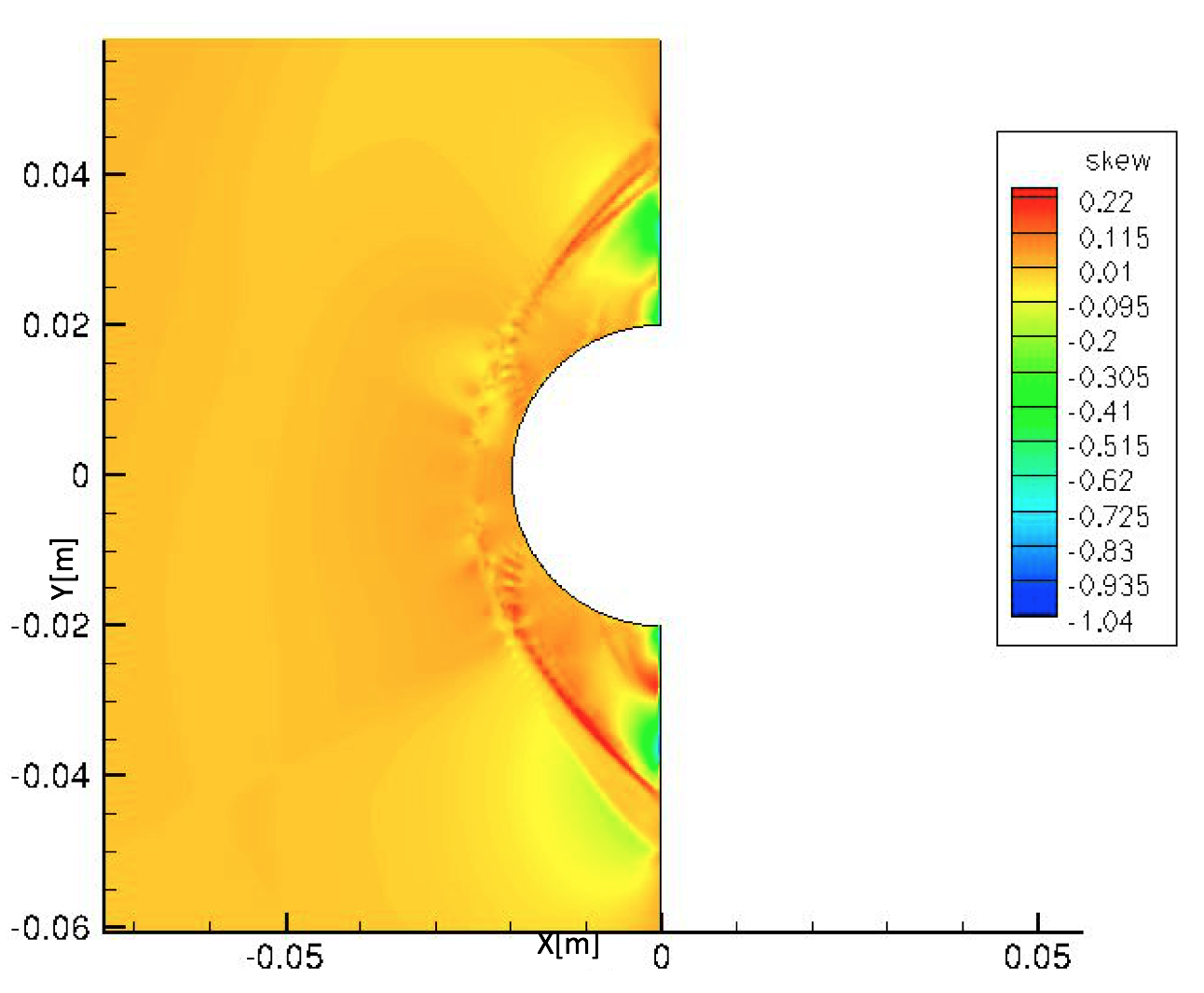}}
\caption{$\mathcal{MQI}$ applied to the Hornung test case--Skewness based}
\label{fig:skew_Hornung}
\end{figure}

\begin{figure}[H]
\centering{\includegraphics[scale=0.4]{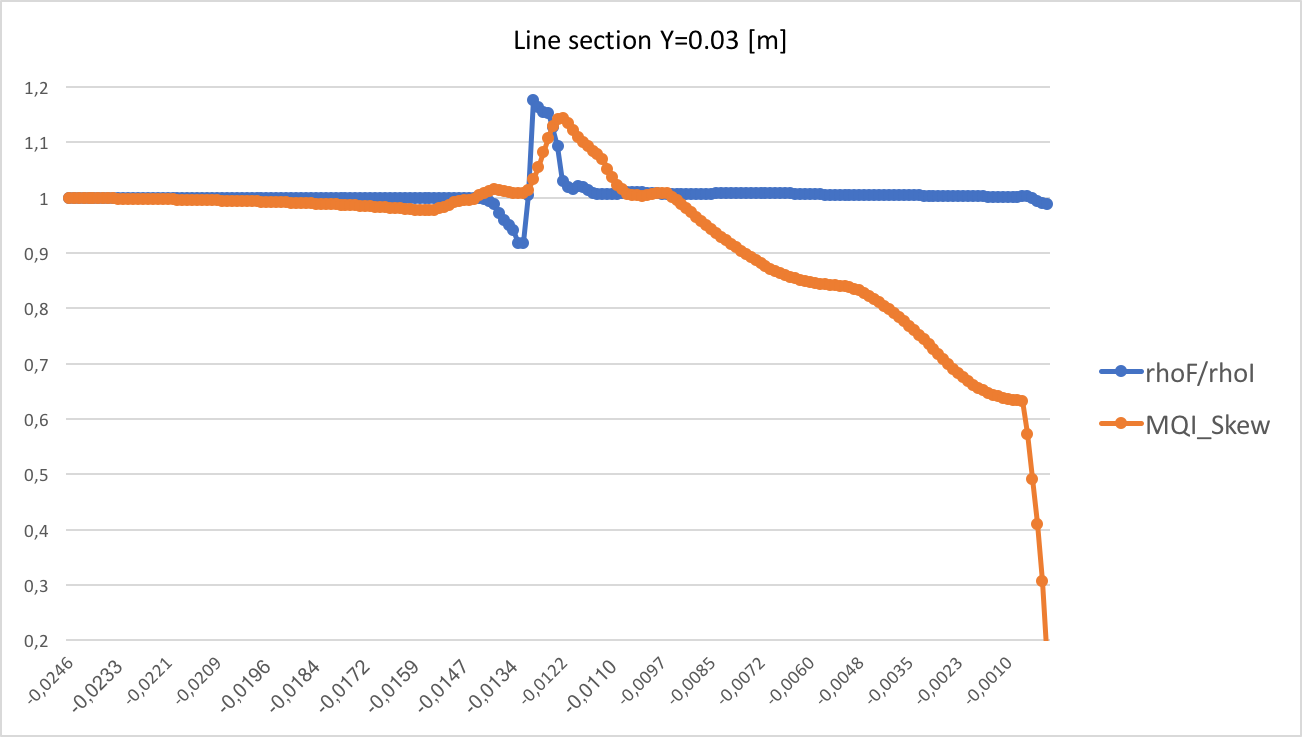}}
\caption{$\mathcal{MQI}$ applied to the Hornung test case $Y=0.03[m]$ }
\label{fig:skew_Hornung_sec}
\end{figure}

Fig.\ref{fig:skew_Hornung_sec} shows the $\mathcal{MQI}$ based on $\Theta$ for the Hornung test case. In order to have the same constant $\mathcal{C}$ for the free stream flow for all the cases, an improved $\mathcal{MQI}$ based on the skewness of the quadrilateral element is proposed:
\begin{equation}
     \mathcal{MQI}= 1+ \Delta \Theta ~ \frac{\mathcal{S}_{final}}{\mathcal{S}_{init}}.
\end{equation}

Fig.\ref{fig:skew_Hornung} presents more accurate interpretation of the $\mathcal{MQI}$ with respect to the curves in Fig.\ref{fig:AR_003}. In fact, the $\mathcal{MQI}$ shows a peak at the first density jump and skewed elements at the post-shock region indicating the cell alignment with the density field.\\

\begin{figure}[H]
\centering{\includegraphics[scale=0.43]{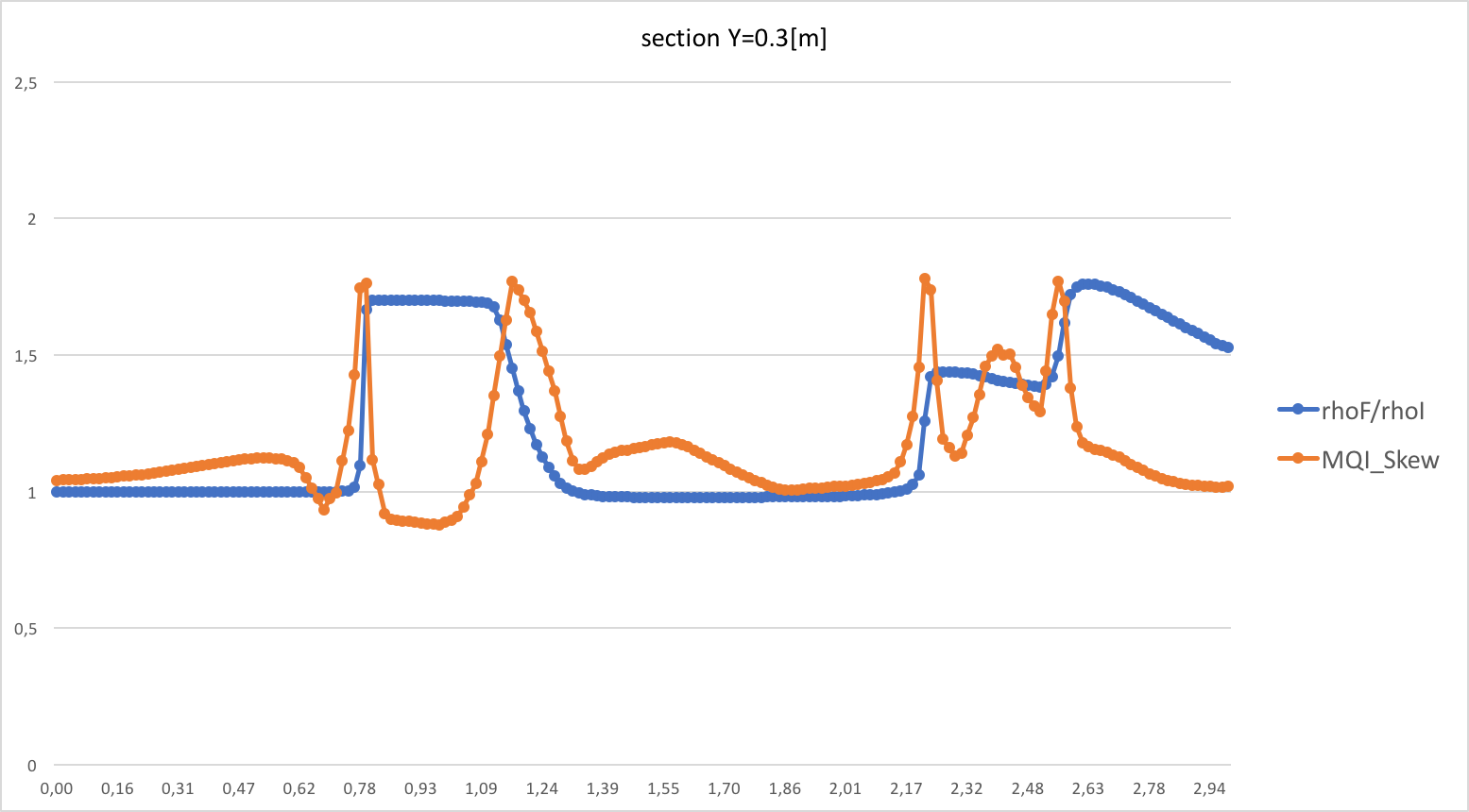}}
\caption{$\mathcal{MQI}$ applied to the double wedge test case-- $Y=0.3[m]$}
\label{fig:wedgeQ03}
\end{figure}

\begin{figure}[H]
\centering{\includegraphics[scale=0.5]{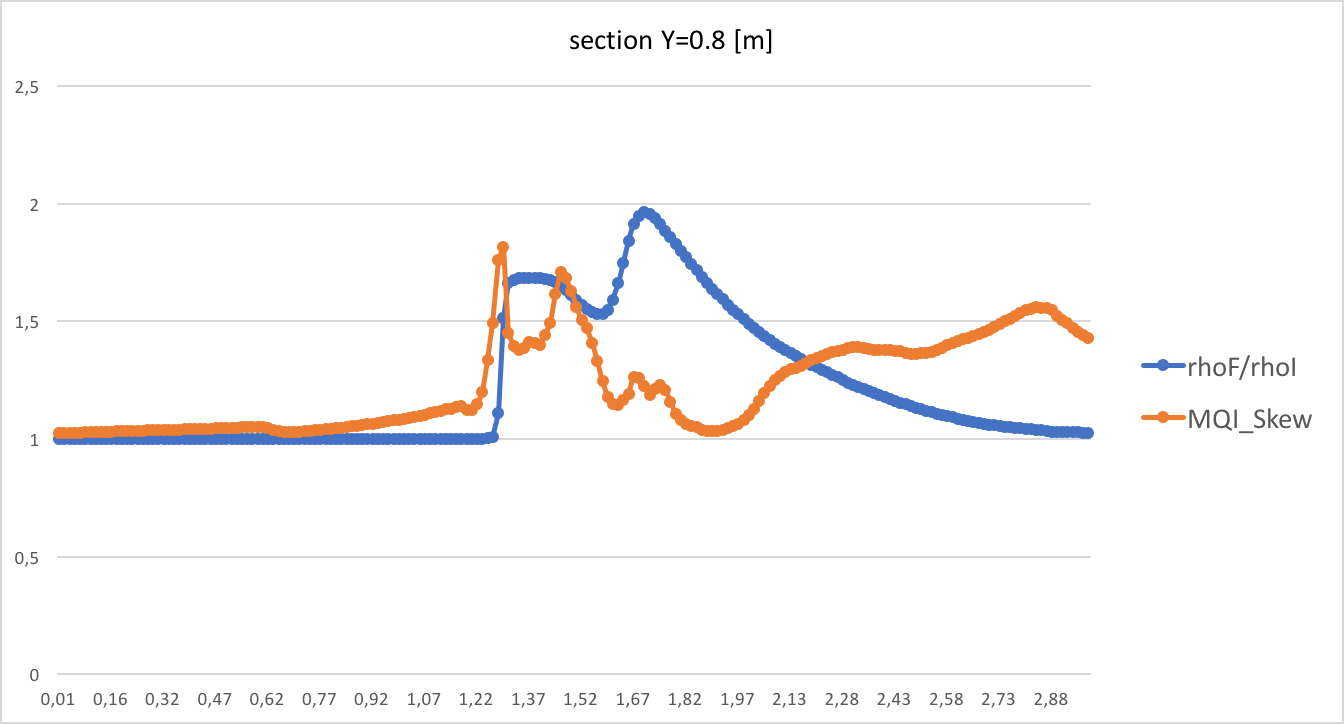}}
\caption{$\mathcal{MQI}$ applied to the double wedge test case-- $Y=0.8[m]$}
\label{fig:wedgeQ08}
\end{figure}

\underline{Double wedge quadrilateral mesh}\\
Fig.\ref{fig:wedgeQmesh} shows the mesh refinement final results based on the linear spring analogy where the flow conditions are presented in Tab.\ref{tab:DWflowchar}. The mesh quality indicator based on skewness element is discussed hereby:
\begin{figure}[H]
\centering{\includegraphics[scale=0.4]{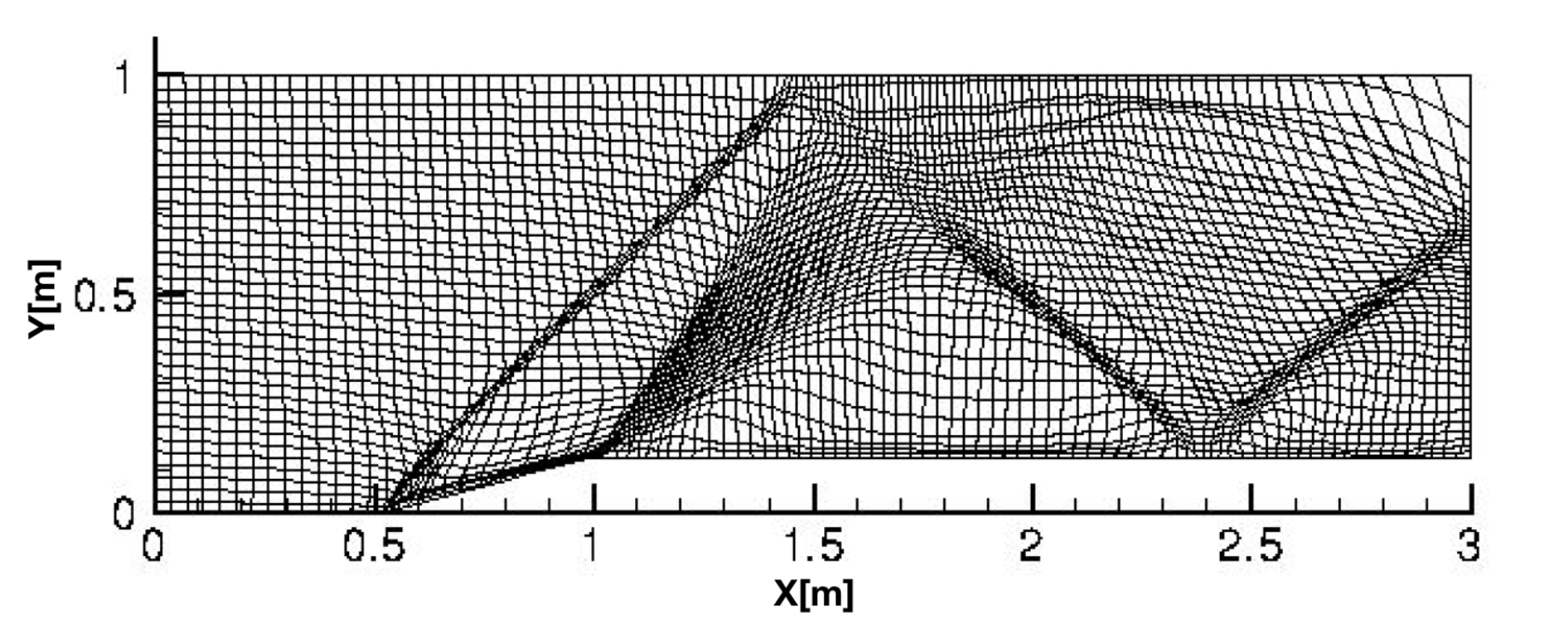}}
\caption{Final mesh- double wedge quadrilateral mesh}
\label{fig:wedgeQmesh}
\end{figure}
The free stream flow provides a $\mathcal{MQI}$=$0$. Fig.\ref{fig:wedgeQ03} and Fig.\ref{fig:wedgeQ08} show peaks at density discontinuities. The increase of $\mathcal{MQI}$ for nodes $\in$ [2, 3] at the section $Y=0.8[m]$ can be explained by the fact that the nodes are contributing to the refinement of the expansion shock and second oblique shock reflection near the outlet boundary.

\begin{figure}[H]
\centering{\includegraphics[scale=0.4]{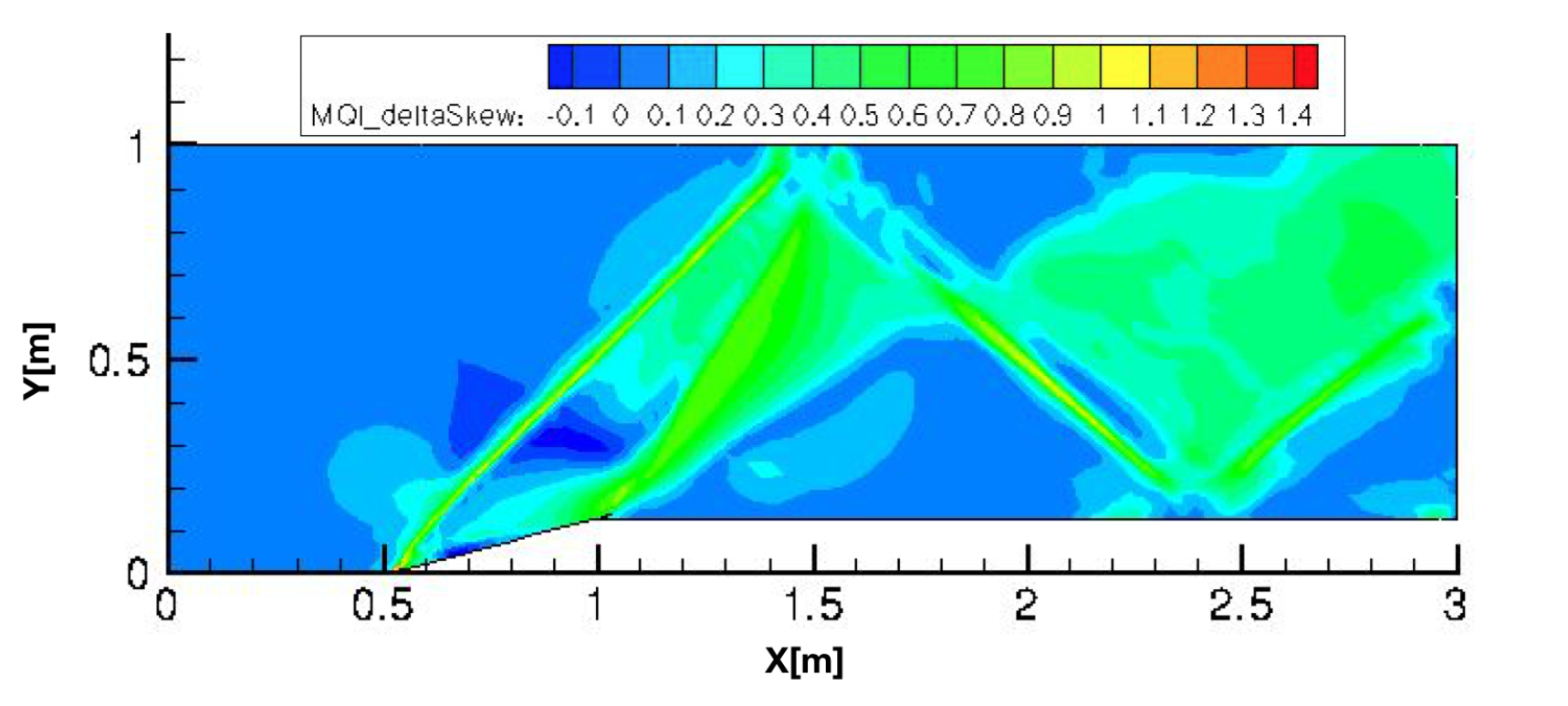}}
\caption{$\mathcal{MQI}$ applied to the double wedge test case}
\label{fig:wedgeQMQI}
\end{figure}


\subsection{MQI applied to 3D meshes}
\subsubsection{3D tetrahedral}
The concept of the inserted circle in a triangle is extended to the 3D tetrahedral element. The cell distortion criteria is defined as the radius of the inscribed sphere inside the tetrahedron, denoted as $\mathcal{R}^{S}$.\\
Eq.\ref{eq:quality} will be transformed into:
\begin{equation}
    \mathcal{MQI} = \frac{\mathcal{R}^{S}_{final}}{\mathcal{R}^{S}_{init}} ~ \frac{\mathcal{S}_{final}}{\mathcal{S}_{init}}.
\end{equation}
The ratio $\frac{\mathcal{R}^{S}_{final}}{\mathcal{R}^{S}_{init}}$ is investigated in the following:
$$
   \frac{\mathcal{R}^{S}_{final}}{\mathcal{R}^{S}_{init}} \left\{
    \begin{array}{ll}
         =1,  \mbox {      if the cell keeps the same shape; }\\
         <1,  \mbox {      if the cell becomes smaller;}\\
        >1,  \mbox {      if the cell is enlarged.}
    \end{array}
\right.
$$\\
For a tetrahedral element $sijk$, the in-radius $\mathcal{R}^{S}$ is determined through the following Eq.\ref{eq:SRadius}, as expressed in \cite{Rin}:
\begin{equation}
    \label{eq:SRadius}
    \mathcal{R}^{S}=\frac{3 V_{sijk}}{A_{ijk}+A_{sij}+A_{sik}+A_{sjk}},
\end{equation}
where  $A_{sijk}$ denotes the volume of the $sijk$ and $A_{ijk}$ denotes the area of the triangle $ijk$. The extrapolation to a nodal value is done by averaging all the in-sphere radius of the $N$ tetrahedron attached to considered the vertex $i$:
\begin{equation}
\label{Extrapolation2}
    \mathcal{R}^{S}_{i}=\frac{1}{N}\sum_{m=1}^{N}\frac{3 V_{sijk}}{A_{ijk}+A_{sij}+A_{sik}+A_{sjk}}.
\end{equation}
\underline{Hemisphere}\\
The $\mathcal{MQI}$ values are shown in Fig.\ref{fig:MQI000}:
\begin{figure}[H]
\centering{\includegraphics[scale=0.45]{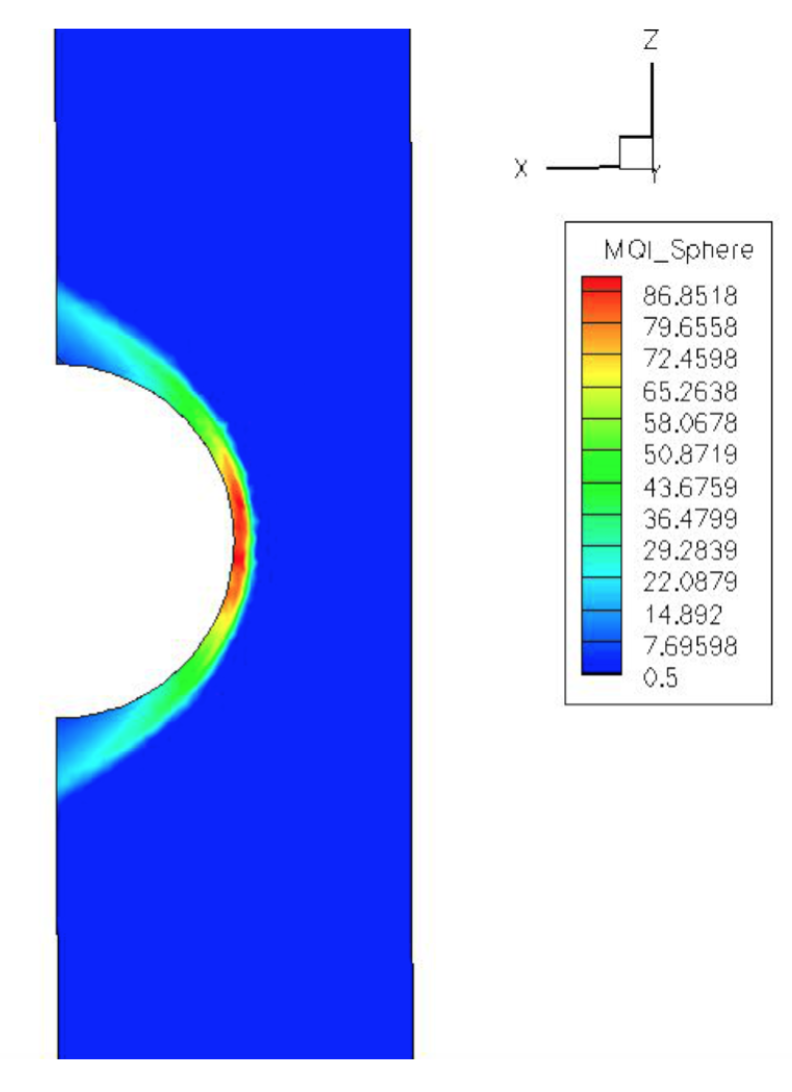}}
\caption{$\mathcal{MQI}$ applied to Hemisphere test case section $Y=0$}
\label{fig:MQI000}
\end{figure}

\begin{figure}[H]
\centering{\includegraphics[scale=0.45]{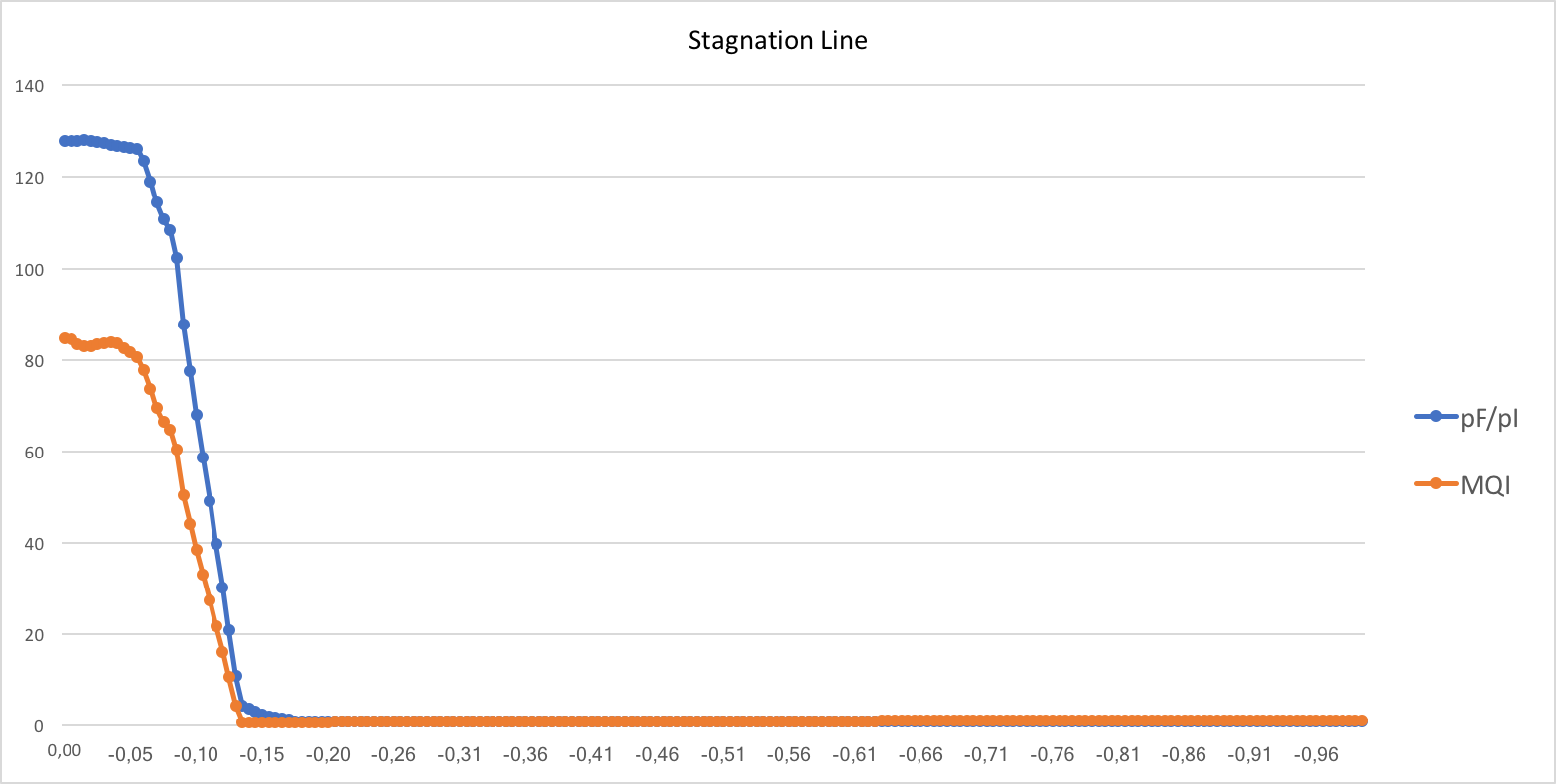}}
\caption{Line section $Y=0, Z=0$}
\label{fig:MQI_Hemisp}
\end{figure}

\begin{figure}[H]
\centering{\includegraphics[scale=0.4]{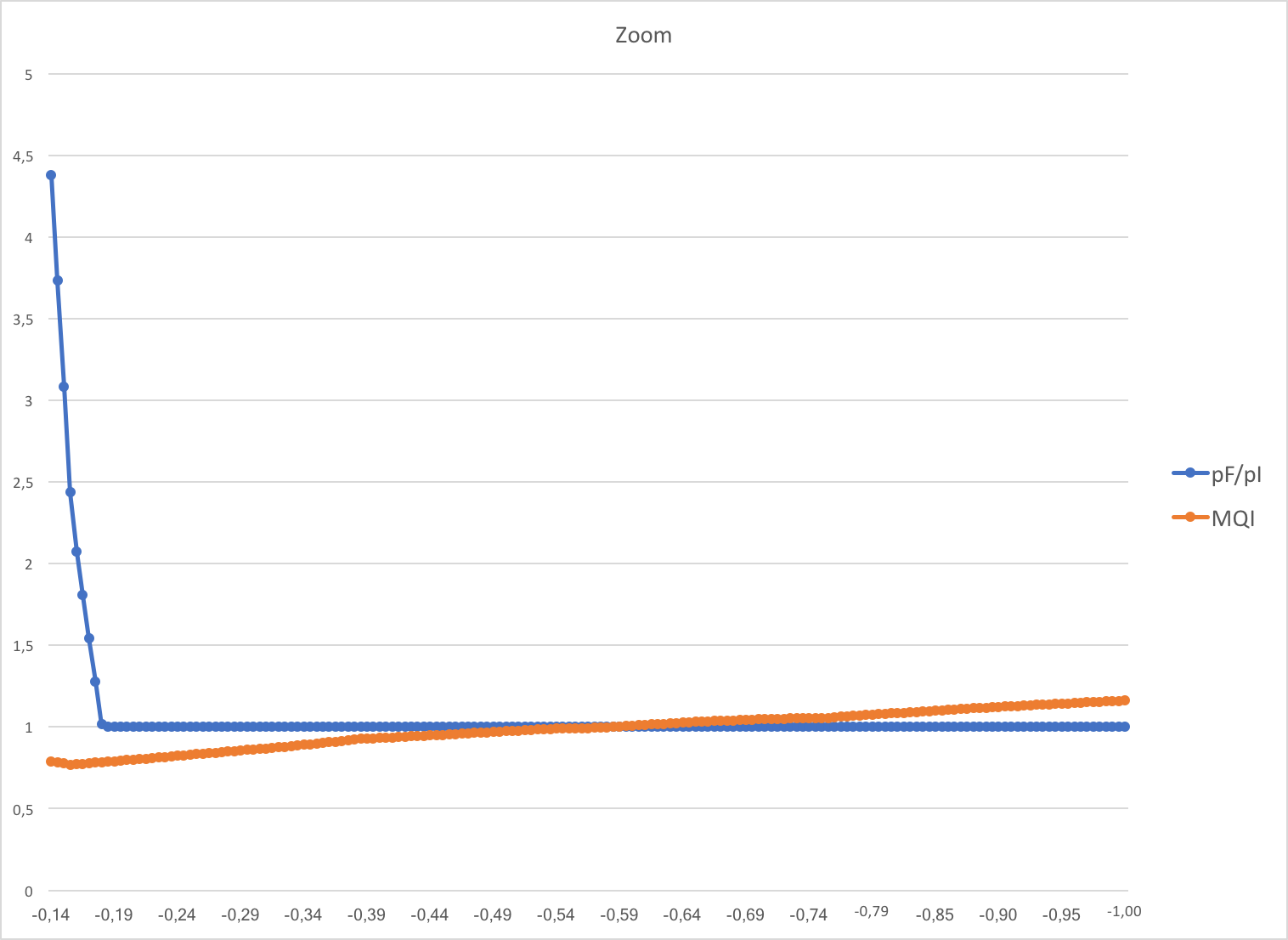}}
\caption{Line section $Y=0, Z=0$ - zoom}
\label{fig:MQI_Hemip_zoom}
\end{figure}
The $\mathcal{MQI}$ free stream value $\approx$ 1 (see Fig.\ref{fig:MQI_Hemisp}). In fact, the cells, in the free stream, are enlarged to contribute to the refinement of the bow shock. From Fig.\ref{fig:MQI_Hemisp}, one can observe that the cells close to discontinuities becomes smaller to increase the local mesh node density.
\subsection{Advantages and Drawbacks of MQI}
The $\mathcal{MQI}$ presents several advantages, to name only a few:
\begin{itemize}
    \item Combines both local physical and geometrical properties;
    \item Applicable to 2D and 3D;
    \item $\mathcal{MQI}$=1 at free stream flow;
    \item  Allows for grading an adapted mesh and can be used to capture shocks;
    \item The $\mathcal{MQI}$ peaks return the shock intensity and reflect, in general, the cells mesh distortion.
\end{itemize}
Also, the $\mathcal{MQI}$ presents some drawbacks:
\begin{itemize}
    \item Mesh-type dependent;
    \item While the solution states and distortions criteria are cell-based for the Finite Volume solver, the $\mathcal{MQI}$ is nodal based and the extrapolation can introduce some errors.
\end{itemize}


\section{Refinement Stop Indicator}
\label{sec:RSI}
This section presents a new method to help decide whether or not terminate the mesh refinement process qualitatively.
In r-adaptive steady-state simulations, the residuals (i.e. $L_{2}$ norms of some monitor quantities which are used to determine the iterative convergence of the flow solver) will be affected by some fluctuations due to nodal re-positioning implying an increase in the computational time and memory cost. At each mesh fitting process, the flow solver is seeing a new mesh, and therefore the residuals will increase when applying the mesh fitting. 
The Author's key idea is based on defining a certain user-defined tolerance on the mesh movement, denoted $\epsilon$, and a measure of the relative mesh movement criterion, denoted $\delta$. The Refinement Stop Indicator, denoted $\mathcal{RSI}$, will be a function of $\delta$.
The $\epsilon$ and  $\delta$ definitions yield to:
$$
   \mathcal{RSI}= f(\delta) \left\{
    \begin{array}{ll}
          > \epsilon,  \mbox { Continue the mesh refinement $\Rightarrow$ the mesh is not stable;}\\
         \le \epsilon,  \mbox { The mesh can be considered stable $\Rightarrow$ Stop the mesh refinement.}\\
    \end{array}
\right.
$$\\
Many challenges arise when defining the $\mathcal{RSI}$ and $\delta$ due to the complex nature of the problem: number of moving nodes, relative displacement magnitude etc...\\
The following empirical formula is proposed
\begin{equation}
\label{RSI complex}
    \mathcal{RSI}=\frac{1}{N} (\sum_i^{m_{up}} A_i^{x_i} \sqrt(\delta_i^{up}) + \sum_i^{m_{down}} B_i \delta_i^{down}),
\end{equation}
where:
\begin{itemize}
    \item $N$ is the number of mesh nodes,
    \item $\delta_i^{up}$ the relative displacement of nodes > $\epsilon$,
    \item $\delta_i^{down}$ the relative displacement of nodes $\le \epsilon$,
    \item $A_i$ the number of nodes having a relative displacement $\delta_i^{up}$,
    \item $m_{up}$ (resp. $m_{down}$) the number of nodes having a relative displacement >$\epsilon $ (resp. $\le$ $\epsilon $),
    \item $B_i$ the number of nodes having a relative displacement $\delta_i^{down}$,
\item $x_i$: the numerical contribution of the value $A_i$ will be affected by this exponent:
\begin{enumerate}
    \item $x_i$ needs to increase when $A_i$ increases and $\delta_i^{up}$ decreases yet still > $\epsilon$. \label{1}
    \item $x_i$ needs to further increase with respect to point (\ref{1}) when $A_i$ is small and $\delta_i^{up}$ is high.
    \item $x_i$ needs to be small (resp. high) when  $A_i$ and $\delta_i^{up}$ are small (resp. high).
    \end{enumerate}
\end{itemize}
Hence, the exponent $x_i$ becomes:
\begin{equation}
    x_i= \sqrt A_i \delta_i^{up}.
\end{equation}
To simplify Eq.\ref{RSI complex}, one can define the following parameters:
\begin{itemize}
    \item $A$ the number of all nodes having a relative displacement > $\epsilon$,
    \item $B$ the number of all nodes having a relative displacement < $\epsilon$,
    \item $\delta^{up}$ the average of $\delta_{i}^{up}$,
    \item $\delta^{down}$ the average of $\delta_{i}^{down}$,
    \item $x= \sqrt A \delta^{up}$.
\end{itemize}
Therefore, Eq.\ref{RSI complex} becomes:
\begin{equation}
    \label{eq:RSI average}
    \mathcal{RSI}=\frac{1}{N} (A^{x} \sqrt(\delta^{up}) +B \delta^{down}).
\end{equation}
To make this approach more robust, the user also defines the iteration (a.k.a. Trigger $\mathcal{RSI}$) at which the $\mathcal{RSI}$ computation will start. The purpose of such value is to distinguish between the case of a stable mesh and a case where shocks are slowly developing and detaching. This value will depend on how fast the simulation is developing.\\

\subsection{RSI applied to 2D mesh}

\subsubsection{2D Quadrilateral mesh}
The relative displacement will be based on a cell distortion criterion specific to a quadrilateral element. The choice is made based on $\mathcal{AR}$ to compute $\mathcal{RSI}$ since one is only interested in the nodal displacement and not the distortion of a cell. In addition, the $\mathcal{AR}$ will be extrapolated to a nodal value $i$ at time steps $n$ and $n+m$, where the mesh is updated every m field flow iteration:
\begin{equation}
    \delta_i = \frac{|\mathcal{AR}_i^{n+1}-\mathcal{AR}_i^{n}|}{\mathcal{AR}_i^{n}}.
\end{equation}\\

\begin{figure}[H]
\centering
        \captionsetup{justification=centering}

\begin{minipage}{.45\linewidth}
  \includegraphics[width=\linewidth]{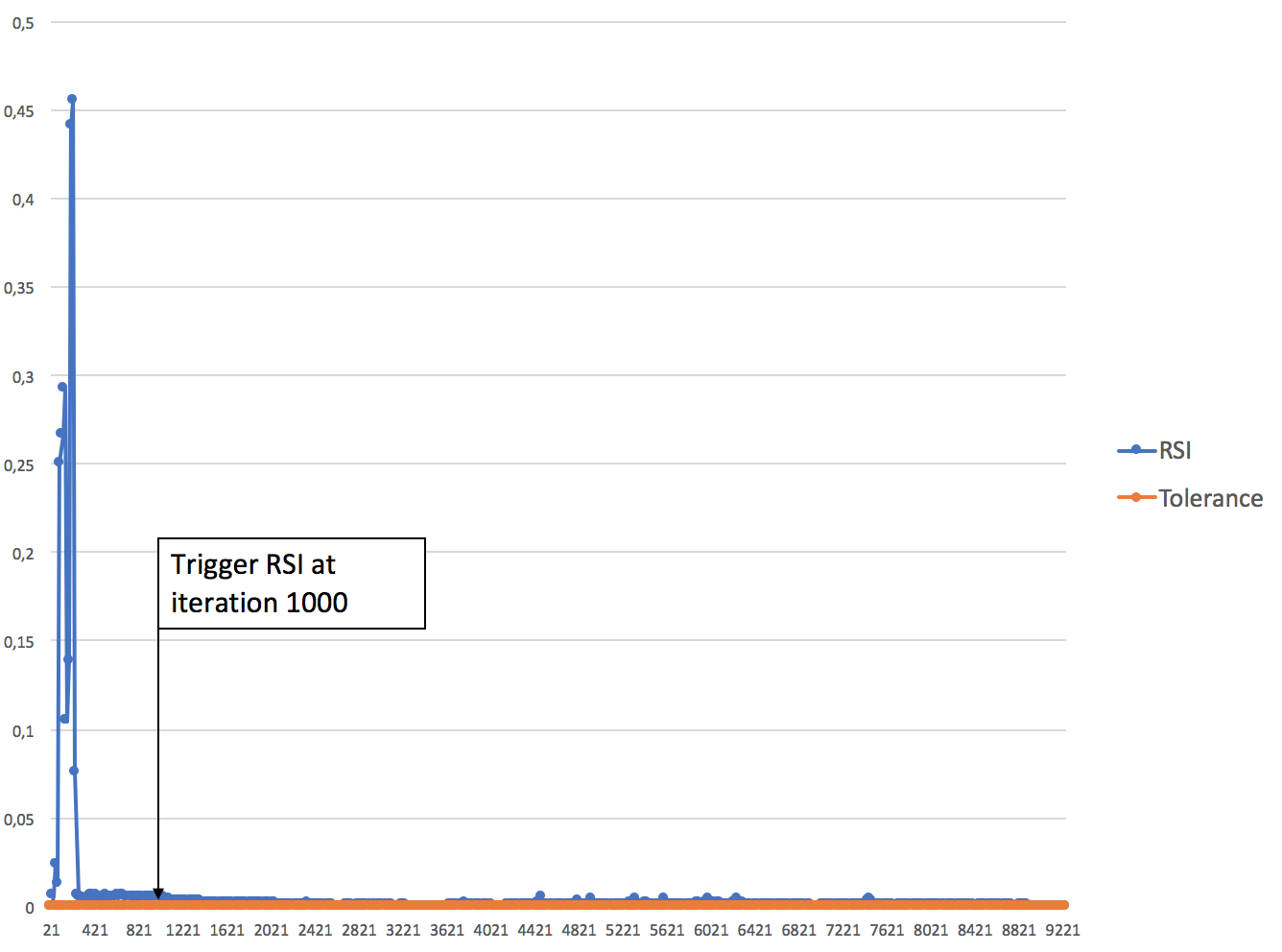}
  \caption{$\mathcal{RSI}$ in function of the number of iterations}
  \label{RSIquads}
\end{minipage}
\hspace{.05\linewidth}
\begin{minipage}{.45\linewidth}
  \includegraphics[width=\linewidth]{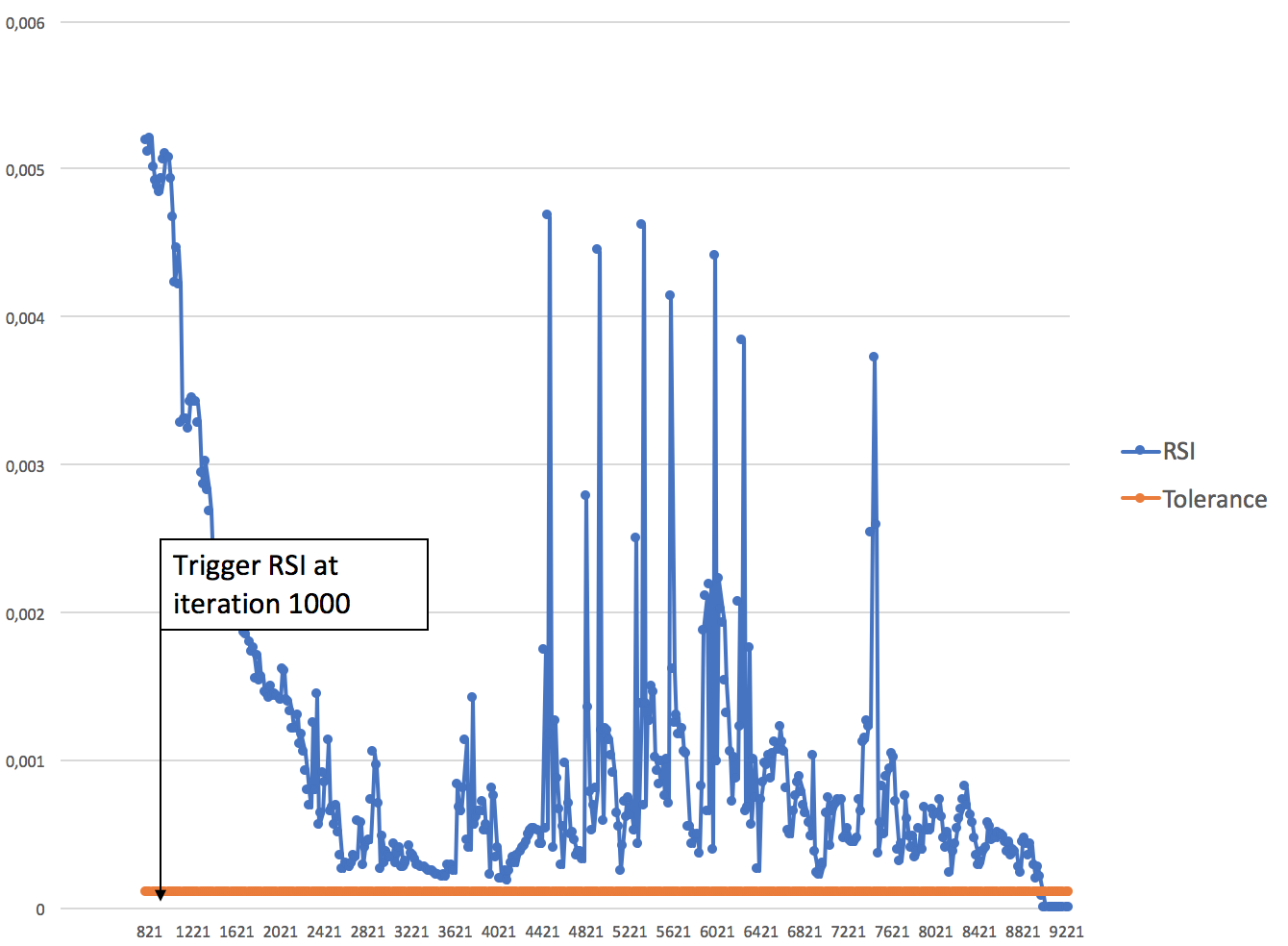}
  \caption{$\mathcal{RSI}$ in function of the number of iterations- zoom}
  \label{zoomRSIquads}
\end{minipage}
\end{figure}

\underline{2D double wedge quadrilateral test case}\\
The user-defined tolerance is chosen to be equal to $\epsilon$=0.01\%.
The mesh fitting process continues till the iteration 2981, as Fig.\ref{RSIquads} and Fig.\ref{zoomRSIquads} show. When $\mathcal{RSI}$<$\epsilon$, the mesh fitting process stops enabling a fast convergence. In the convergence history, we can observe that the oscillation disappears when stopping the refinement.

\begin{figure}[H]
        \captionsetup{justification=centering}

    \begin{minipage}[t]{6cm}
        \centering
        \includegraphics[width=6cm]{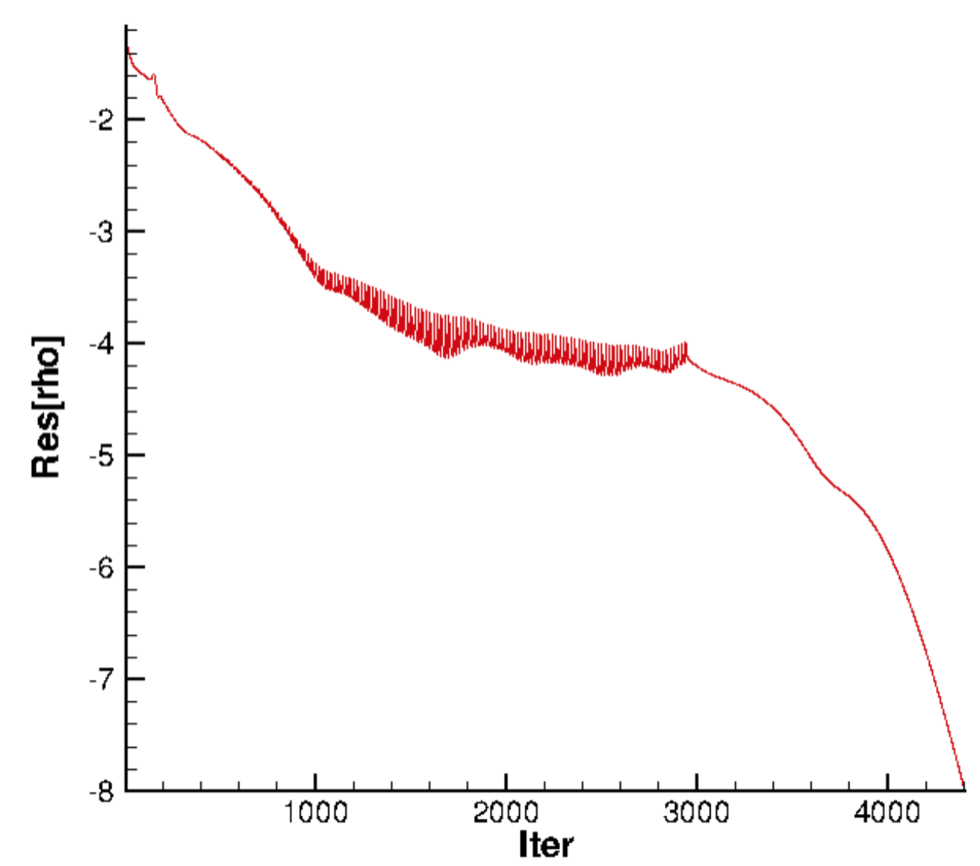}
        \caption{Convergence history with $\mathcal{RSI}$}
        \label{fig:convQuads}
    \end{minipage}
    \begin{minipage}[t]{6cm}
        \centering
        \includegraphics[width=6cm]{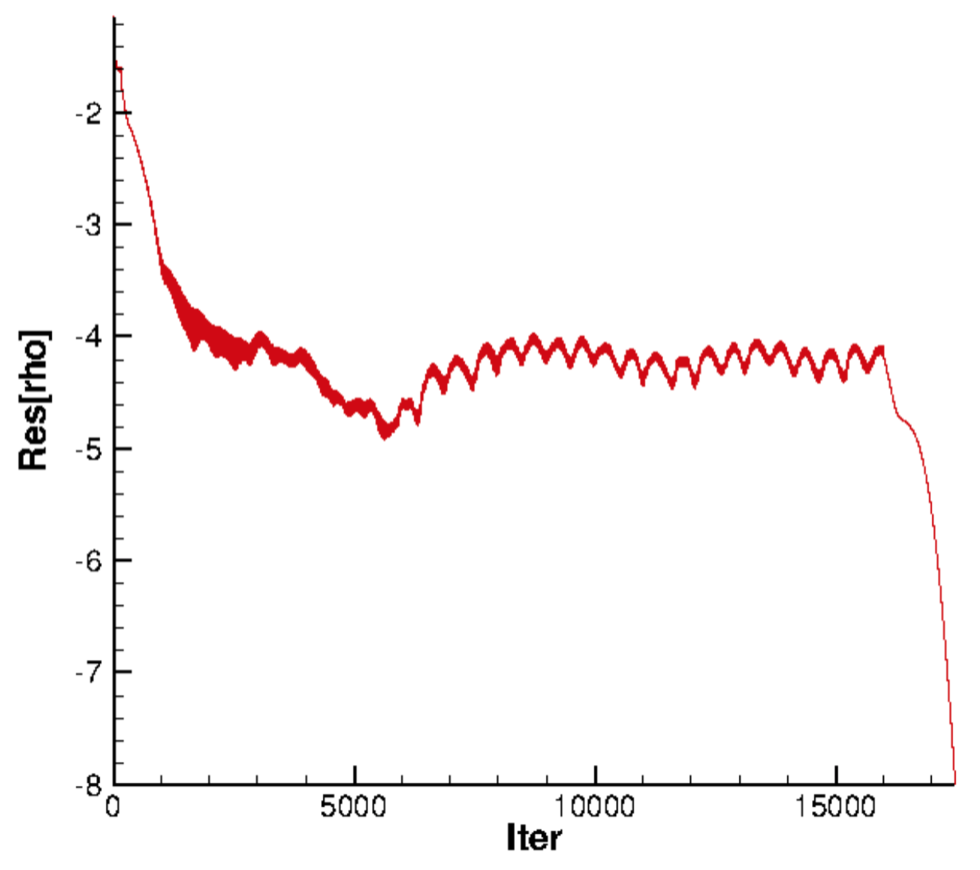}
        \caption{Convergence history with a classical stop condition iter=16000}
        \label{fig:conQuads16000}
    \end{minipage}
\end{figure}

Fig.\ref{fig:convQuads} shows the gain in convergence when using $\mathcal{RSI}$ compared to the same simulation using a classical stop condition (see Fig.\ref{fig:conQuads16000}). The implementation of the refinement stop indicator clearly influences the convergence rate positively.

\subsubsection{2D triangular mesh}
\underline{2D double wedge triangular test case }\\
The relative displacement is based on the in-circle radius of the triangular element extrapolated to the nodal value $i$ at mesh fitting time steps $n$ and $n+m$.
\begin{equation}
\label{eq:delta_i_RSI}
    \delta_i = \frac{|\mathcal{R}_i^{n+m}-\mathcal{R}_i^{n}|}{\mathcal{R}_i^{n}}.
\end{equation}

\begin{figure}[H]
        \captionsetup{justification=centering}

    \begin{minipage}[t]{6cm}
        \centering
        \includegraphics[width=6cm]{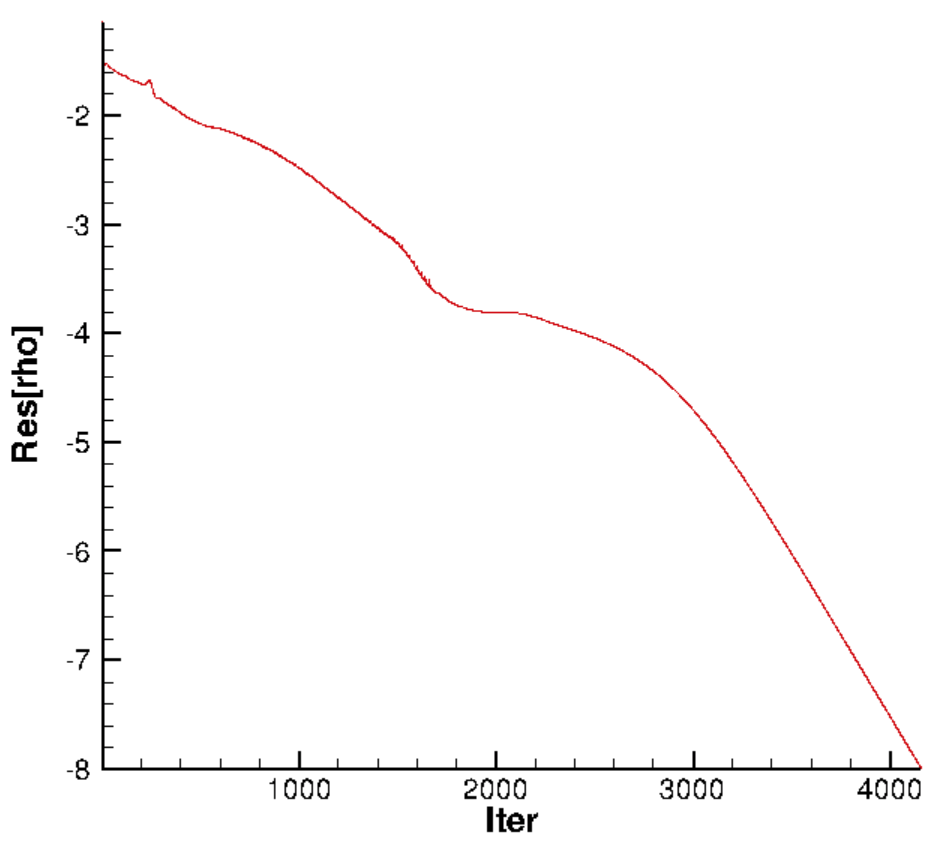}
        \caption{Convergence history with $\mathcal{RSI}$}
        \label{fig:convWithRSI}
    \end{minipage}
    \begin{minipage}[t]{6cm}
            \captionsetup{justification=centering}

        \centering
        \includegraphics[width=6cm]{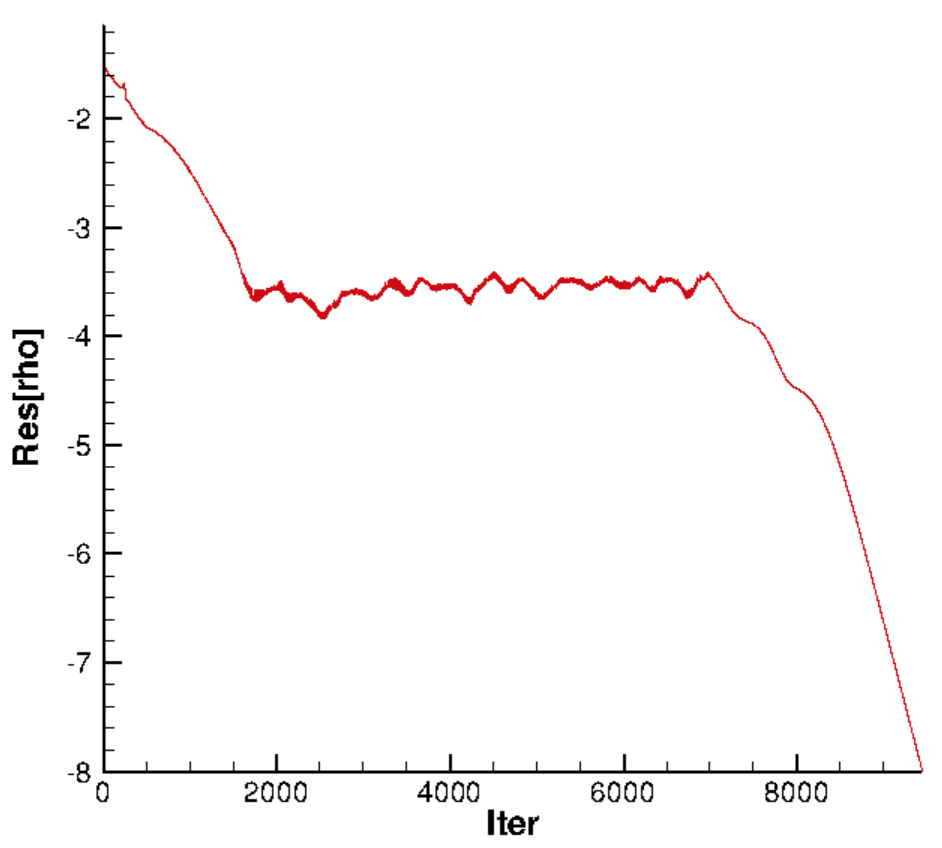}
        \caption{Convergence history with a classical stop condition iter=7000}
        \label{fig:StopTriangular7000}
    \end{minipage}
\end{figure}

Fig.\ref{fig:convWithRSI} shows the advantage of using the $\mathcal{RSI}$ as a stop condition for the mesh fitting process. The convergence is reached twice faster when $\mathcal{RSI}$ is applied (see Fig.\ref{fig:StopTriangular7000}), also thanks to the disappearance of the fluctuations due to the small nodal displacement.

\subsubsection{Advantages and Drawbacks of RSI}
The $\mathcal{RSI}$ presents several advantages, for instance:
\begin{itemize}
    \item Provides only one value, therefore being easy to monitor;
    \item Allows for reducing the run-time cost while automatizing the refinement process till convergence;
    \item Accelerates the convergence to steady state and limits the fluctuations of the convergence history;
    \item Can be finely tuned by a user-defined tolerance.
\end{itemize}
Also, the $\mathcal{RSI}$ presents some drawbacks:
\begin{itemize}
    \item Mesh-type dependent;
    \item Relies on an empirical formula that may need further adjustment of the coefficients depending on the case.
\end{itemize}

\section{Conclusion}
\label{sec:conclusion}
A novel physics-based r-refinement has been developed and, in combination with an existing Finite Volume CFD solver, successfully applied to several high-speed and space plasmas test cases based on multiple spring concepts mainly linear, semi-torsional and orth-semi-torsional spring analogies for two- and three-dimensional flows. Our AMR solver showed its ability to resolve different flow features depending on the user-defined monitored variable. This work also introduced and showed the potential of a newly defined mesh quality indicator to grade an adapted mesh qualitatively. Finally, computational improvements and simulation speed-up have been demonstrated through the use of a proposed refinement stop indicator. 


\end{document}